\documentclass[twocolumn]{aa}

\usepackage{natbib}
\bibpunct{(}{)}{;}{a}{}{,} % to follow the A&A style

\usepackage[table,xcdraw]{xcolor}
\usepackage{float}
\usepackage{amsmath}

\usepackage{graphicx}
\usepackage{subfigure}
\usepackage{color}
\usepackage{fancyvrb}

\definecolor{Blue}{rgb}{0.1,0.1,1.0}
\definecolor{Magenta}{rgb}{1.0,0.1,0.5}
\definecolor{LRed}{rgb}{0.8,0.0,0.0}

\newcommand{\nc}{\newcommand}

\nc{\be}[1]{\begin{equation}\mbox{$\label{#1}$}}
\nc{\bea}[1]{\begin{eqnarray} \mbox{$\label{#1}$}}
\nc{\Section}[2]{\section{#2}\label{#1}}
\nc{\Bibitem}[1]{\bibitem{#1}}
\nc{\Label}[1]{\label{#1}}

\nc{\eea}{\end{eqnarray}}
\nc{\ee}{\end{equation}}

\nc{\bc}{\begin{center}}
\nc{\ec}{\end{center}}
\nc{\ea}{\end{array}}
\nc{\bab}{\begin{abstract}}
\nc{\eab}{\end{abstract}}
\nc{\btab}{\begin{tabular}}
\nc{\etab}{\end{tabular}}
\nc{\bit}{\begin{itemize}}
\nc{\eit}{\end{itemize}}
\nc{\ben}{\begin{enumerate}}
\nc{\een}{\end{enumerate}}
\nc{\bfig}{\begin{figure}}
\nc{\efig}{\end{figure}}

\nc{\arreq}{&\!=\!&}
\nc{\arrmi}{&\!-\!&}
\nc{\arrpl}{&\!+\!&}
\nc{\arrap}{&\!\!\!\approx\!\!\!&}
\nc{\non}{\nonumber}

\def\lsim{\; \raise0.3ex\hbox{$<$\kern-0.75em
      \raise-1.1ex\hbox{$\sim$}}\; }
\def\gsim{\; \raise0.3ex\hbox{$>$\kern-0.75em
      \raise-1.1ex\hbox{$\sim$}}\; }

\nc{\DOT}{\hspace{-0.08in}{\bf .}\hspace{0.1in}}

\nc{\abs}[1]{\left|#1\right|}

\nc{\al}{\alpha}
\nc{\g}{\gamma}
\nc{\Del}{\Delta}
\nc{\e}{\textrm{e}}
\nc{\eps}{\epsilon}
\nc{\lam}{\lambda}
\nc{\Om}{\Omega}
\nc{\ve}{\varepsilon}
\nc{\mn}{{\mu\nu}}
\nc{\vp}{\varphi}

\nc{\mL}{\mathcal{L}}
\nc{\rf}[1]{(\ref{#1})}
\nc{\nn}{\nonumber \\*}

\nc{\etal}{\mbox{\it et al.}}
\nc{\ie}{{\it i.e. }}
\nc{\eg}{{\it e.g. }}

\nc{\diag}{{\textrm{diag}}}

\nc{\I}{\textrm{I}}
\nc{\II}{\textrm{II}}
\nc{\III}{\textrm{III}}

\nc{\bee}{\begin{equation}}
\nc{\ene}{\end{equation}}

\def\fig{Figure~}

\usepackage{graphicx}

\nc{\smica}{\texttt{SMICA}~}
\nc{\csmica}{\texttt{SMICA}}
\nc{\sevem}{\texttt{SEVEM}~}
\nc{\csevem}{\texttt{SEVEM}}
\nc{\nilc}{\texttt{NILC}~}
\nc{\cnilc}{\texttt{NILC}}
\nc{\ruler}{\texttt{CR}~}
\nc{\cruler}{\texttt{CR}}

\def\Planck{\textit{Planck}~}

\begin{document}

%\title{Testing for foreground residuals in the Planck foreground cleaned maps: Challenging the Planck U73 common mask}

\title{Testing for foreground residuals in the Planck foreground cleaned maps: A new method for designing confidence masks}

\author{M.~Axelsson\thanks{e-mail:~\texttt{magnusax@astro.uio.no}} \and  H.~T.~Ihle \and S.~Scodeller \and F.~K.~Hansen}

\institute{Institute of Theoretical Astrophysics, University of
  Oslo, P.O.\ Box 1029 Blindern, N-0315 Oslo, Norway}
  
\date{\today}
 
\abstract{
  We test for foreground residuals in the foreground cleaned \Planck Cosmic Microwave Background (CMB) maps outside and inside U73 mask commonly used for cosmological analysis. The aim of this paper is to introduce a new method to validate masks by looking at the differences in cleaned maps obtained by different component separation methods. By analyzing the power spectrum as well as the mean, variance and skewness of needlet coefficients on bands outside and inside the U73 mask we first confirm that the pixels already masked by U73 are highly contaminated and cannot be used for cosmological analysis. We further find that the U73 mask needs extension in order to reduce large scale foreground residuals to a level of less than $20\%$ of the standard deviation of CMB fluctuations within the bands closest to the galactic equator. We also find 276 point sources in the cleaned foreground maps which are currently not masked by the U73 mask. Our final publicly available extended mask leaves $65.9\%$ of the sky for cosmological analysis. Note that this extended mask may be important for analyses on local sky patches; in full sky analyses the additional residuals near the galactic equator may average out.
} % end abstract

\keywords{cosmic microwave background --- cosmology: observations ---
  methods: numerical}

\titlerunning{Testing for foreground residuals in the Planck foreground cleaned maps}
%\aurhorrunning{M.~Axelsson}

\maketitle

\section{Introduction}

The recent results from ESA's \Planck~\citep{planck:data} experiment have significantly improved
cosmological parameter estimates, and today we understand many of the processes that have formed our
universe. A plethora of phenomena are explained by the best fit $\Lambda$CDM model,
which complies with the cosmological principles of homogeneity and isotropy. For over a decade
it has withstood serious challenges brought forth by confrontation with high precision data
delivered by the WMAP satellite \citep{Bennett:wmap1,Hinshaw:wmap3,Hinshaw:wmap5,Jarosik:wmap7,Bennett:wmap9},
not to mention other numerous experiments such as BOOMERanG \citep{boom:2003}, MAXIMA \citep{Maxi:2001}, DASI \citep{Dasi:2002} 
ACBAR \citep{acbar:2007}, and others.  

Notably, the BICEP2 experiment \citep{BICEP2:exp} might have further
cemented one of the most crucial hypotheses put forth by the standard model, the inflation theory, through
direct detection of B-mode polarization at a significance $>5\sigma$. The signal peaks at the correct angular
scales, and could be an indirect observation of gravitational waves which are, according to inflation theory, produced
by quantum fluctuations in the gravity field. As they travel towards our detectors their wavelengths become stretched, generating
a faint B-mode signal. There is some apparent tension between the tensor-scalar ratio found in \Planck and the corresponding
value predicted from the BICEP 2 experiment which is not fully resolved yet, but it could well be a statistical fluke.  

There is still debate whether the BICEP2 results are valid, as the foreground subtraction procedure is up for 
scrutiny, and it seems the results are also consistent with a model without gravitational waves, but with
a significant dust polarization signal \citep{flauger:2014,Adam:2014bub}. Further analysis of \Planck polarization data, soon to be
released, will hopefully shed more light on the subject.

There are, however, still issues to be resolved, regardless of the BICEP2 results. It seems, the most intriguing 
discrepancies between observed CMB data and the best fit model occurs at the very largest angular scales. The so-called 
hemispherical power asymmetry first reported by \cite{Eriksen:2004,Hansen2004}, and subsequently re-analyzed in a number of papers,
see \eg \cite{Hansen2009} and also observed in \Planck \citep{planck:isotropy}, has been shown to be statistically significant at 
least at the $3.3\sigma$ level. The fact that this curious effect persists in several experiments, argues against an explanation in terms of systematic 
effects, and may pose a challenge to the standard model. 

It is of utmost importance, that any cosmological analysis 
is performed on maps where foreground contaminations are at a minimum, hence, consistency checks should always be performed
whenever possible. In this paper, we aim to shine a bright light on the publicly available \Planck data maps, and especially
examine the level of any residuals, if there are any. Foregrounds have been subtracted from \Planck raw data using
four separate cleaning algorithms: \smica (Spectral Matching Independent Component Analysis), 
\nilc (Needlet Internal Linear Combination), \sevem (Spectral Expectation Via Maximization-Expectation), 
and \texttt{Commander-Ruler}. Common to all methods, is the use of observations at multiple frequencies in order to reduce foregrounds. 
The \smica method has been dubbed the main product in the first release. 

The \smica \citep{smica:2003} method consists of three basic steps. In the first step, spectral statistics are derived from 
a matrix computed from correlations between observations in harmonic space, where each observation is assumed to be a
superposition of individual components. Subsequently, a component model is fitted to the result which is then used 
to estimate a Wiener filter in harmonic space. The filtered spectral components are then transformed back into pixel
space, using the inverse spherical harmonic transform.

The \sevem \citep{sevem:2003} method treats all components, except the CMB signal, as generalized noise. Internal 
templates \citep{wifit} are fitted and subtracted from the frequency maps.

The \nilc \citep{nilc:2011} is a generalization of the WMAP ILC method, which constructs multidimensional filters that
are used to estimate the emission from complex components, spawned by multiple correlated emissions. Hence, from
a given map, which can be thought of as a superposition of components, the CMB is removed, as opposed to the usual
procedure of removing the non-cosmological signals. It is generalized, in the sense that the number of foreground
components is not assumed fixed. The method performs local estimation of the foregrounds, in order to suppress
the instrumental noise levels.  

The \texttt{Commander-Ruler} method \citep{Commander:2008} (henceforth referred to as \cruler) implements Bayesian component separation in pixel space, fitting a parametric model to the data by sampling the posterior distribution. Gibbs sampling is used to fit foreground amplitude and spectral parameters at low resolution (typically $N_{\mathrm{side}}=256$), and subsequently, the amplitudes are converted to high resolution by solving a least squares system of equations in each pixel, with the spectral parameters fixed to their values
from the low-resolution run, while at the same time taking pixelization effects into account in order to avoid sharp boundaries in the high-resolution map.

In the first \Planck release, each method provided its own mask based on the properties of each cleaned CMB map. The available sky fraction in these masks varies from $75\%$ to $93\%$. In most cosmological analyses the so-called U73 mask, the product of all these individual masks, is applied. The aim of this paper is to investigate (1) if the cleaned maps are sufficiently clean outside the U73 mask and (2) if some areas of the sky inside the masked pixels of the U73 maps may be safe for cosmological analysis. Both the galactic mask as well as the point source mask will be investigated. In order to assess these questions, we will (1) study the local power spectra around the galactic plane, (2) study the mean, variance and skewness of needlet coefficients in bands around the U73 cut, both in the fully foreground separated maps as well as in the difference maps between the different methods, and (3) investigate the presence of residual unmasked point sources in the difference maps based on the approach described in \cite{scodeller}.

A large part of the analysis undertaken in this paper is based on needlets. Their localization properties both in pixel- as well as in harmonic space make them particularly suited to locate foreground residuals. 
Wavelets (and in particular needlets) have previously been applied to several aspects of statistical CMB analysis, such as tests for non-Gaussianity and asymmetries \citep{Vielva:2004,Cabella:2004,Wiaux:2006,MceWen:2008,Wiaux:2008,Mar2008,Piet2008,Rud2009}, polarization analysis 
\citep{Cabella:2007}, foreground component separation and reduction \citep{wifit}, point source detection in CMB data \citep{Scodeller2012}, power spectrum estimation \citep{Basak2012}. Also, the cold spot was first detected through wavelet analysis \citep{Cruz:2005}.
For a general introduction to needlets and their properties, see \eg \citep{Baldi:2006,Mari:2011}.

%Needlets have previous been used with great success for 
%many different applications in CMB analysis \citep{Mar2008,Piet2008,Rud2009,Basak2012,Scodeller2012}.{\bf need to insert some references here}.

  The approach which we develop and apply to \Planck temperature data in this paper is a methodology which allows the construction of a common mask based on data cleaned with many different methods. We will here show the importance of applying such a procedure in order to obtain a consistency test of component separation methods as well as in designing a fiducial mask. For the coming release of \Planck polarization data were the foreground properties are less known, such an approach may become even more important.

In section \ref{data} we discuss the data products used in this paper. In section \ref{method} we discuss the details of our methodology and define several tests applied to the cleaned maps. In section \ref{analysis1} we analyze individual maps, whereas in section \ref{analysis2} the analysis is repeated, but this time on difference maps in order to perform consistency checks. In section \ref{mask} we use difference maps in needlet space to manipulate
the mask in order to explore how statistics is affected by either adding, or subtracting, parts of the sky close to the
galactic plane. The point source mask is investigated in section \ref{ps} and we discuss our findings in section \ref{conclusion}.

\section{Data}
\label{data}
In this paper we use the publicly available \cnilc, \csmica, \sevem and \ruler foreground cleaned maps as well as their beam functions and accompanying FFP6 simulation sets. The sixth round full focal plane (FFP6) simulations have been passed through the component separation pipeline and therfore have beam and noise properties similar to the foreground cleaned maps. The method specific masks as well as the common mask based on the product of these are used. We also create jack-knife maps based on the difference between half-ring maps of the data. The advantage of jack-knife maps is that they have noise properties very close to the noise properties of the actual data. As described in detail below, they are used to adjust the noise level in the simulations, in order to obtain best possible agreement with the noise properties in the data.
%such that they are as close as possible to the properties of the data.

\section{Method}
\label{method}

In order to study the variation of possible foreground residuals with distance from the galactic plane, we construct 7 bands in each hemisphere starting from the borders of the U73 mask proceeding out toward the polar caps. We will number these bands from 1 to 7, each "band" consists of the sum of the corresponding bands in both hemispheres.  The northern and southern bands are combined in order to increase statistics. Band 1 consists of the two bands closest to the U73 mask, band 7 consists of the polar caps (see \fig\ref{bands}). The bands are constructed by smoothing the U73 mask with a large beam, then including all pixels below a certain threshold. This process is repeated for each band. The sky fractions covered by bands 1 to 7 are 0.145, 0.138, 0.126, 0.11, 0.096, 0.078 and 0.041 respectively. A further division of the first bands will be necessary as detailed below.

\bfig[!tb]
\includegraphics[width=0.65\linewidth,angle=90]{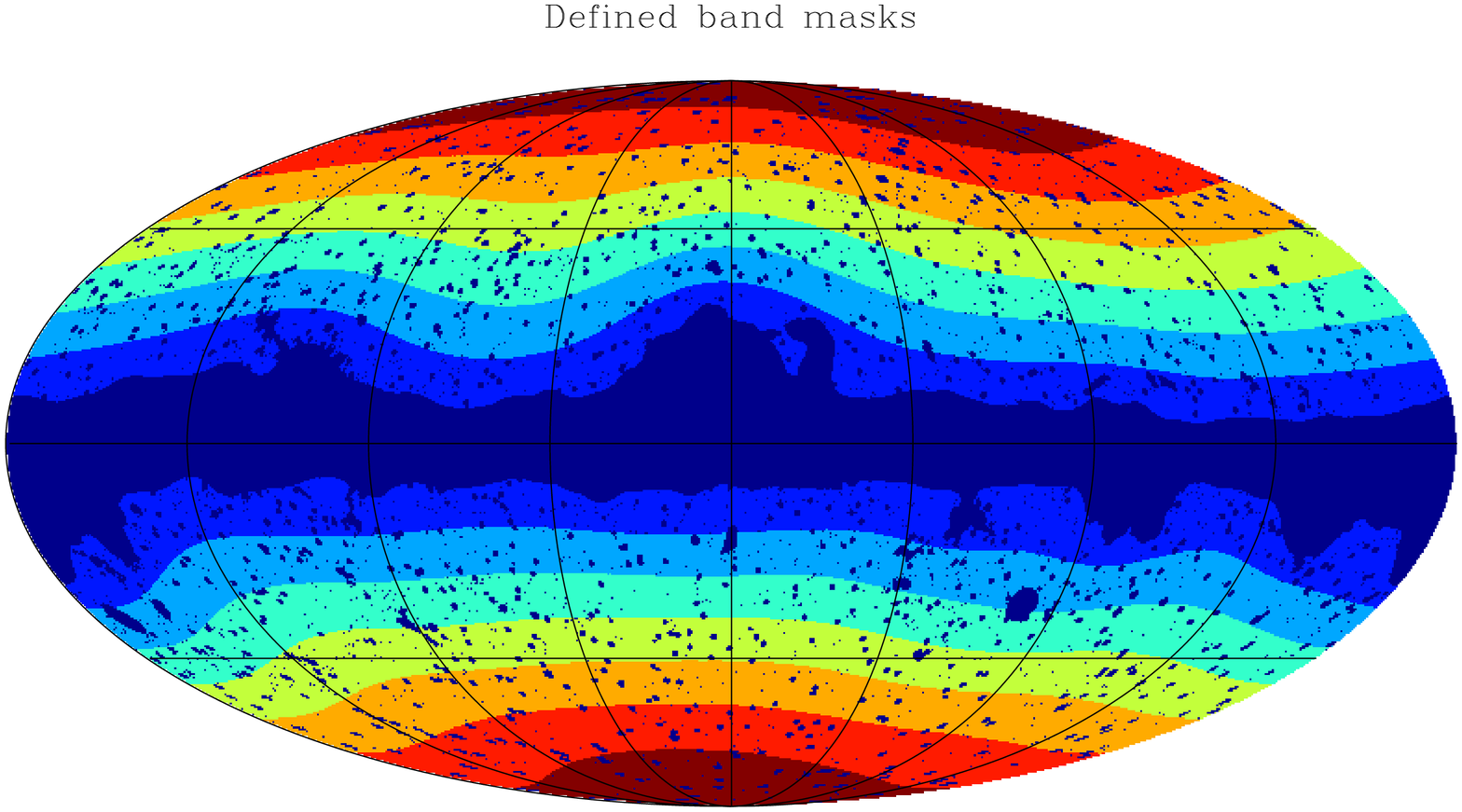}
%%  \mbox{\epsfig{figure=figs/bandmasks_grat.eps,angle=90,width=\linewidth,clip=}}
  \caption{The 7 band masks on which the analysis is performed. A single band mask consists of a set of pixels on both
    the northern and southern galactic hemispheres as indicated by the matching color schemes. \label{bands}}
\efig

\begin{figure}[t]
\includegraphics[width=0.56\linewidth,angle=90]{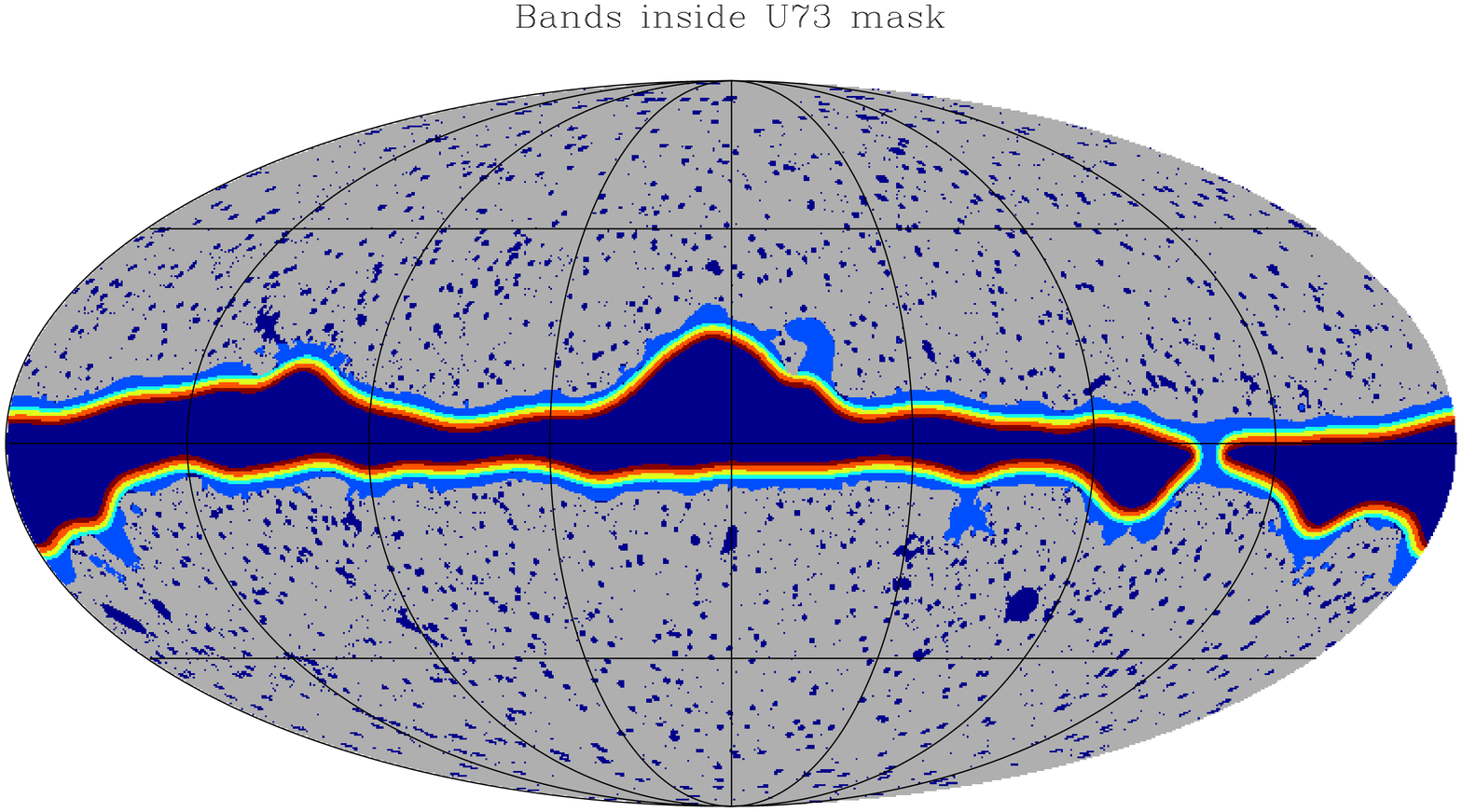}
%%\mbox{\epsfig{figure=figs/CandidateBands.eps,angle=90,width=\linewidth,clip=}}
  \caption{ Constructed bands in the interior of the U73 mask, prior to data reduction. }
\label{inside}
\end{figure}

Furthermore, using the same approach as described above to construct bands outside U73, we have also constructed five bands inside the U73 mask. This in 
order to test whether some of these areas appear sufficiently clean for cosmological analysis. In \fig\ref{inside} we show these inside bands. The LFI 
and HFI \Planck point source masks \citep{planck:ps} are used to ensure that no pixels in the inside bands are contaminated by point sources. 

We have estimated a set of quantities in each of these bands and compared to the corresponding quantities within the same band on simulated maps. The 
indicators of foreground residuals which will be used are the following:

\begin{enumerate}
\item We have estimated the power spectrum within each band using the MASTER approach \citep{hivon2002}. Due to the small sky fraction available to each band, we needed to bin the resulting spectra in bins of 10 multipoles.
\item We have calculated the mean, variance and skewness of needlet coefficients for each band. In this process each needlet coefficient is weighted by the inverse of its CMB+noise variance. We use standard needlets with needlet base $B=1.8393$ and scales $j=[2,11]$ which correspond to multipoles in the range $\ell=[2,1500]$.
\end{enumerate}

\begin{figure}%[h!]
\includegraphics[width=\linewidth,angle=0]{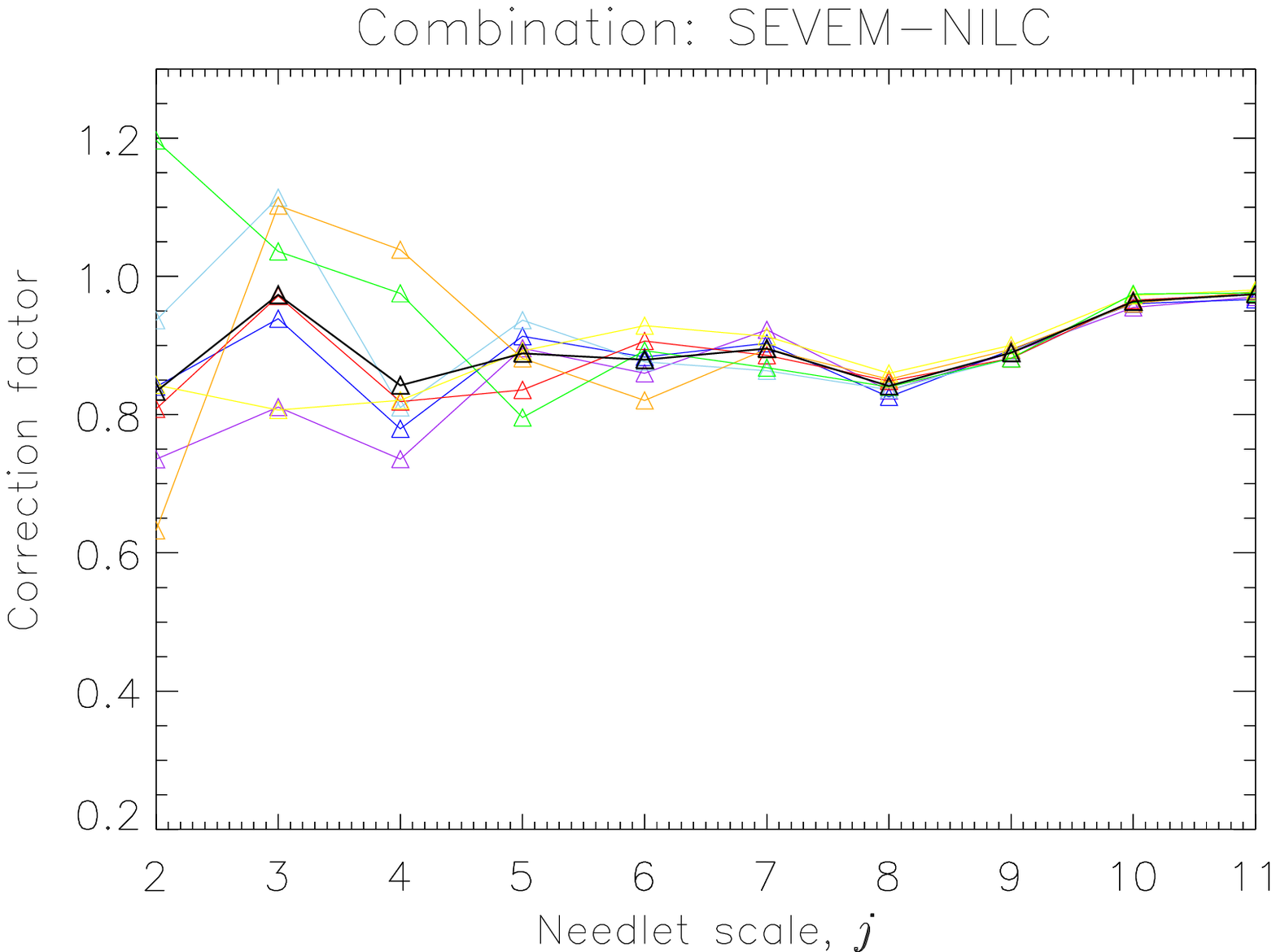}
\includegraphics[width=\linewidth,angle=0]{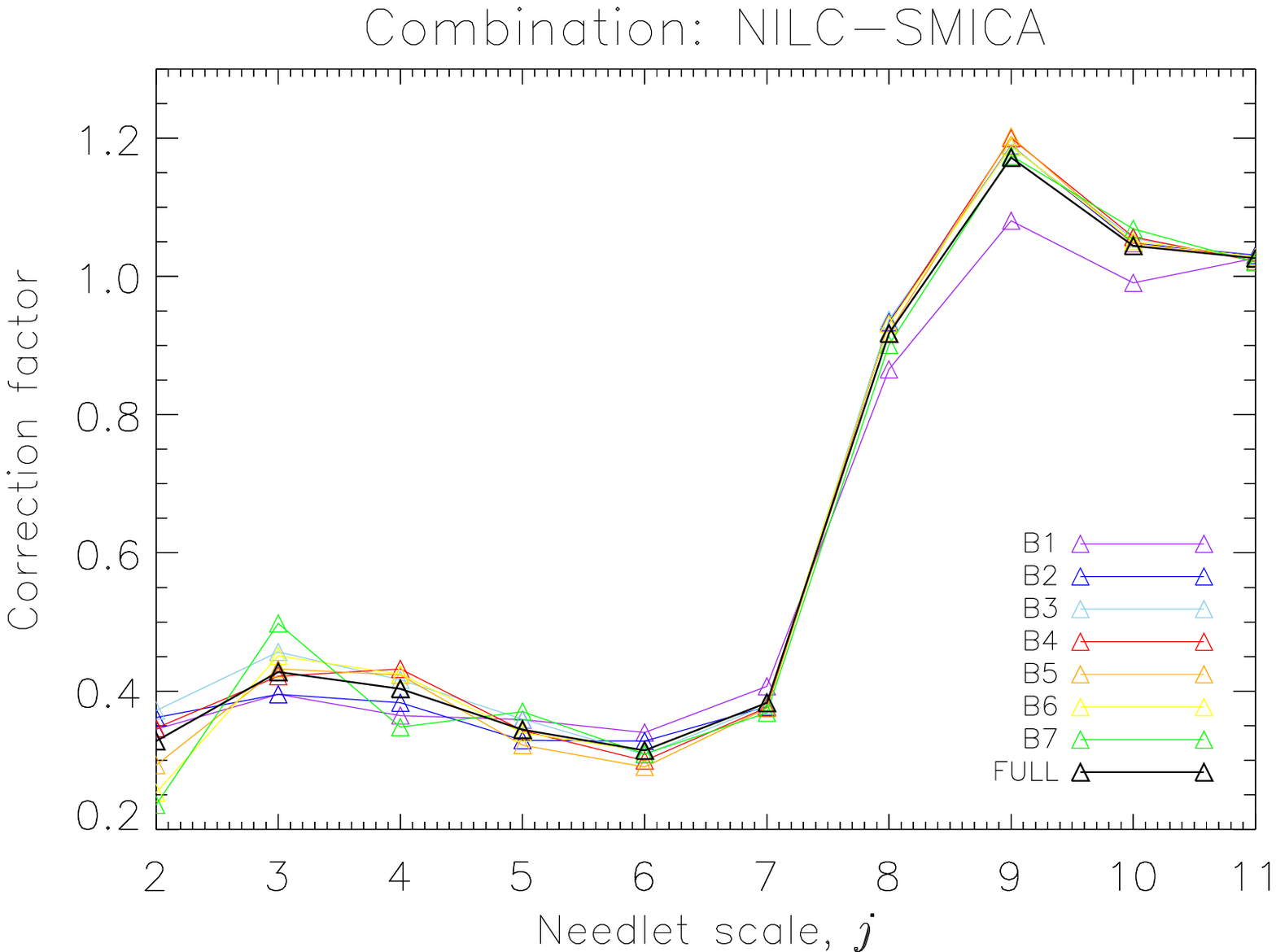}
\includegraphics[width=\linewidth,angle=0]{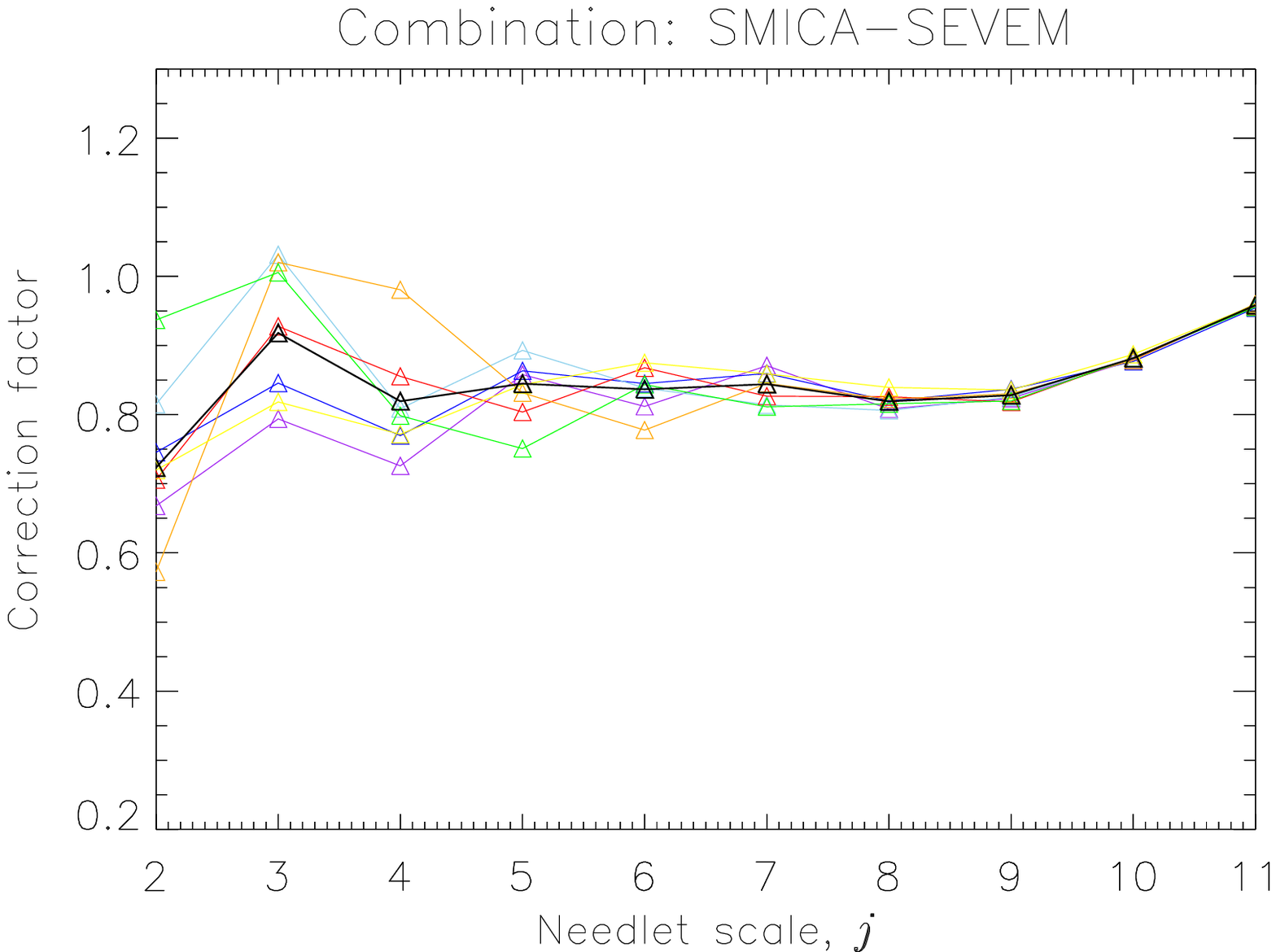}
\caption{Plot of bias correction factors in each band outside the U73 mask for selected difference maps, see \fig\ref{bands}. 
  \emph{Top}: Correction factors in pixel space for \csevem-\cnilc.
  \emph{Middle}: Correction factors for \cnilc-\csmica,
  \emph{Bottom}: Correction factors applied to \csmica-\csevem. 
 \label{corfacsU73}}
\end{figure}

These indicators have been calculated on two sets of maps:
\begin{enumerate}
\item The officially released \csmica, \csevem, \nilc and \ruler cleaned \Planck maps.
\item On difference maps between pairs of cleaned maps. For each difference map, we smooth the maps to a common resolution and subtract. In the difference maps, the CMB cancels out and only noise as well as differences in foreground residuals are present. We found that the noise properties of the data difference maps deviate significantly from the simulated difference maps. We used the jack-knife difference maps for the data to fit and adjust an amplitude correction factor to the noise levels in the simulated maps, scale by scale and band by band (although the variation with band is very small). After this correction we found a very good agreement between the noise level in the simulated maps and in the jack-knife maps. The correction factors for some difference maps are shown in \fig\ref{corfacsU73}.
\end{enumerate}

We further applied the approach in \cite{scodeller} to amplify point sources in the difference maps. We found that the needlet scales $B=1.5$ and $j=17$ gave the largest increase in point source amplitudes.

We will assess significance of our results by studying deviations by plots of $(x-\langle x\rangle)/\sigma$ where $x$ is any of the aforementioned indicators, $\langle x\rangle$ and $\sigma$ are their corresponding mean and standard deviation from simulations.

\section{Single map analysis}
\label{analysis1}

\begin{figure}[th]
   \includegraphics[width=\linewidth,angle=0]{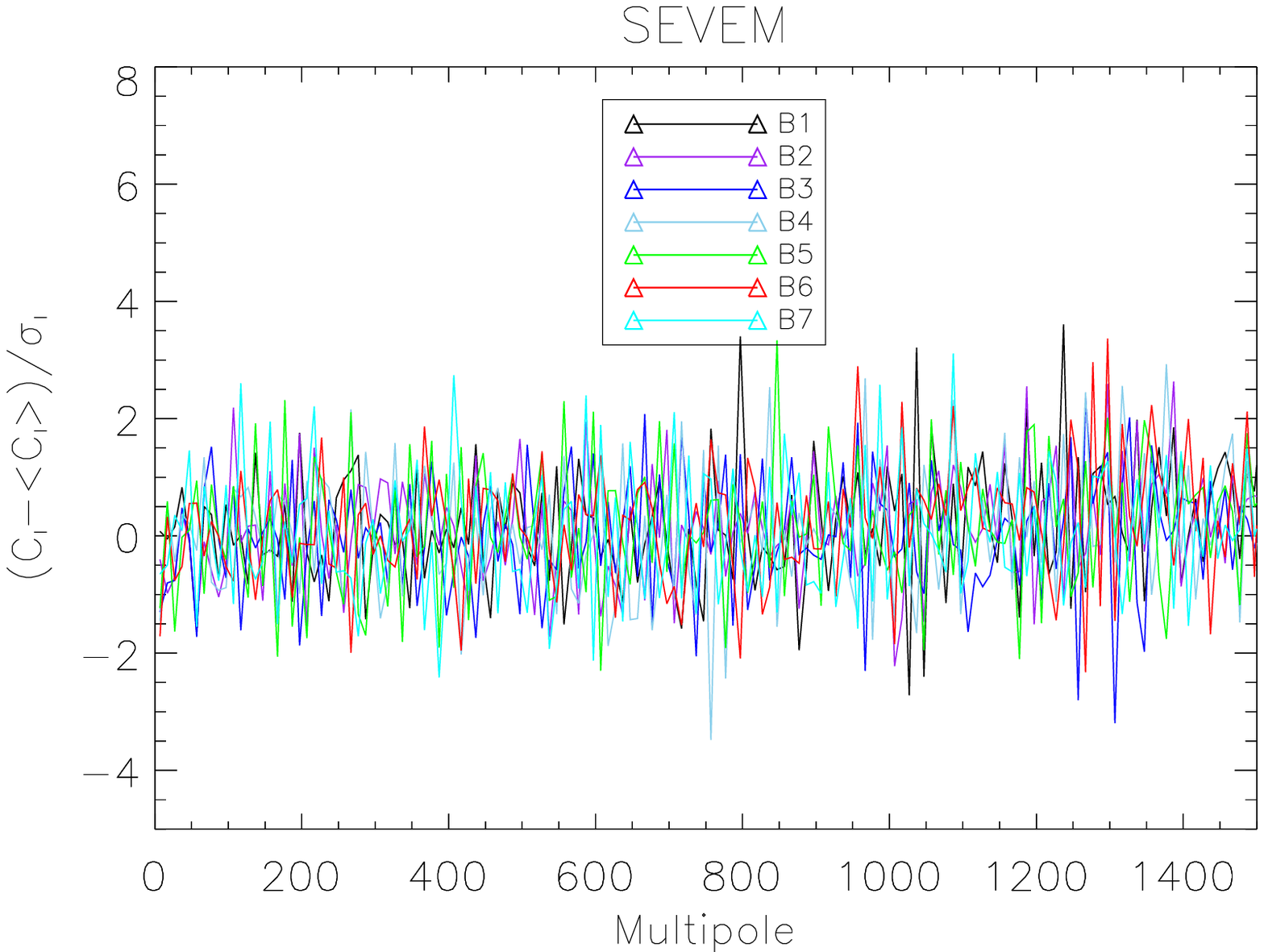}
  \caption{ Plot of $(C_\ell-\langle C_\ell\rangle)/\sigma_\ell$ obtained from the \sevem map. The data have been binned in $\Delta \ell = 10$ sized bins in order to avoid singular matrices. 
The legend label "BX" refers to band number "X", as defined above. The corresponding plots for the other methods are very similar and not shown. \label{clchis}}
\end{figure}

\begin{figure*}[!tb] 
  \begin{minipage}[b]{0.33\linewidth}
    \centering
    \includegraphics[width=\linewidth]{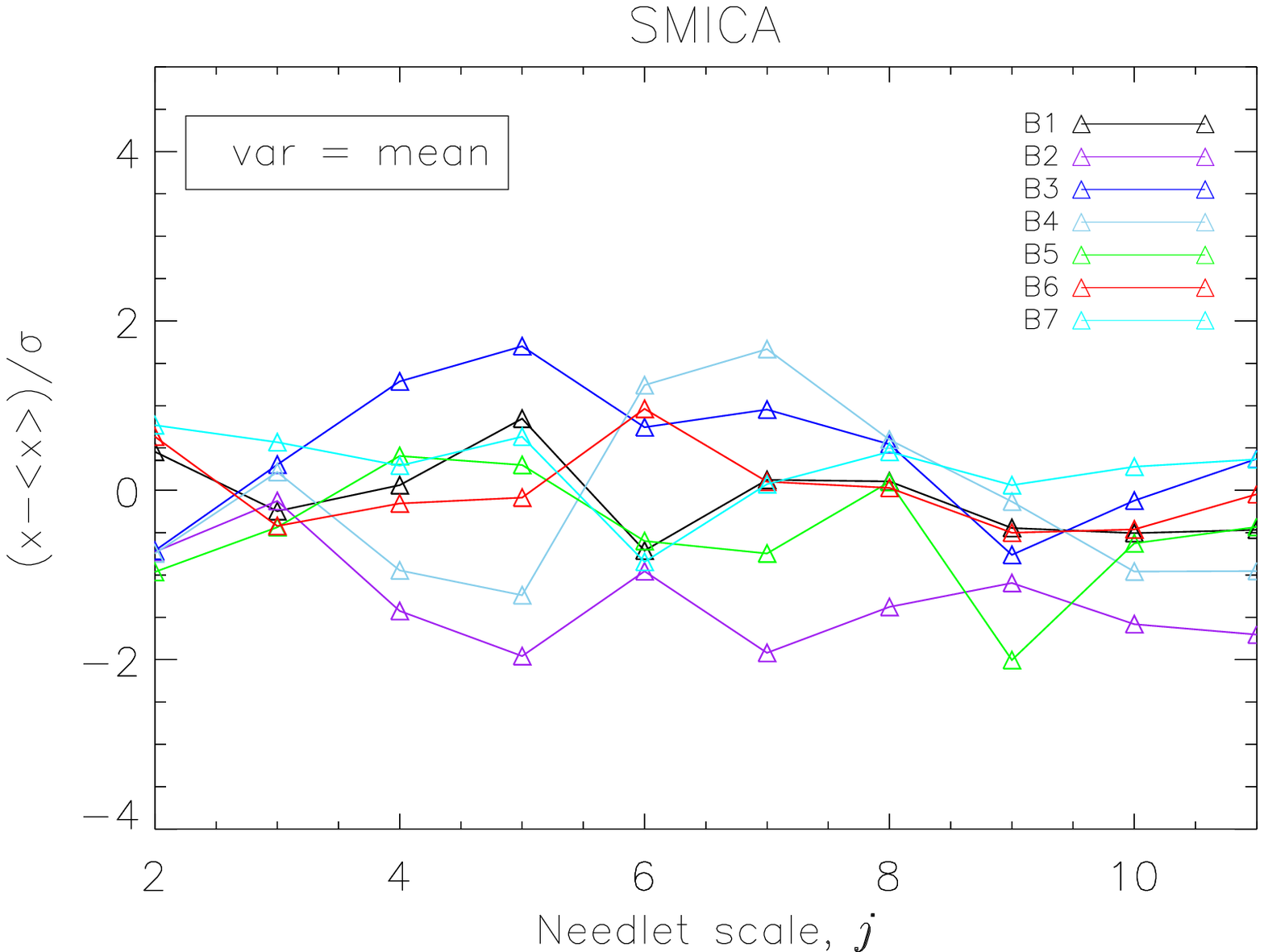}
       %\vspace{4ex}
  \end{minipage}
  \begin{minipage}[b]{0.33\linewidth}
    \centering
    \includegraphics[width=\linewidth]{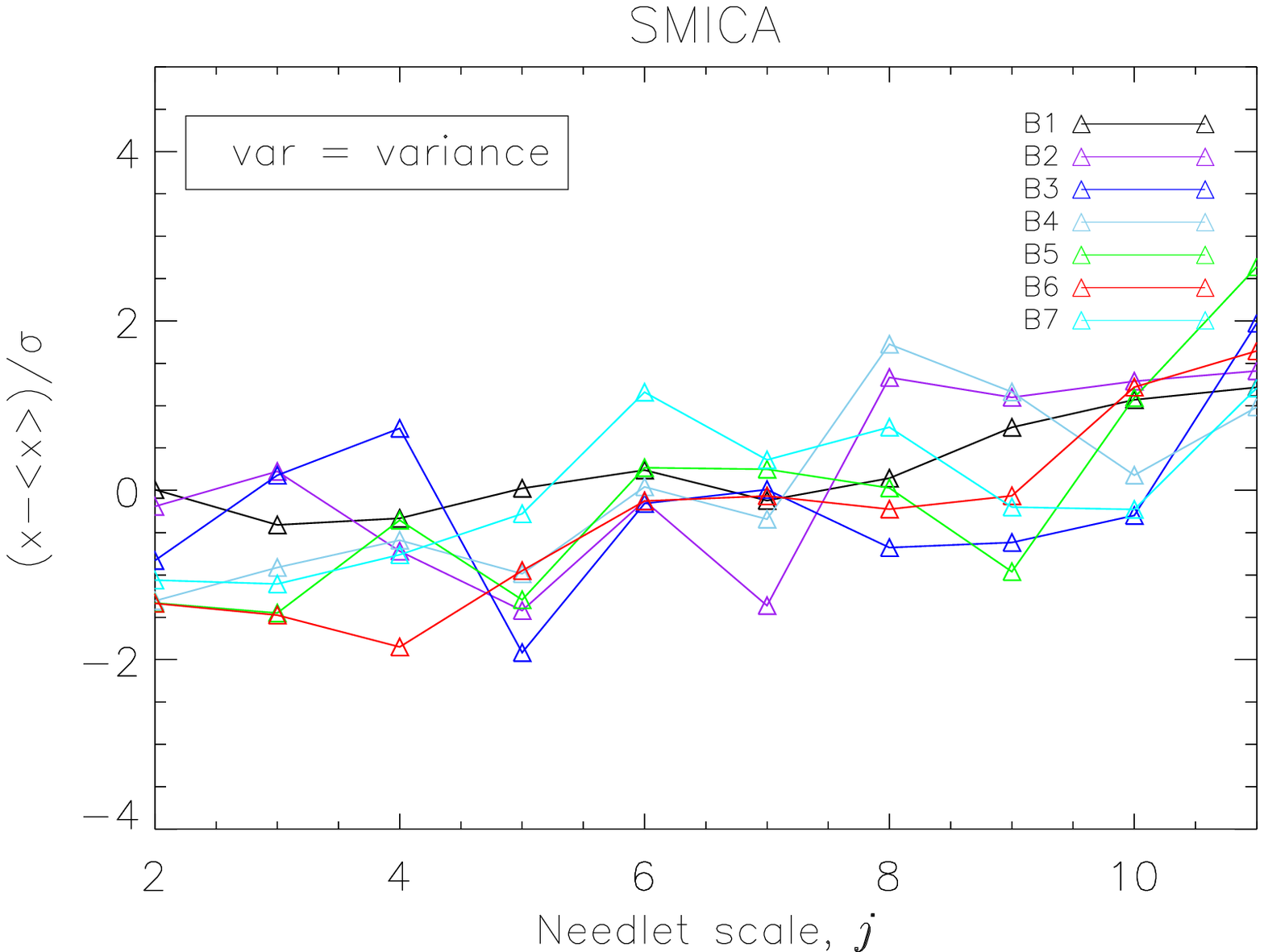}
    %\vspace{4ex}
  \end{minipage} 
  \begin{minipage}[b]{0.33\linewidth}
    \centering
    \includegraphics[width=\linewidth]{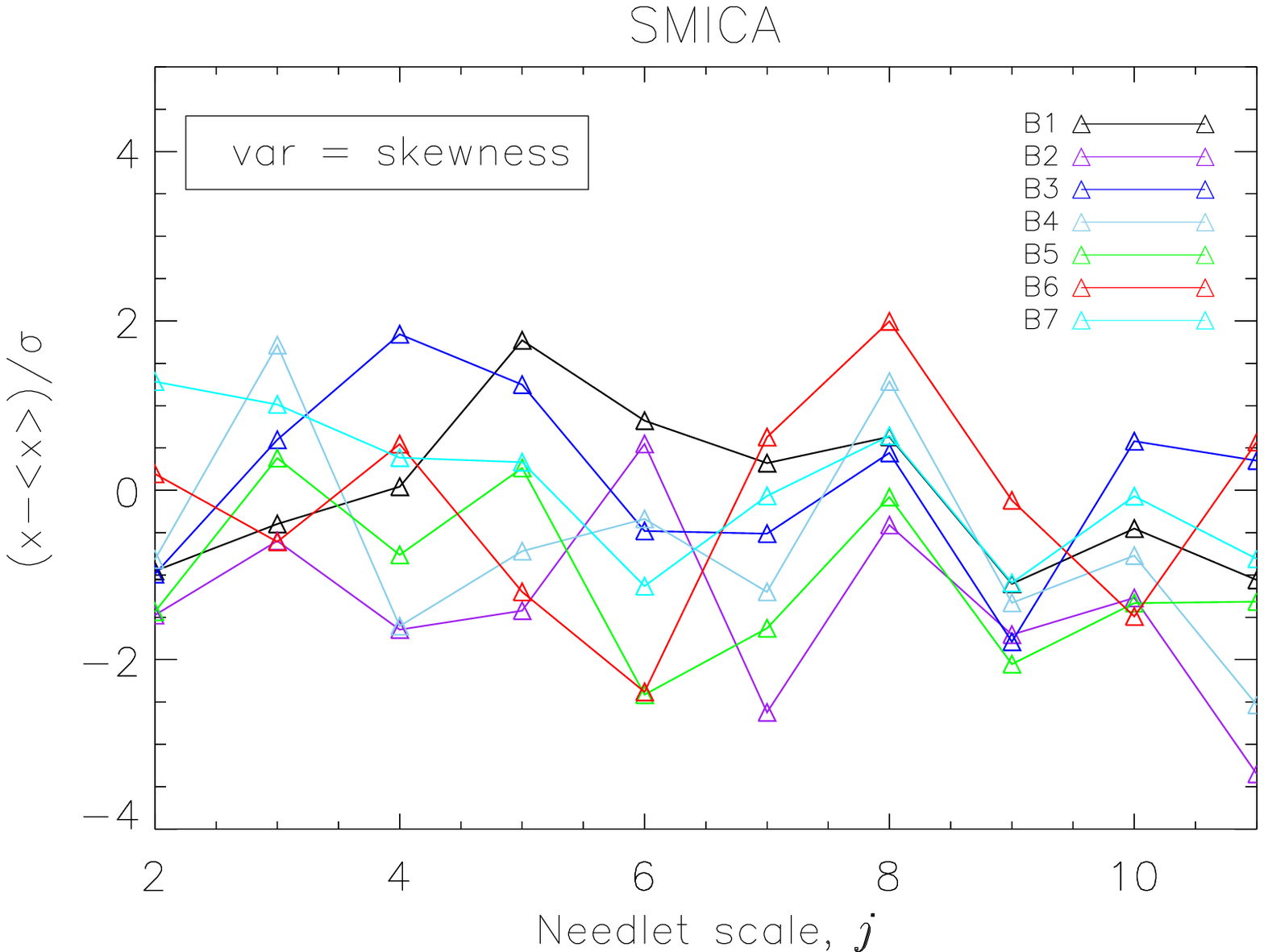}
    %\vspace{4ex}
  \end{minipage}%% 
\\ % needed for AA format (for some reason)
  \begin{minipage}[b]{0.33\linewidth}
    \centering
    \includegraphics[width=\linewidth]{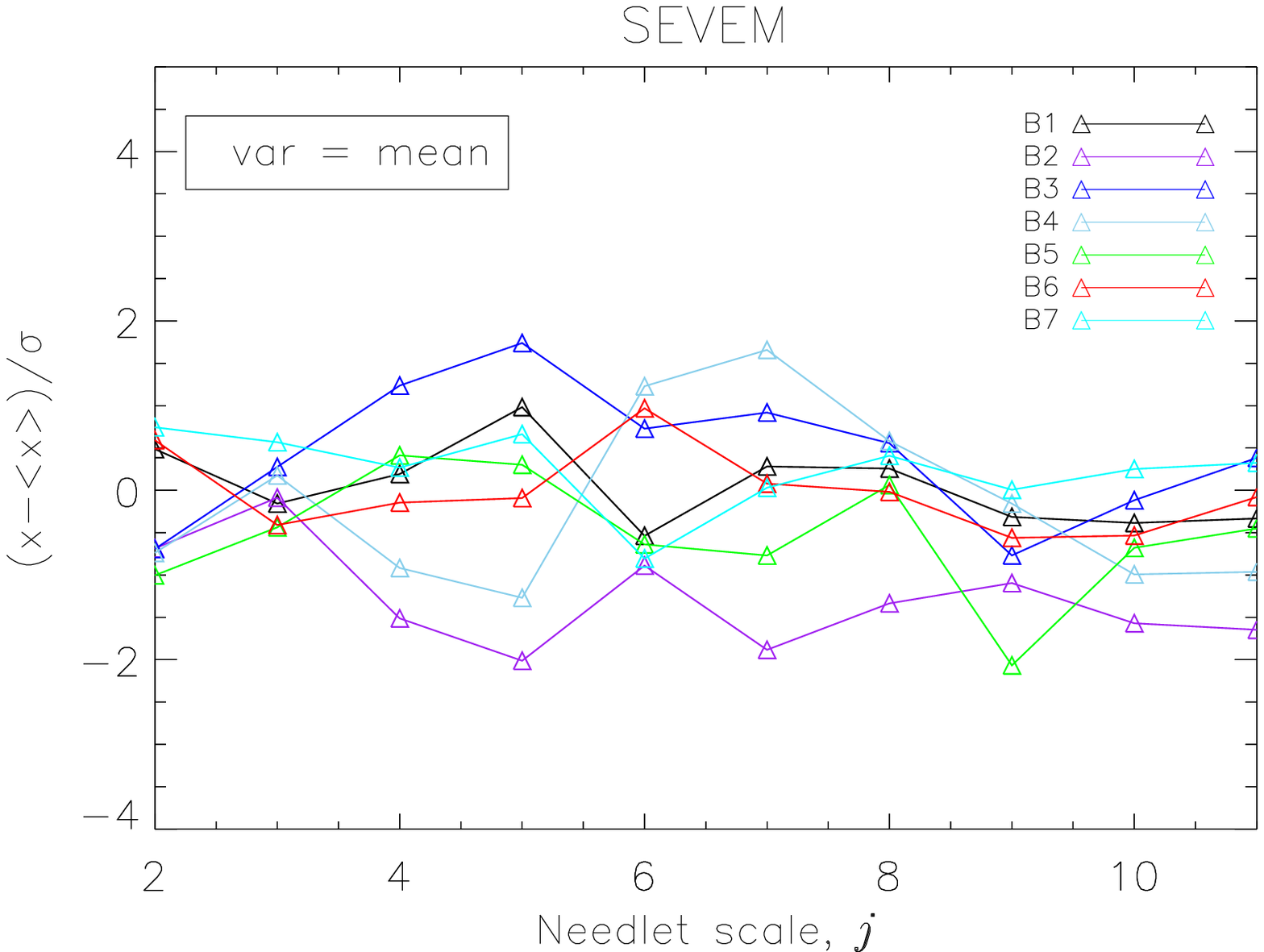}
    %\vspace{4ex}
  \end{minipage}
   \begin{minipage}[b]{0.33\linewidth}
    \centering
    \includegraphics[width=\linewidth]{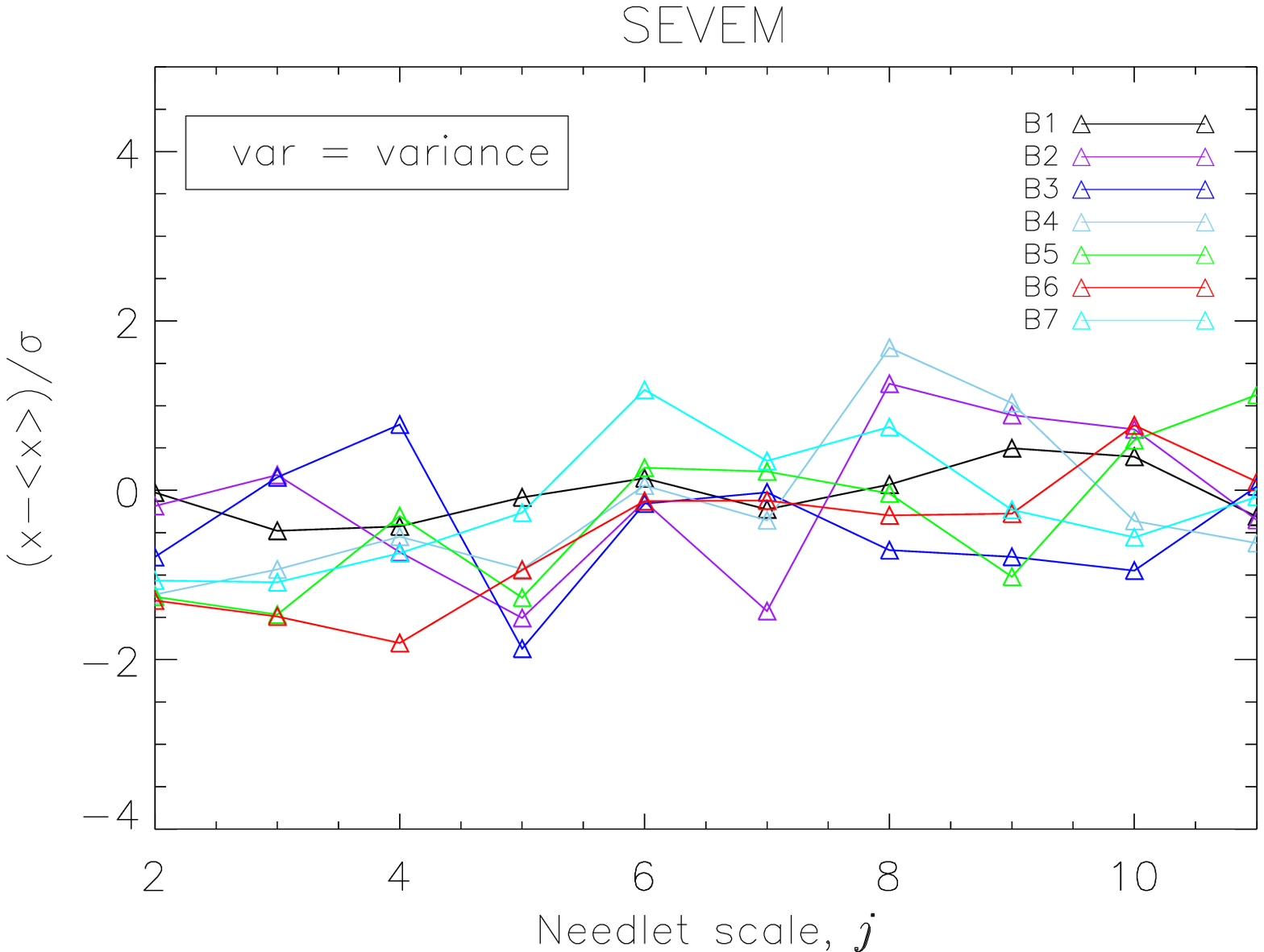}
    %\vspace{4ex}
  \end{minipage}
    \begin{minipage}[b]{0.33\linewidth}
    \centering
    \includegraphics[width=\linewidth]{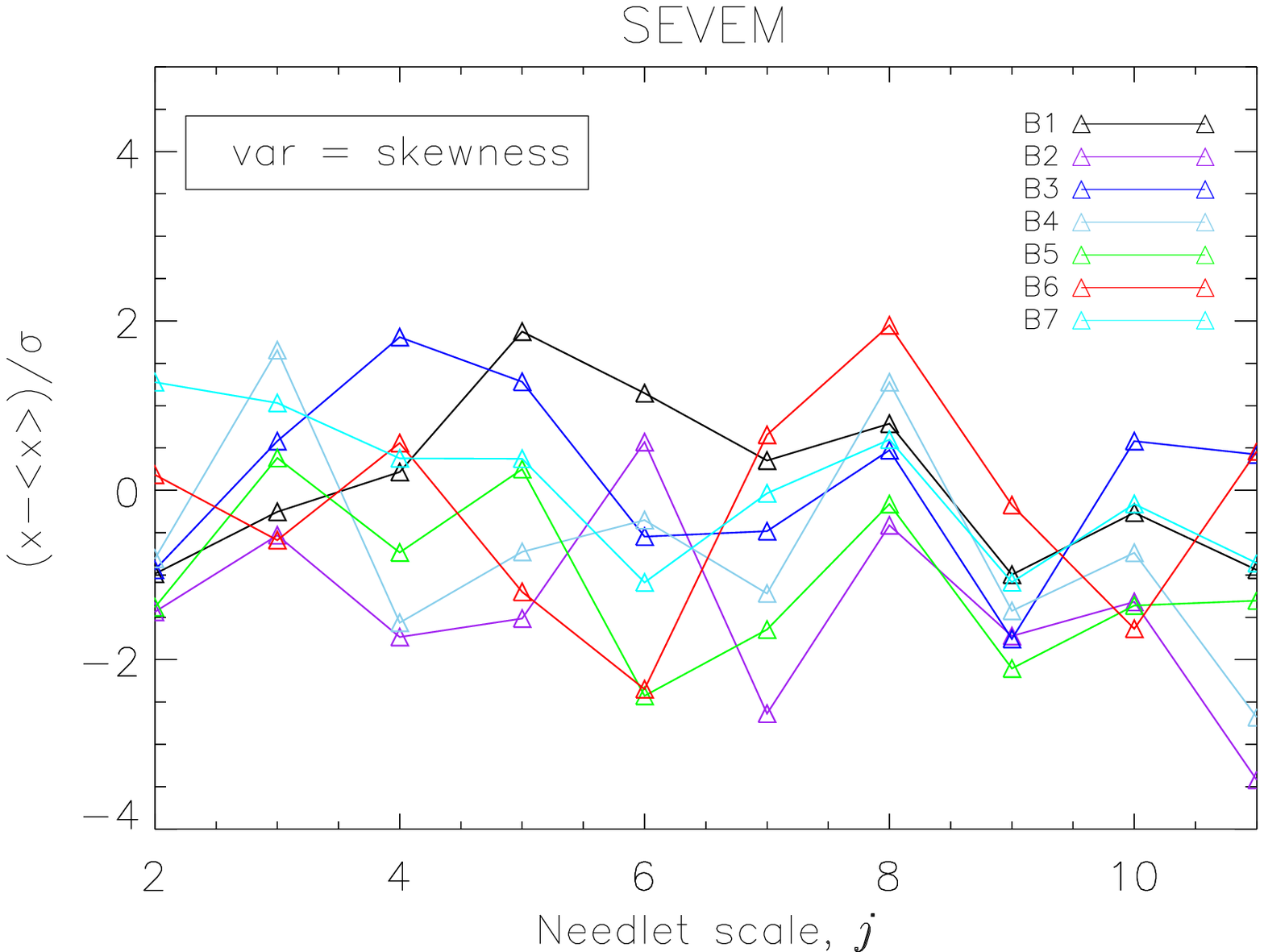}
    %\vspace{4ex}
  \end{minipage}
     \begin{minipage}[b]{0.33\linewidth}
    \centering
    \includegraphics[width=\linewidth]{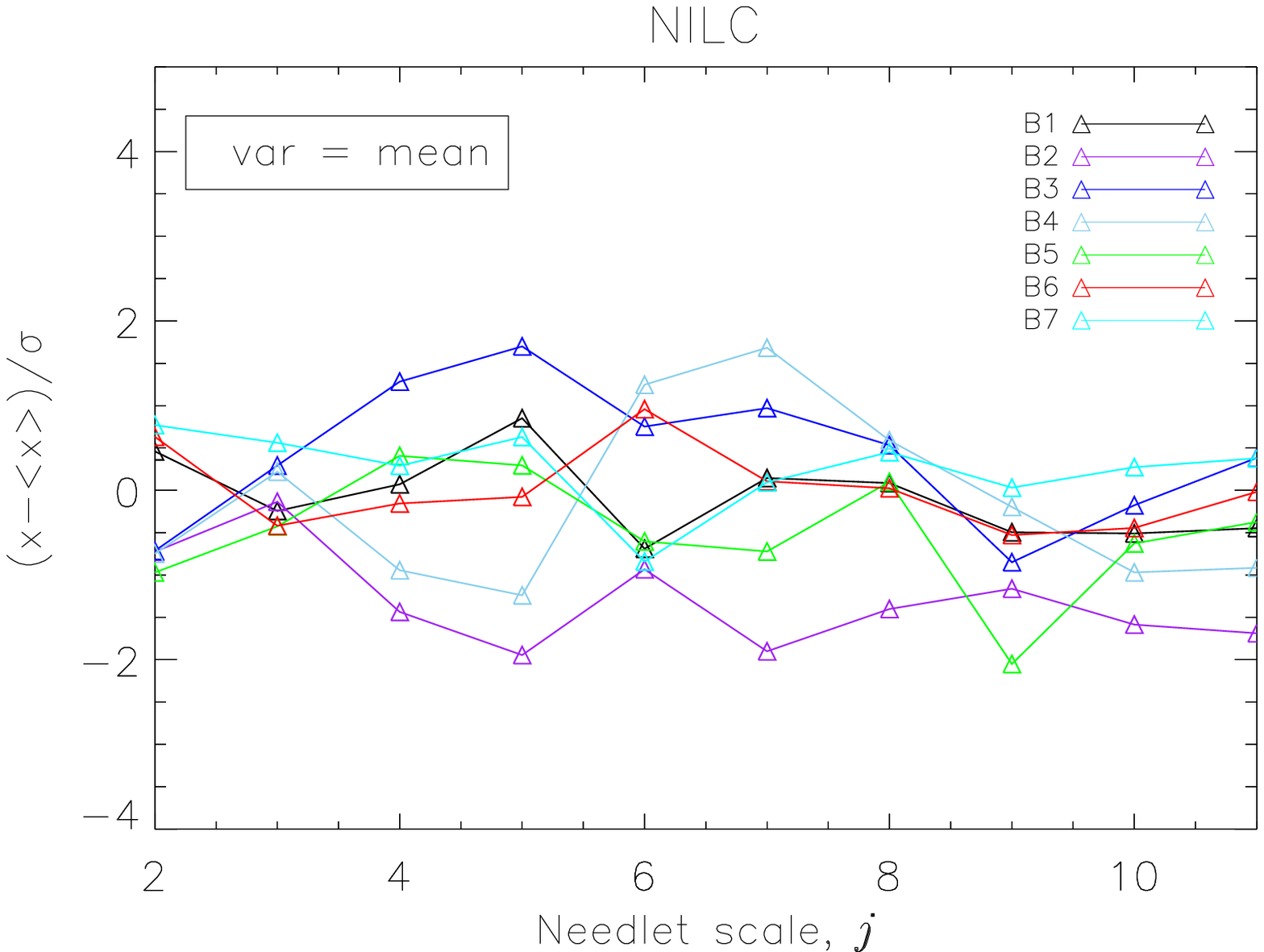}
    %\vspace{4ex}
  \end{minipage}
 \begin{minipage}[b]{0.33\linewidth}
    \centering
    \includegraphics[width=\linewidth]{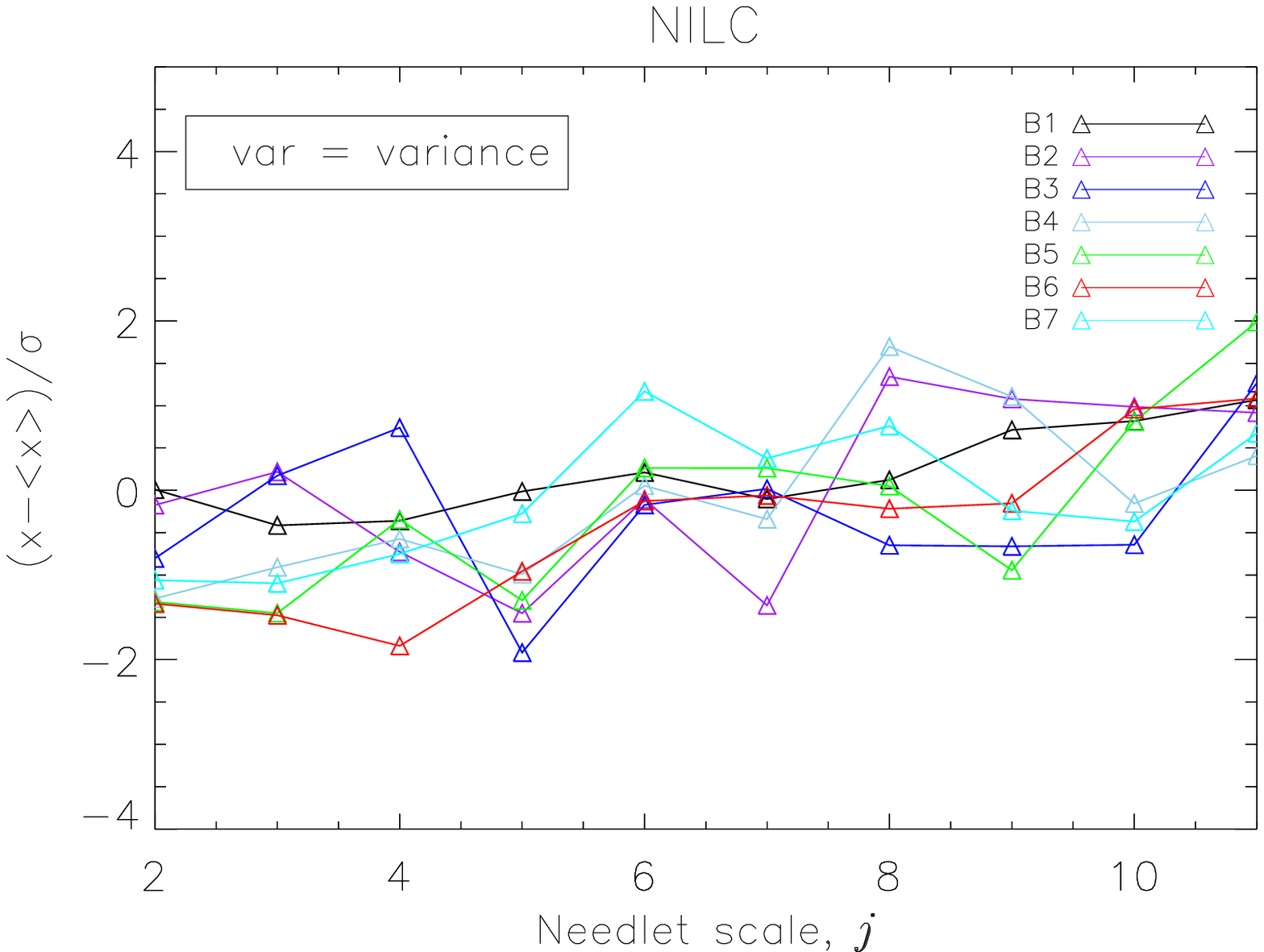}
    %\vspace{4ex}
  \end{minipage}
  \begin{minipage}[b]{0.33\linewidth}
    \centering
    \includegraphics[width=\linewidth]{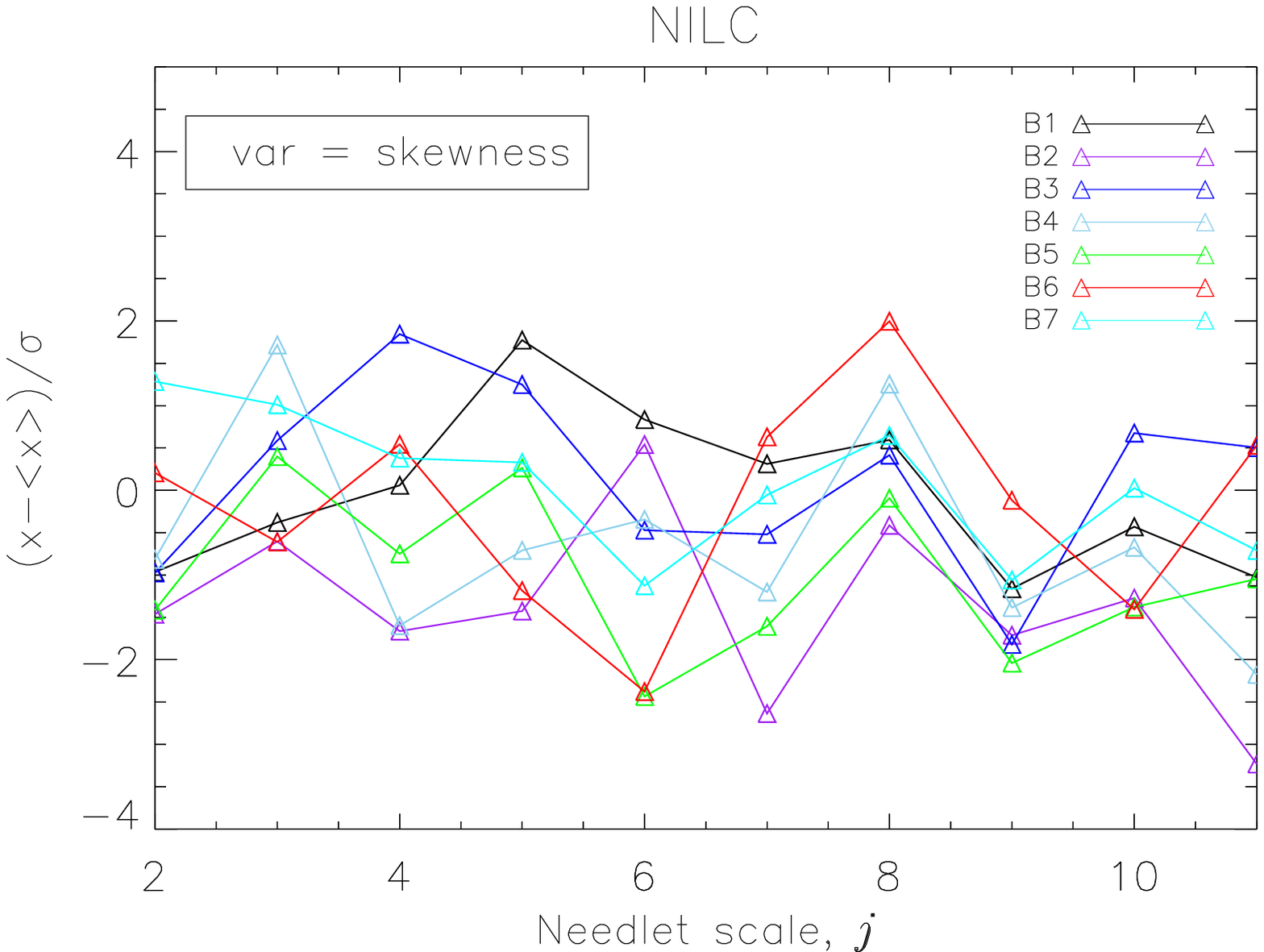}
    %\vspace{4ex}
 \end{minipage}
   \begin{minipage}[b]{0.33\linewidth}
    \centering
    \includegraphics[width=\linewidth]{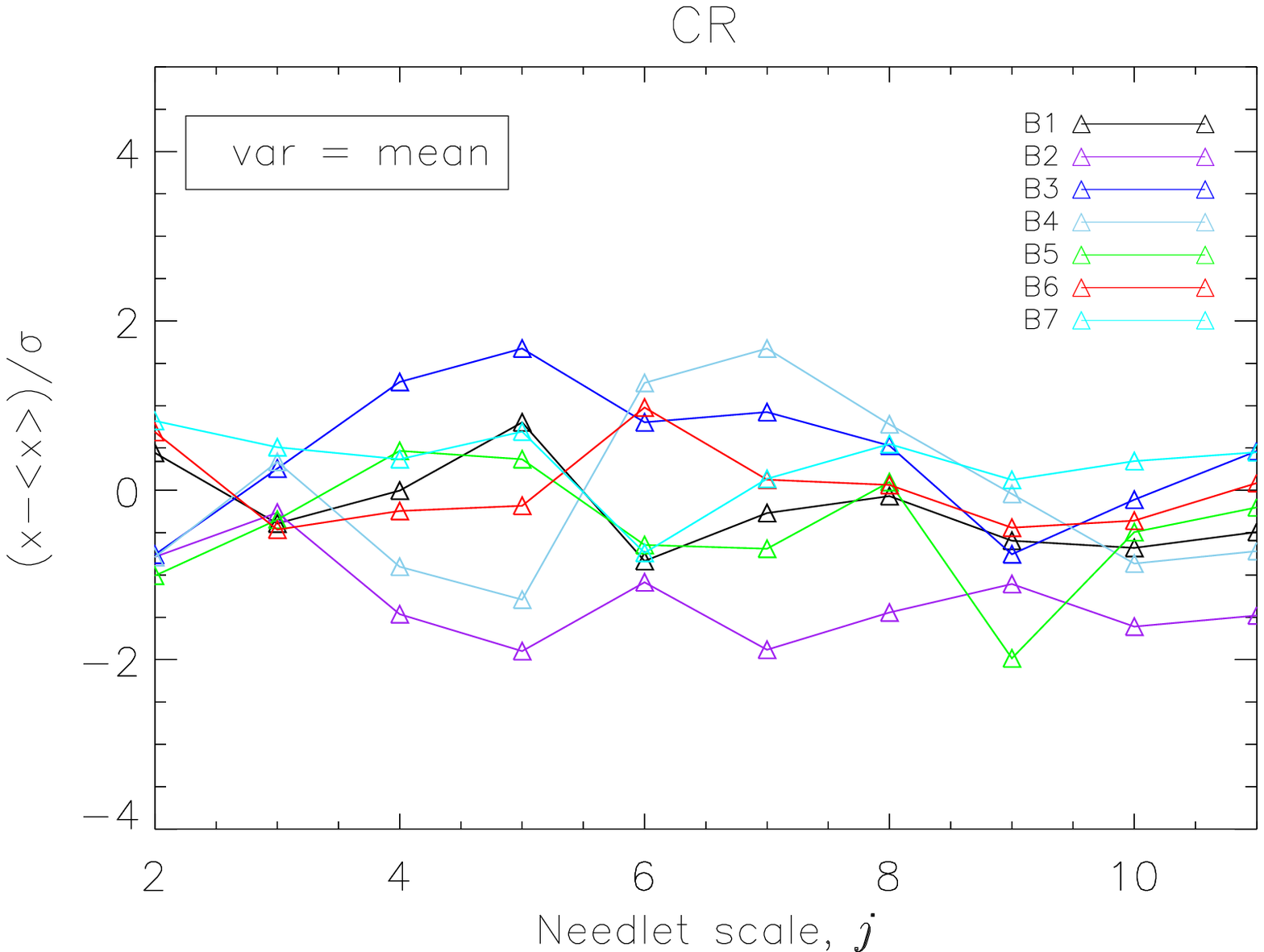}
    %\vspace{4ex}
  \end{minipage}
 \begin{minipage}[b]{0.33\linewidth}
    \centering
    \includegraphics[width=\linewidth]{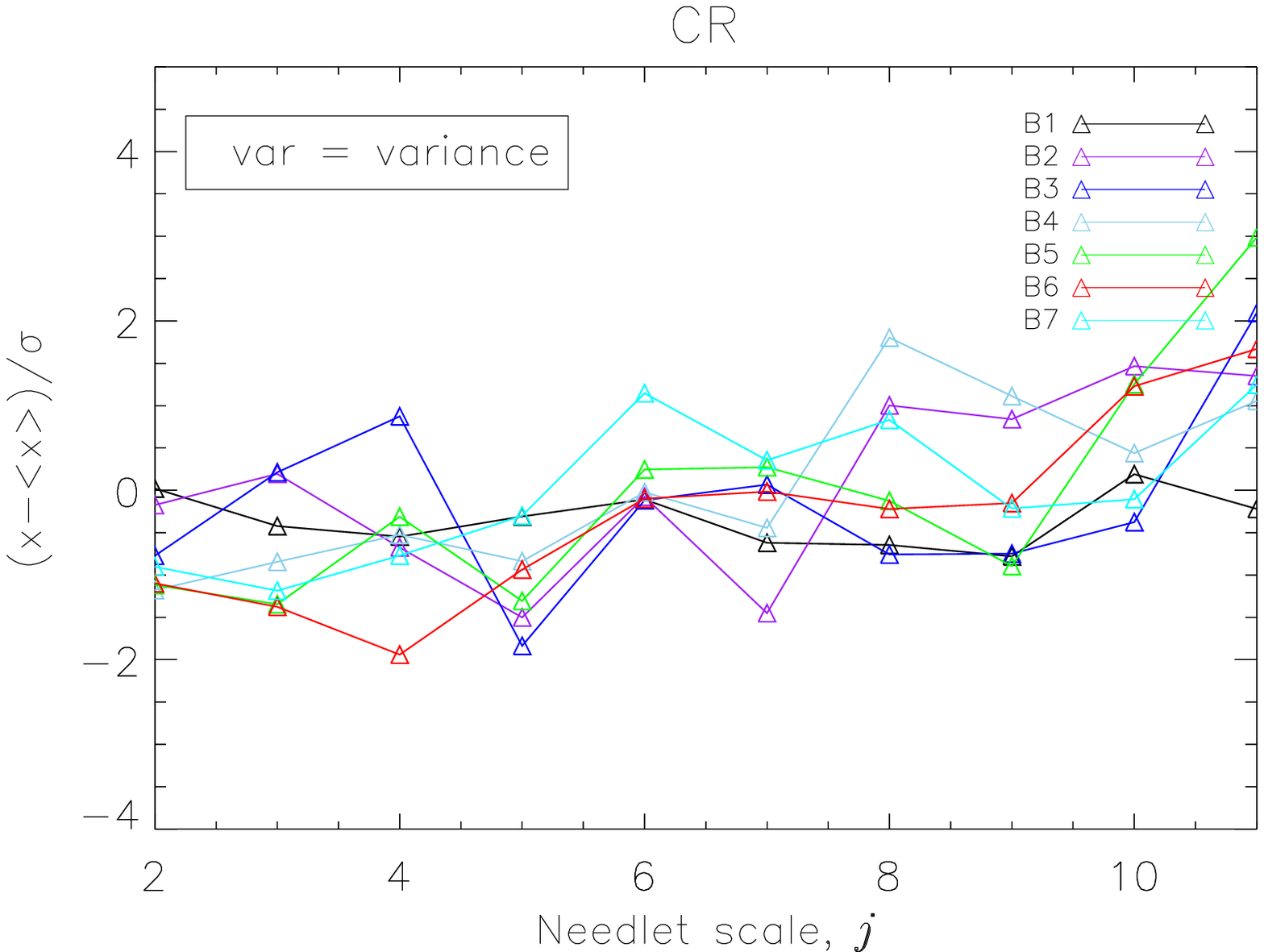}
    %\vspace{4ex}
  \end{minipage}
  \begin{minipage}[b]{0.33\linewidth}
    \centering
    \includegraphics[width=\linewidth]{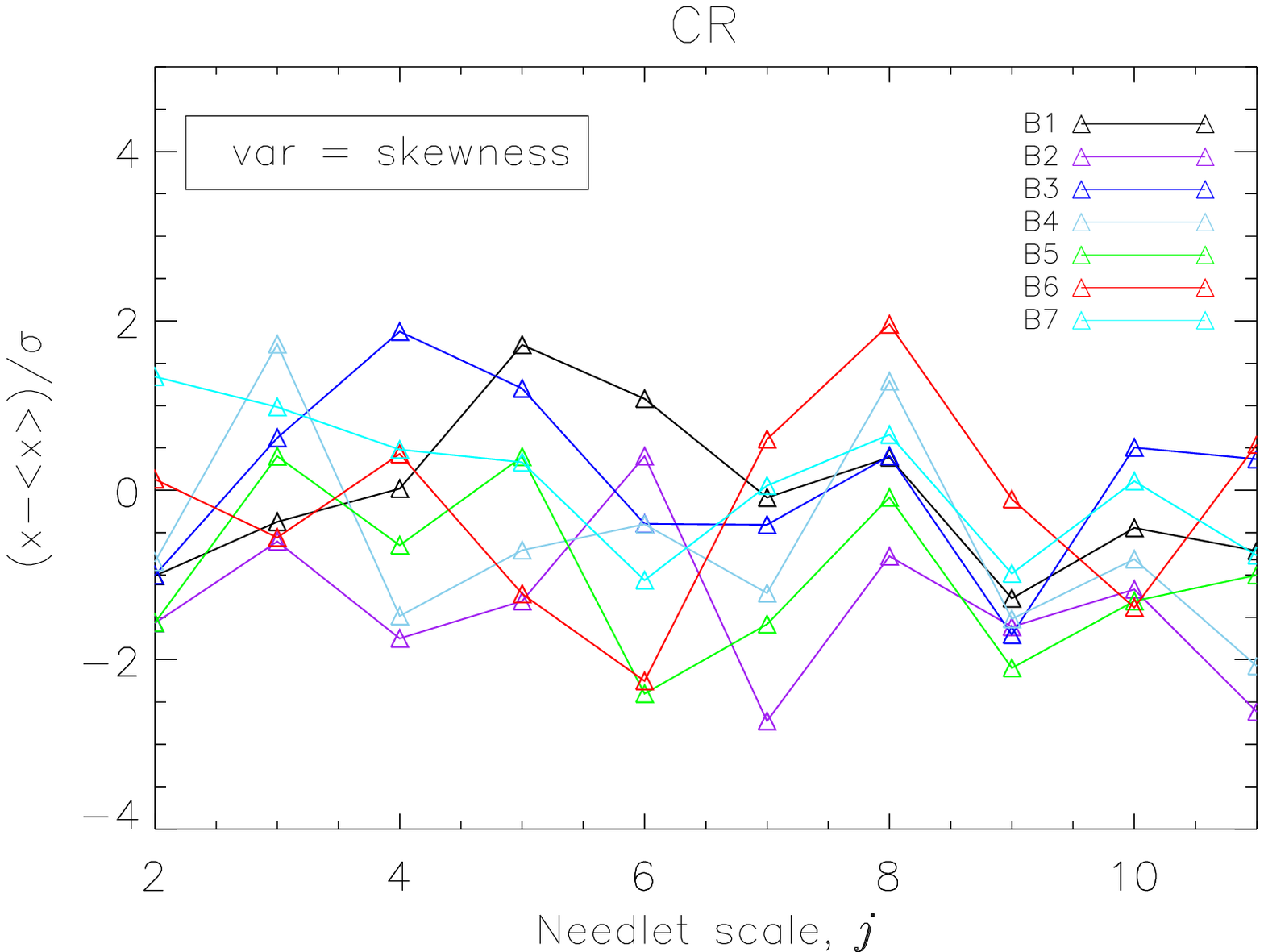}
    %\vspace{4ex}
  \end{minipage}
  \caption{Plot of $(x-\langle x\rangle)/\sigma$ where $x$ corresponds to mean (left column), variance (middle column) and skewness (right column) of needlet coefficients 
    computed from \csmica, \csevem, \nilc and \ruler maps. The various bands, B1 to B7 are shown in \fig\ref{bands}.\label{needchis}}
\end{figure*}

\begin{figure}[htb] 
  \begin{minipage}[b]{0.8\linewidth}
    \centering
    \includegraphics[width=\linewidth]{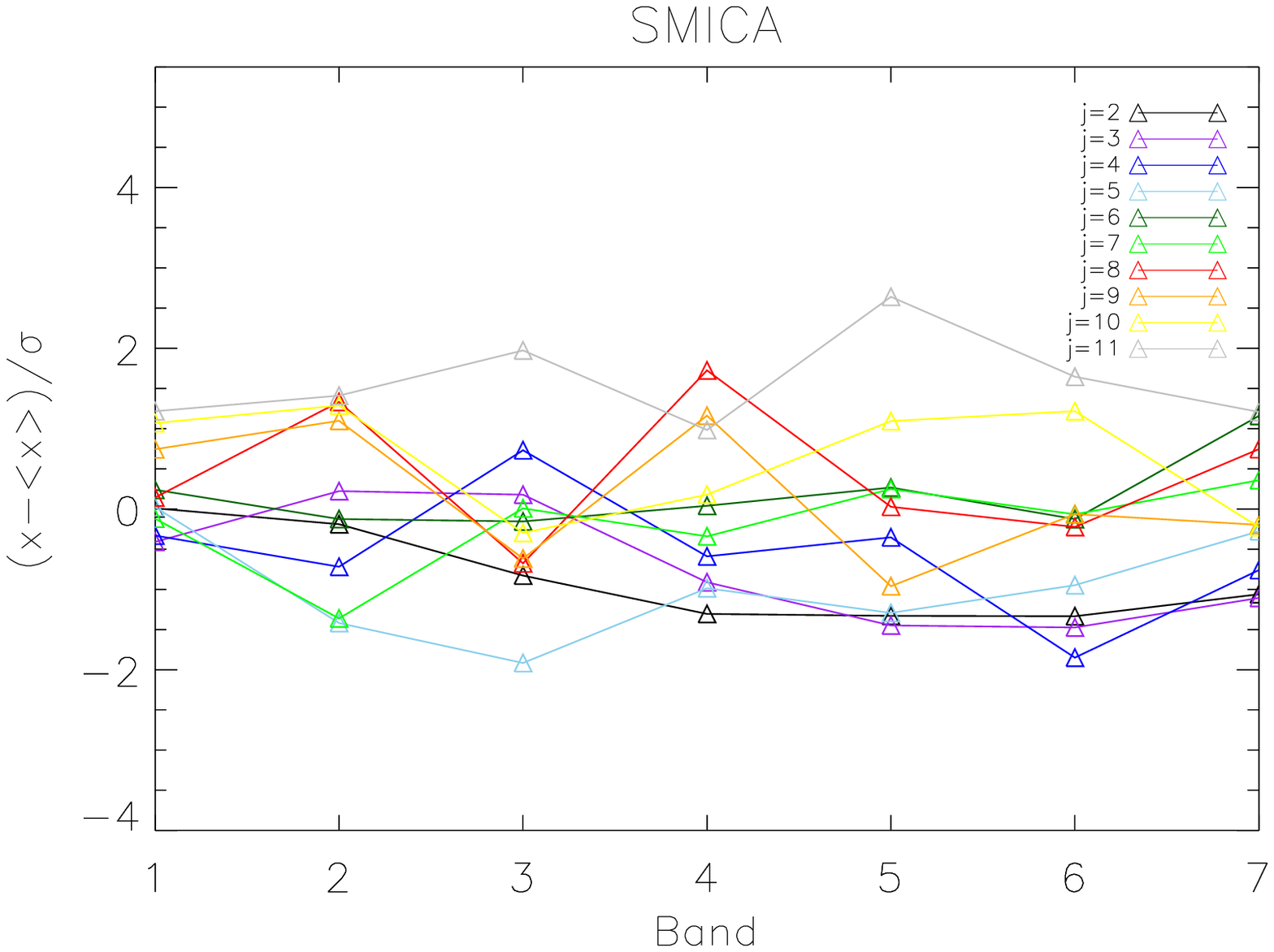}
    %\vspace{4ex}
  \end{minipage} 
   \begin{minipage}[b]{0.8\linewidth}
    \centering
    \includegraphics[width=\linewidth]{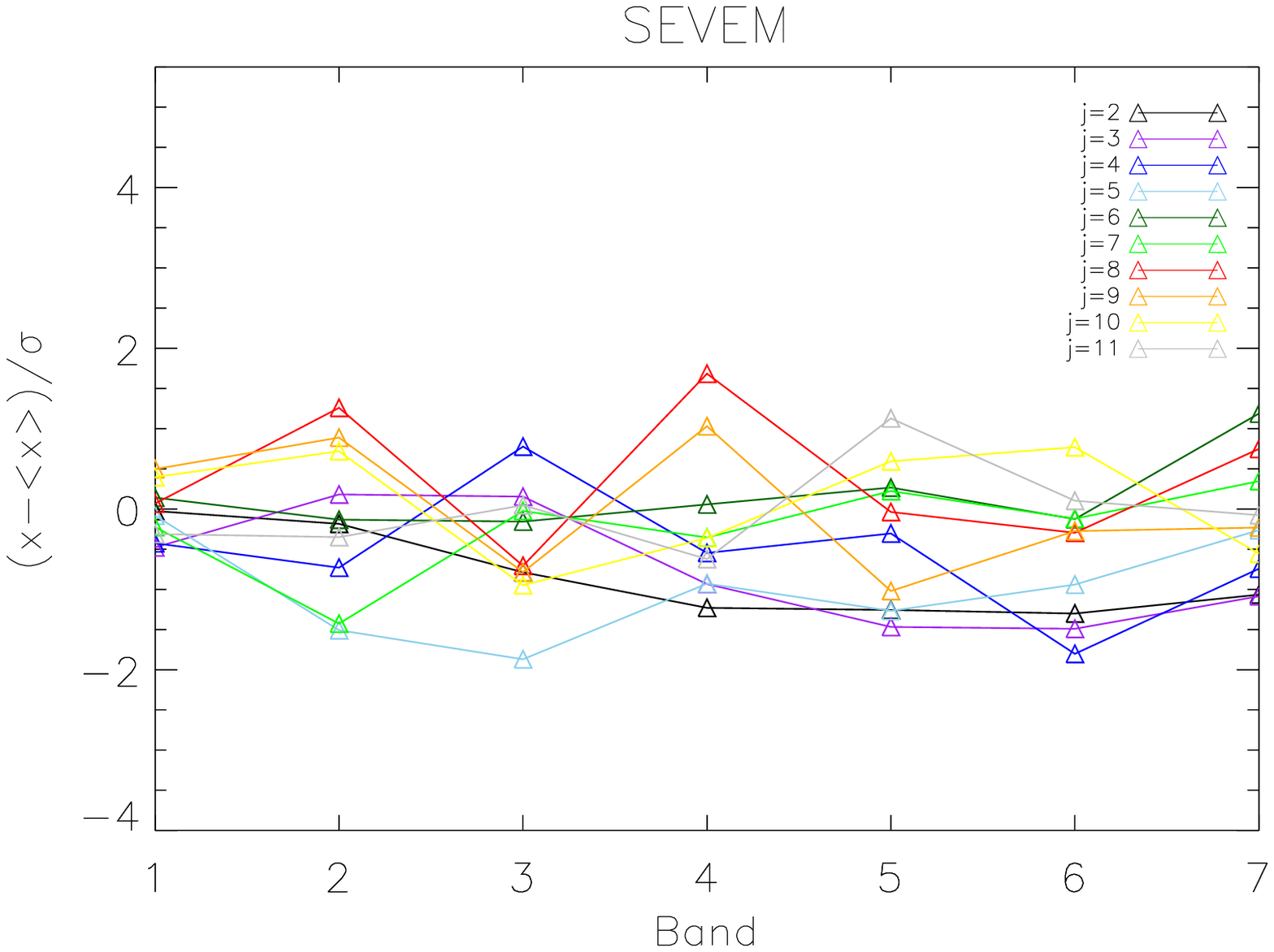}
    %\vspace{4ex}
  \end{minipage}
    \begin{minipage}[b]{0.8\linewidth}
      \centering
      \includegraphics[width=\linewidth]{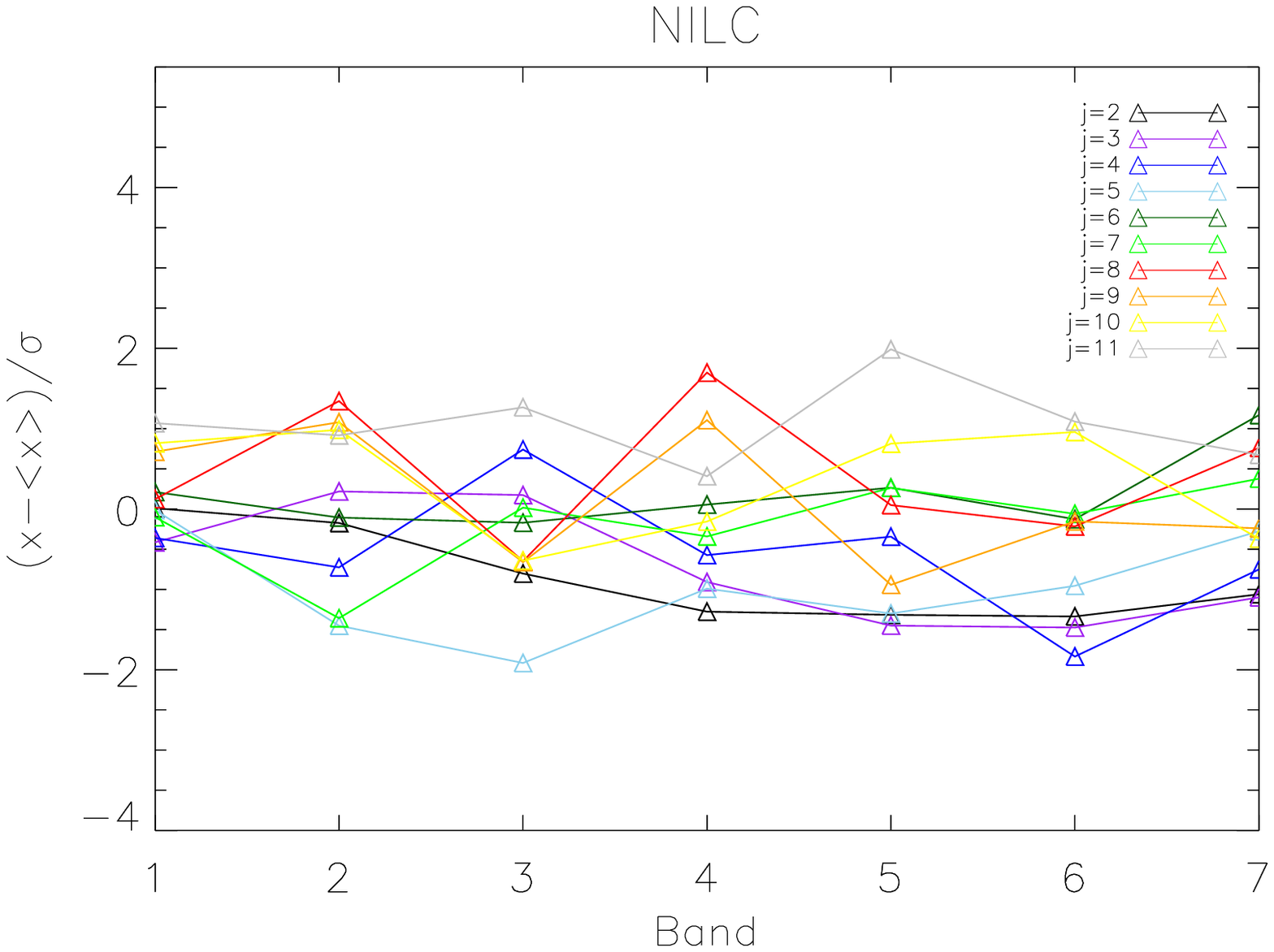}
      %\vspace{4ex}
    \end{minipage}
    \begin{minipage}[b]{0.8\linewidth}
      \centering
      \includegraphics[width=\linewidth]{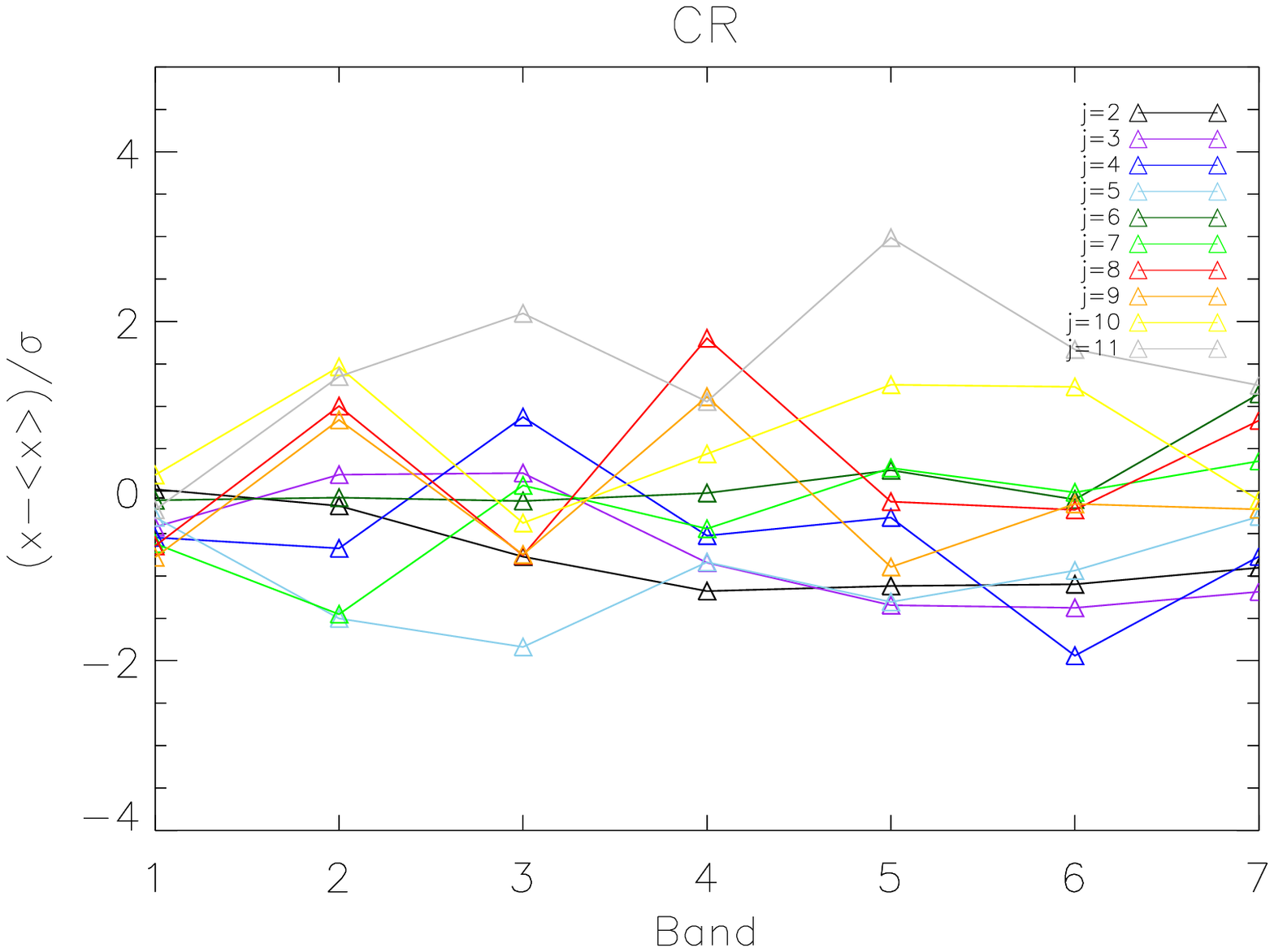}
      %\vspace{4ex}
    \end{minipage}
    \caption{Same as \fig\ref{needchis} for variance only but now plotted with the band number on the x-axis and with color codes indicating needlet scales. \label{needchis_j}}
\end{figure}

In this section we present the results for each individual foreground cleaned map. These maps have both CMB and noise present, although the noise is sub-dominant on most scales. In  \fig\ref{clchis}, \ref{needchis} and \ref{needchis_j} we show results on the power spectrum as well as needlet mean, variance and skewness, band by band and scale by scale. While \fig \ref{needchis} shows the results as a function of needlet scale, \fig \ref{needchis_j} shows the same but as a function of band on the x-axis. The purpose of the former is to show the scale dependence, the purpose of the latter is to show whether there is an increase towards band 1 (galactic plane) which could indicate foreground residuals.

We find very good agreement between data and simulations, although from \fig \ref{needchis} one can clearly see, in particular for the variance of the needlet coefficients, the effect of unresolved point sources on small scales (high $j$). This effect is seen even clearer in the difference maps presented in the next section. Note also that both the mean and skewness of band 2 appears systematically below zero over most scales. 
In the simulated data we found that in $30\%$ of the cases, the mean lies below zero on all scales in at least one band. For skewness this occurred
in $12\%$ of the simulations. Therefore we conclude that the behavior of band 2 can be well explained as a statistical fluctuation. 
%In the simulated maps we found that this occurred for the mean in at least one band in $30\%$ of the simulations and for skewness in $12\%$ of 
%the simulations. We therefore conclude that the behavior of band 2 can be well explained as a statistical fluctuation.
Note further in \fig \ref{needchis_j} that for the three largest scales there is a clear increase towards the galactic plane in all methods. This is particularly seen in bands 1-3, the ones closest to the galactic equator. This is only seen in the variance of the needlet coefficients. The variance can only increase with foregrounds (while the mean and skewness can increase or decrease), as foreground residuals would generally not subtract power from the map. This increase in variance towards the galactic plane, although the increase is towards the expected variance, can therefore be interpreted as an increasing level of foreground residuals. This is further supported by the fact that this increase disappears with an extended mask as we will show later. We do not show mean and skewness in this figure as no signs of residuals were seen in those cases.

\section{Difference map analysis}
\label{analysis2}

In this section we analyze inter-method consistency of the component separation by looking at difference maps between six pairs of the four available foreground cleaned maps. These difference maps consist only of noise and differences in foreground residuals between the methods. Since the CMB has been eliminated the difference maps are much more sensitive to foreground residuals and we will use these maps to quantify to which degree the foreground cleaned maps are reliable. Then in the next section we will use these results in order to suggest an improved common mask.

Due to the higher sensitivity of the difference maps to foreground residuals, we find residuals in most bands and scales for most of the 
computed quantities. 
Knowledge of whether these residuals may bias cosmological results is of very high interest. 
%Of highest interest is knowledge of whether these residuals may bias cosmological analysis of the CMB. 
We therefore plot $(x-\langle x\rangle)/\sigma_\mathrm{CMB}$ instead of $(x-\langle x\rangle)/\sigma$  where $\sigma_\mathrm{CMB}$ is the standard deviation 
%including CMB 
derived from maps with both CMB and noise in them. On the other hand $\sigma$ is the expected noise standard deviation of the difference maps. In this way we measure the residuals in units of fraction of the standard deviation of CMB fluctuations. If the residuals are larger than $0.2\sigma_\mathrm{CMB}$ it means that they may bias cosmological results by the order of $0.2\sigma_\mathrm{CMB}$. For needlet skewness the residuals are still small, so in this case we show $(x-\langle x\rangle)/\sigma$ as previously.

\begin{figure}%[h!]
\includegraphics[width=0.8\linewidth,angle=0]{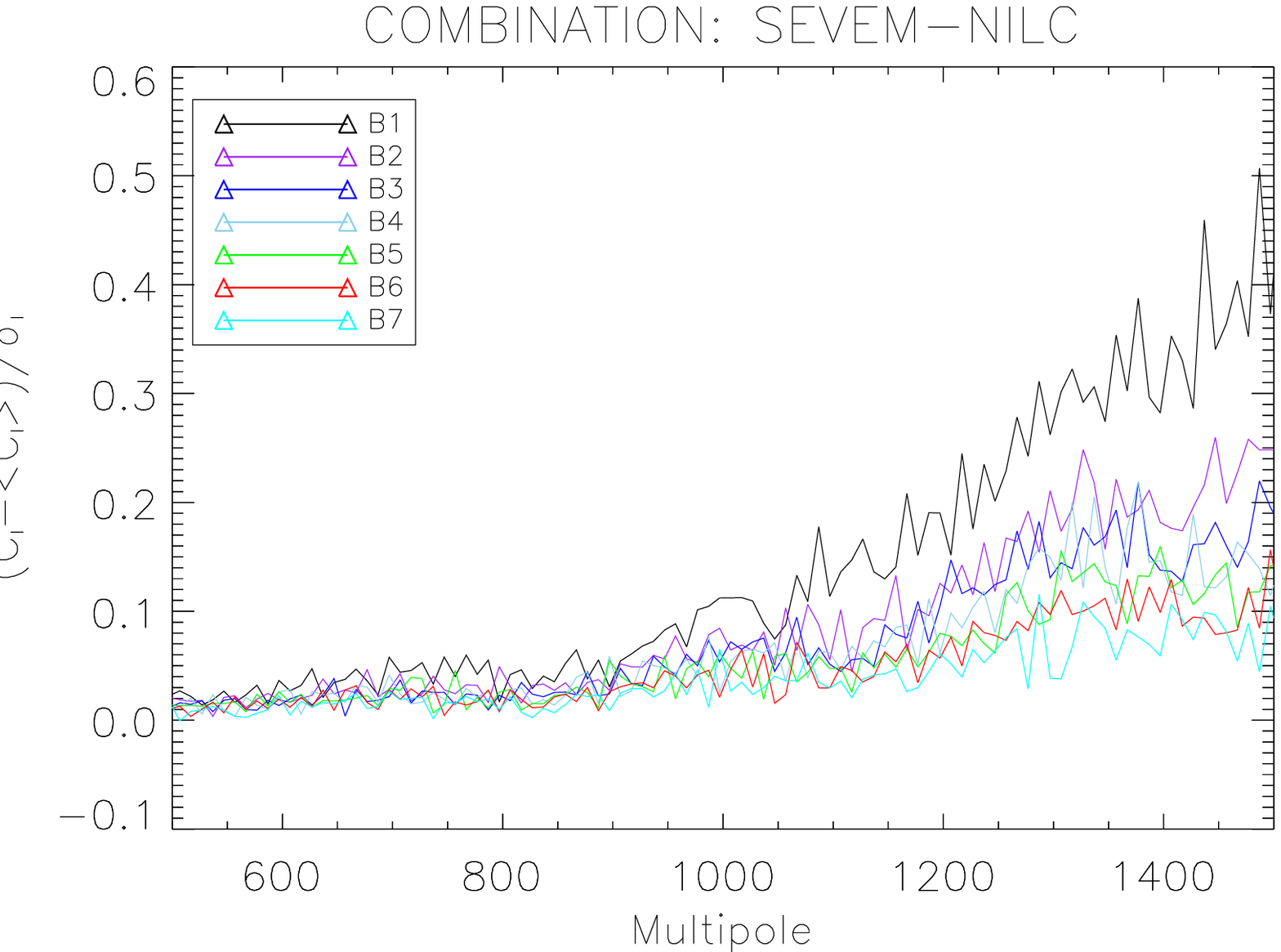}
\includegraphics[width=0.8\linewidth,angle=0]{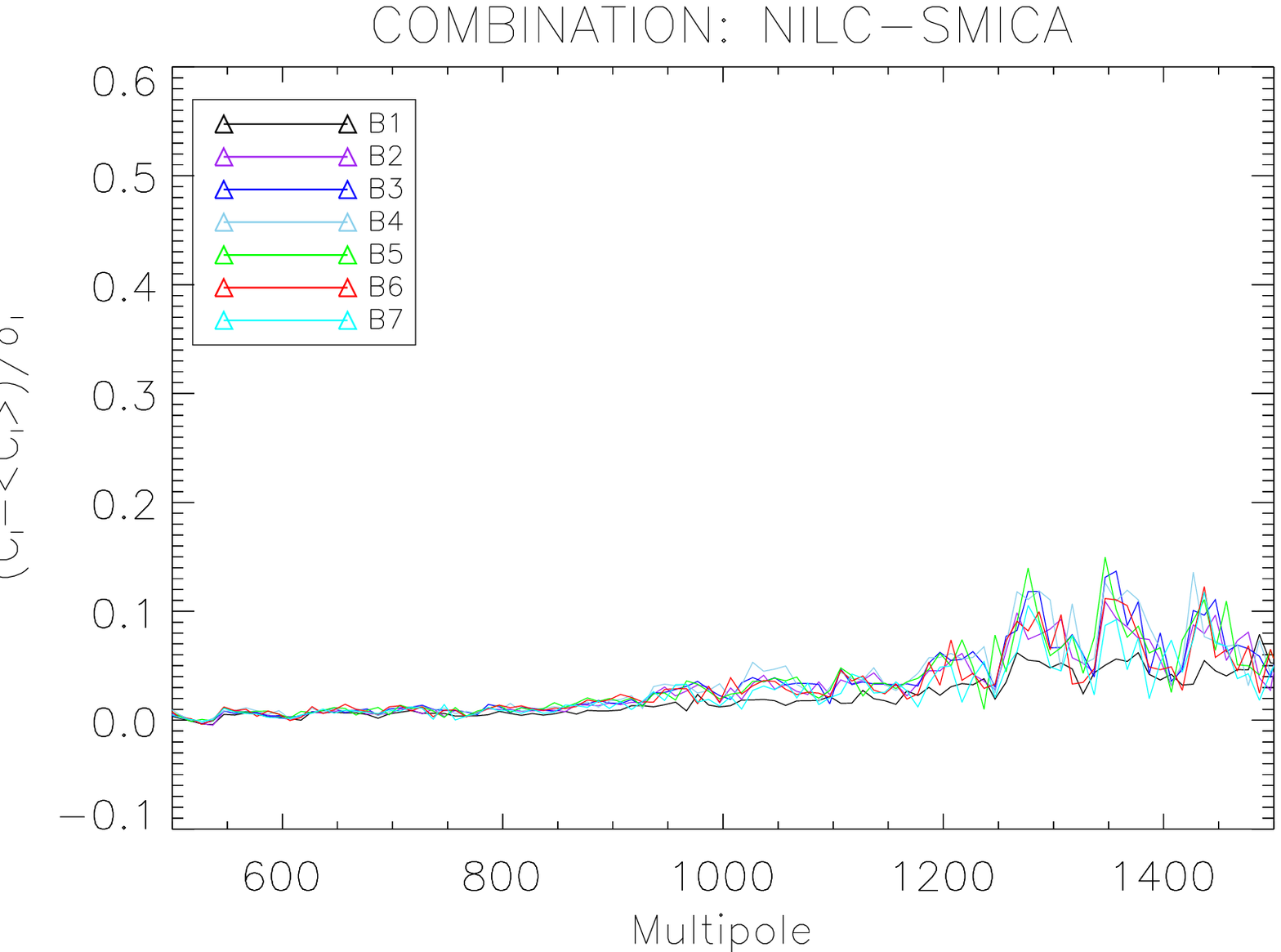}
\includegraphics[width=0.8\linewidth,angle=0]{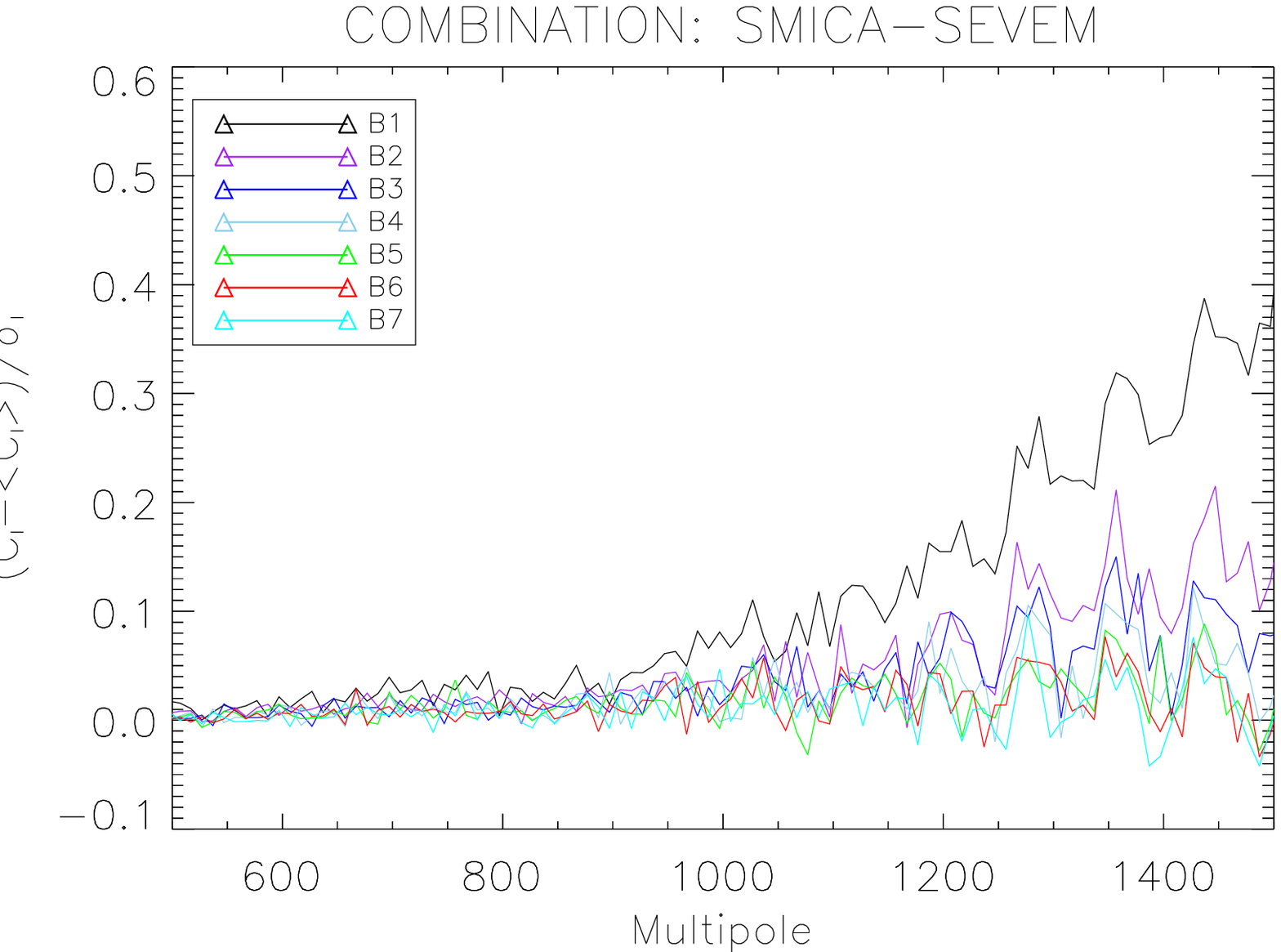}
\includegraphics[width=0.8\linewidth,angle=0]{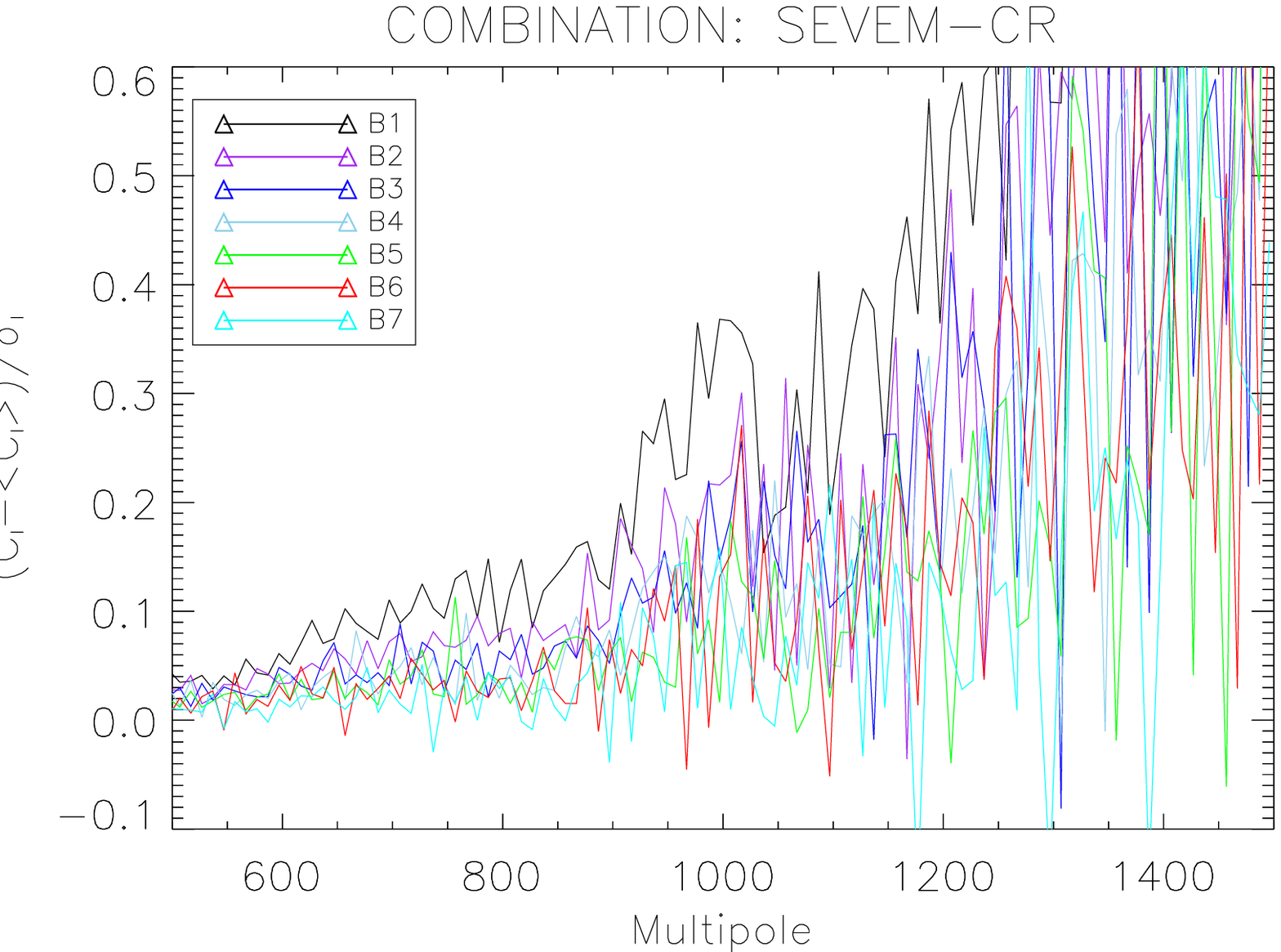}
\caption{$(C_\ell-\langle C_\ell\rangle)/\sigma_\mathrm{CMB}$ for \csevem-\cnilc, \cnilc-\csmica, \csmica-\sevem and \csevem-\ruler derived from MASTER estimated power spectra.
  \label{chival}}
\end{figure} 

The results for the power spectrum are shown in \fig\ref{chival}, for the multipole interval $\ell \in [500,1500]$. For values $[0,500]$ the agreement
between simulations and data is perfect, and hence not shown. We will first consider the \csmica, \nilc and \sevem maps: First note the general increase towards smaller scales from unresolved point sources visible in all bands. As we approach $\ell = 1500$ we notice that the difference increases,
especially for the difference maps including \csevem.  This increase is particularly large in the two bands close to the galactic plane where most foreground residuals are expected. The difference \cnilc-\smica is generally much smaller than the differences including 
\sevem suggesting residuals which are either present only in \sevem or common for both \nilc and \csmica.
In order to obtain further information, we continue with wavelet space analysis.

\begin{figure*}[htb] 

  \begin{minipage}[b]{0.33\linewidth}
    \centering
    \includegraphics[width=\linewidth]{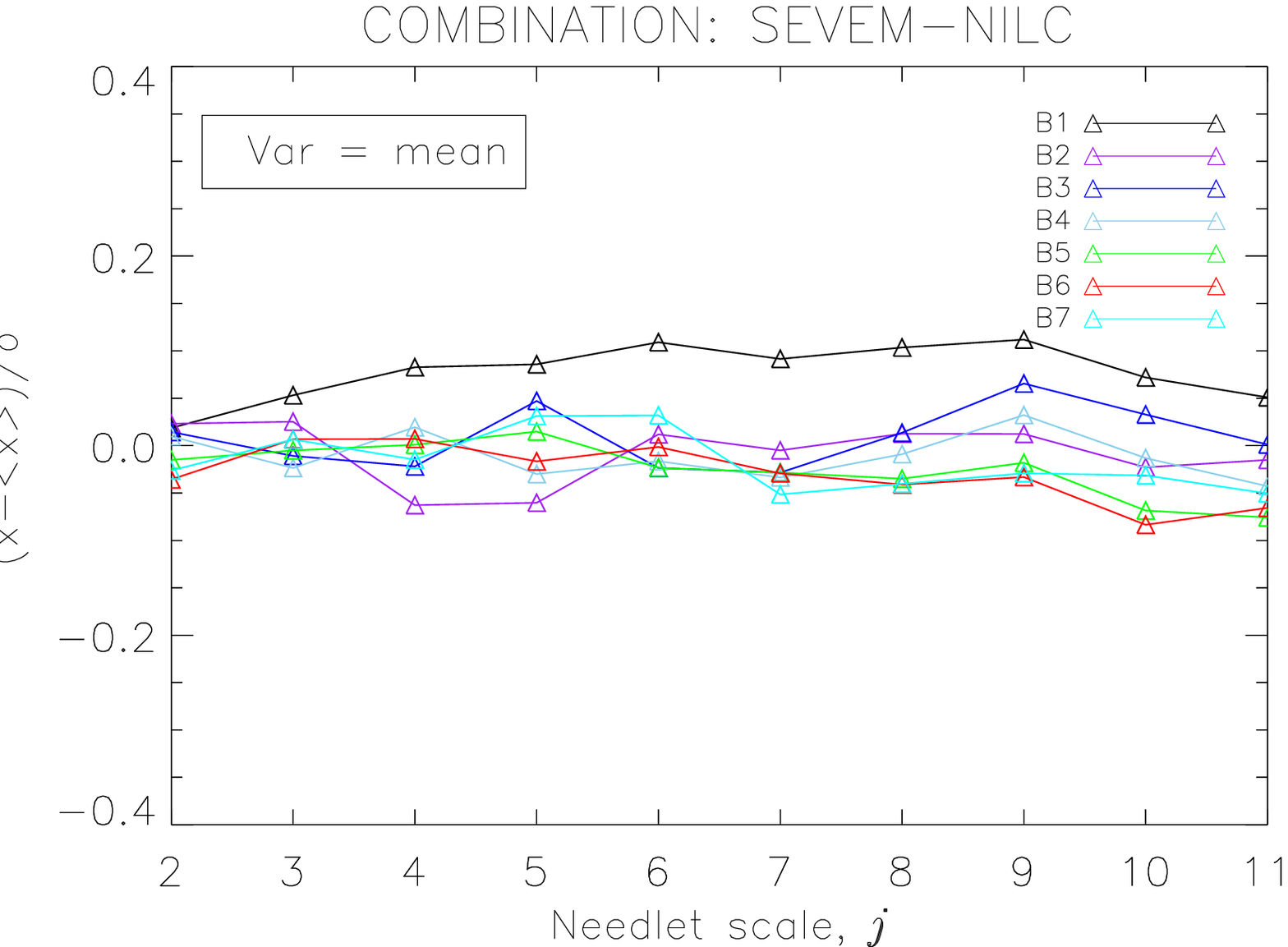}
       %\vspace{4ex}
  \end{minipage}
  \begin{minipage}[b]{0.33\linewidth}
    \centering
    \includegraphics[width=\linewidth]{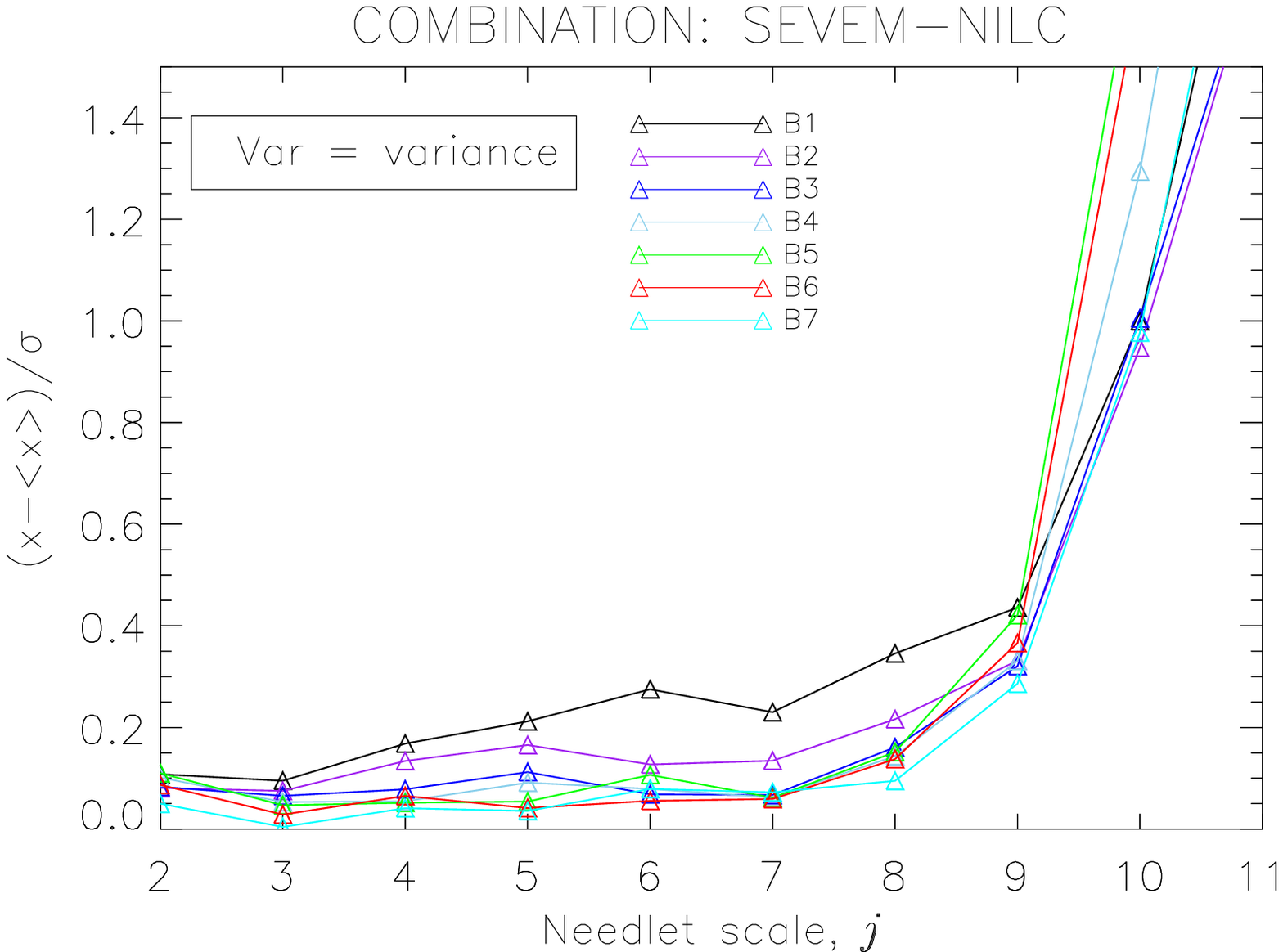}
    %\vspace{4ex}
  \end{minipage} 
  \begin{minipage}[b]{0.33\linewidth}
    \centering
    \includegraphics[width=\linewidth]{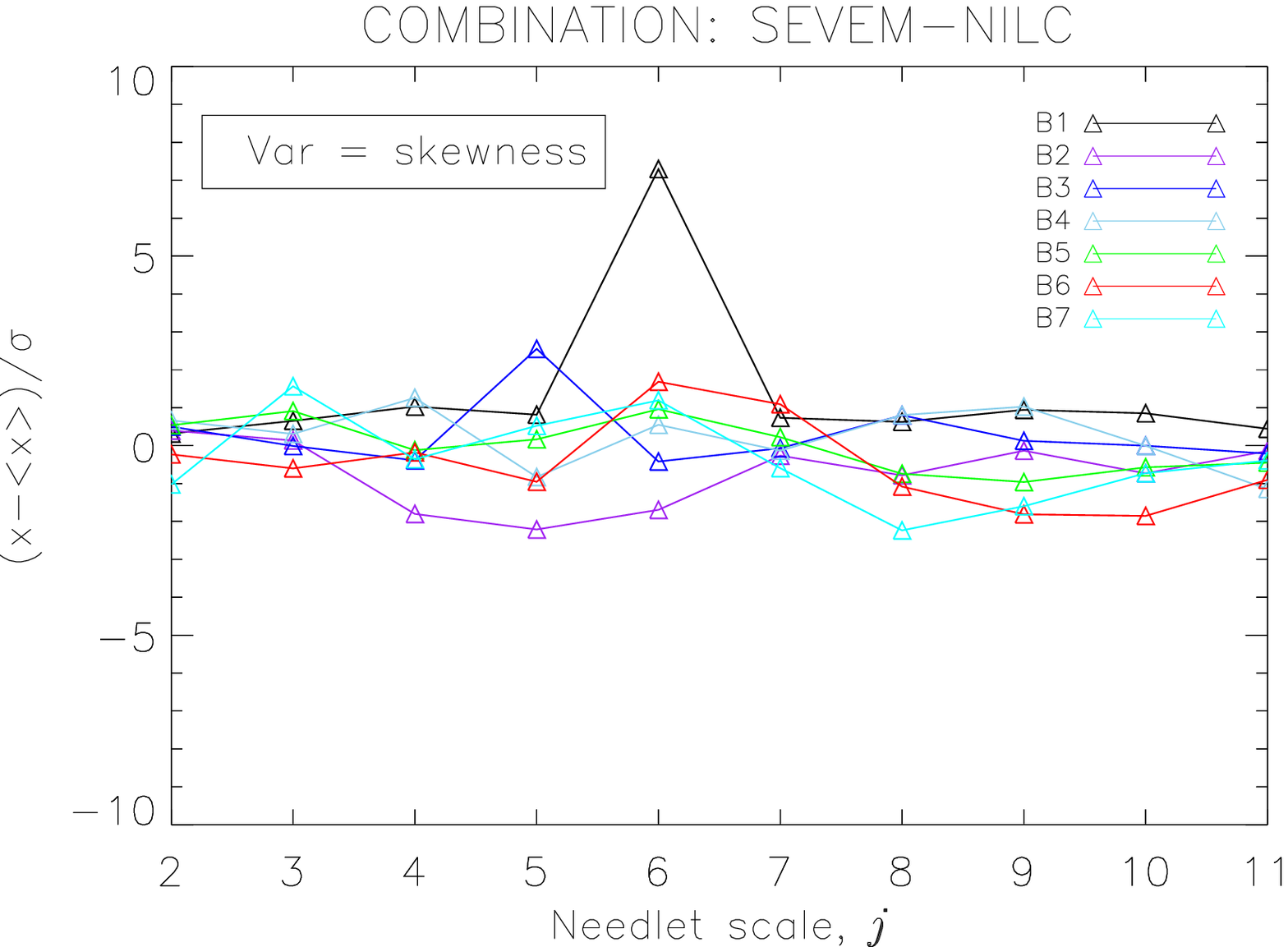}
    %\vspace{4ex}
  \end{minipage}%% 
\\ % needed for AA format (for some reason)
  \begin{minipage}[b]{0.33\linewidth}
    \centering
    \includegraphics[width=\linewidth]{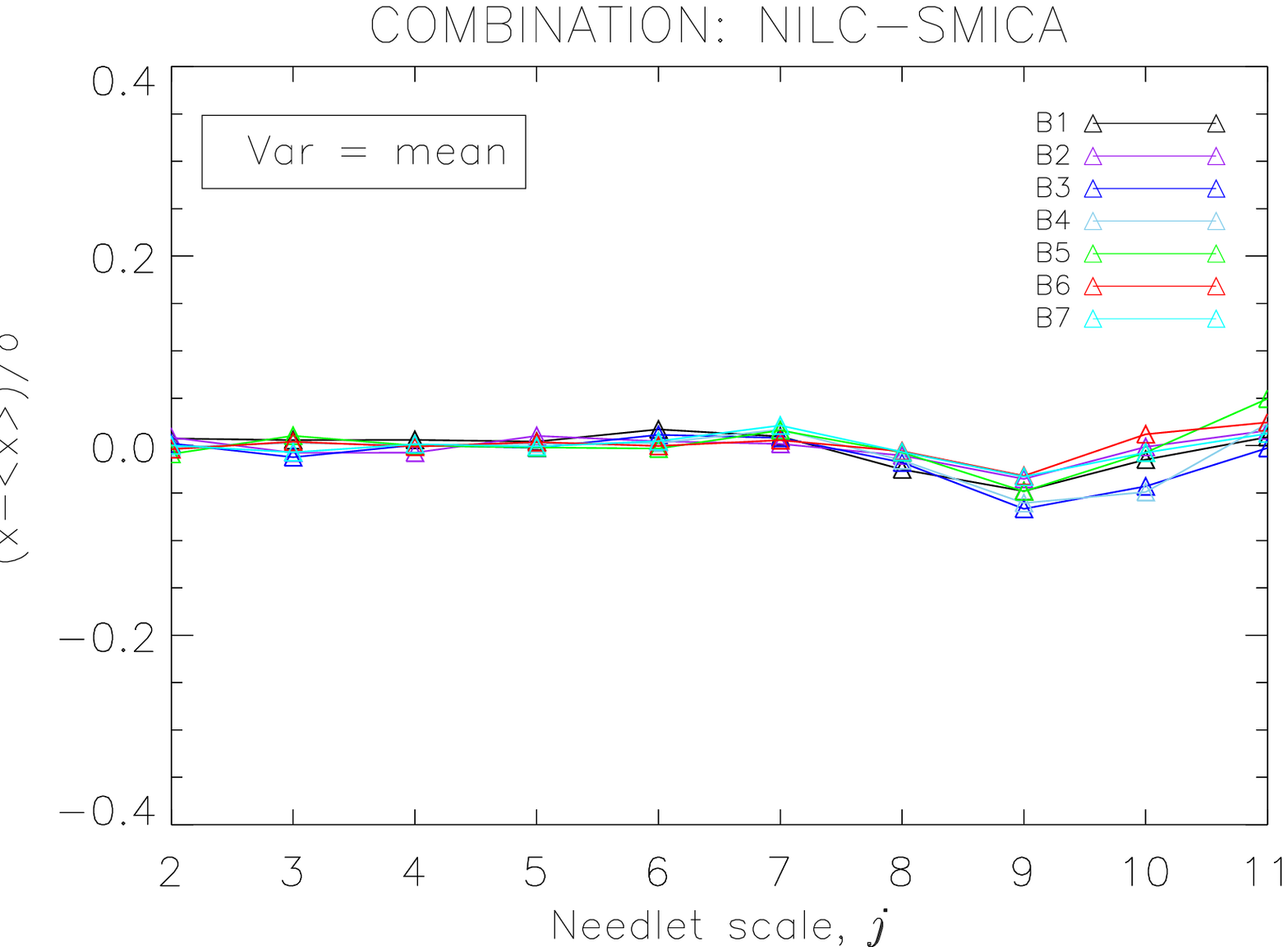}
    %\vspace{4ex}
  \end{minipage}
   \begin{minipage}[b]{0.33\linewidth}
    \centering
    \includegraphics[width=\linewidth]{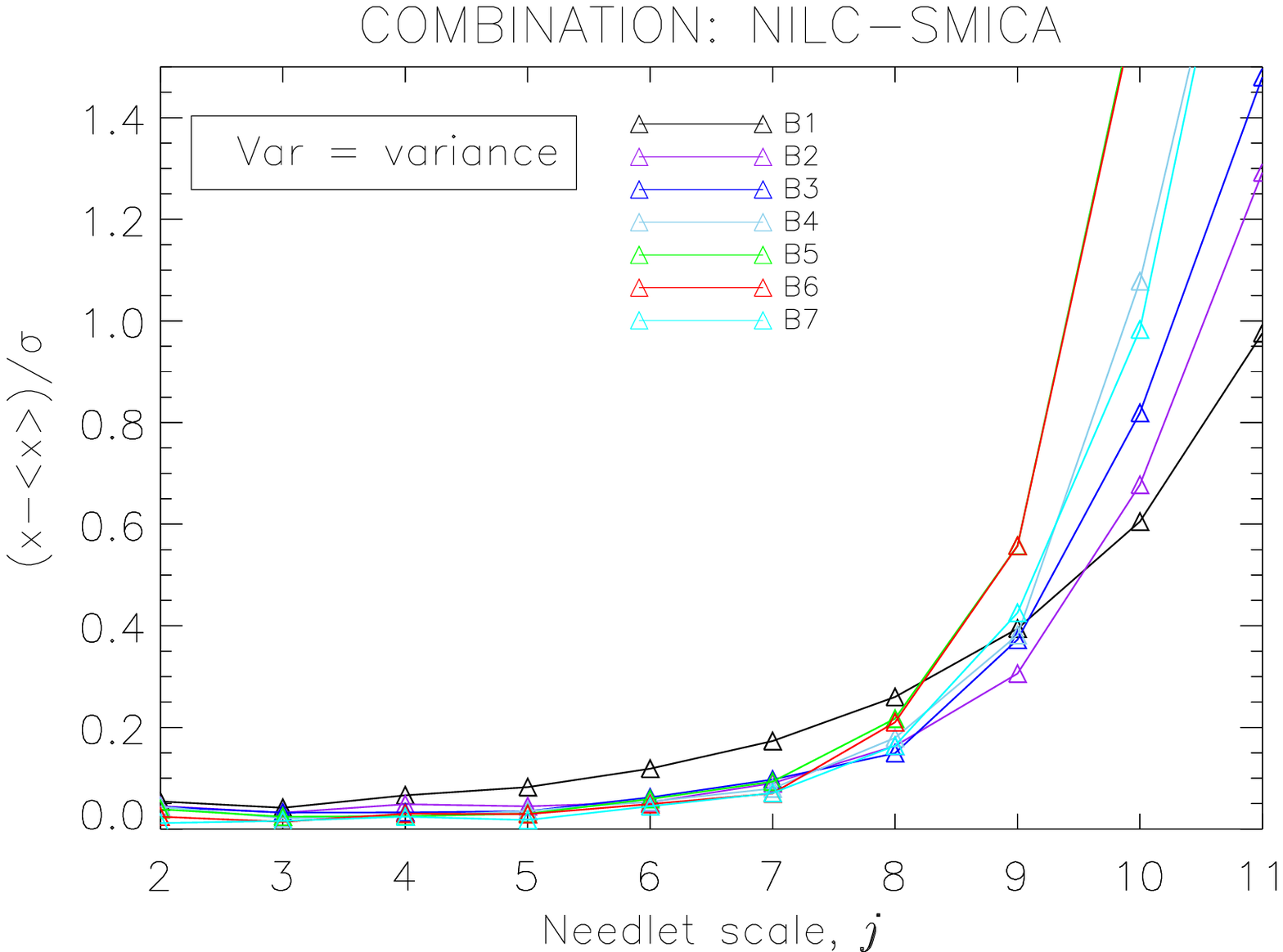}
    %\vspace{4ex}
  \end{minipage}
    \begin{minipage}[b]{0.33\linewidth}
    \centering
    \includegraphics[width=\linewidth]{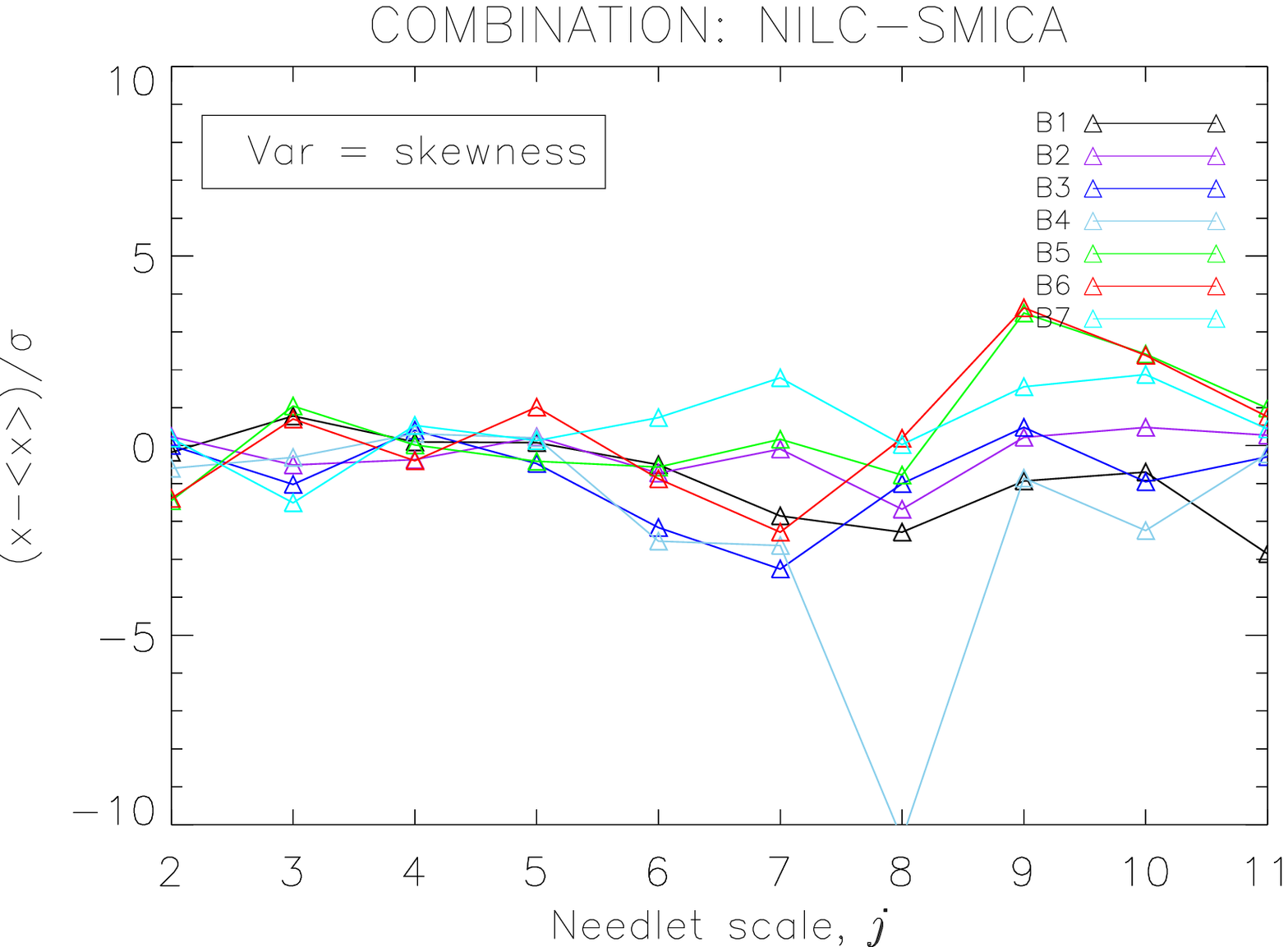}
    %\vspace{4ex}
  \end{minipage}
    \begin{minipage}[b]{0.33\linewidth}
      \centering
      \includegraphics[width=\linewidth]{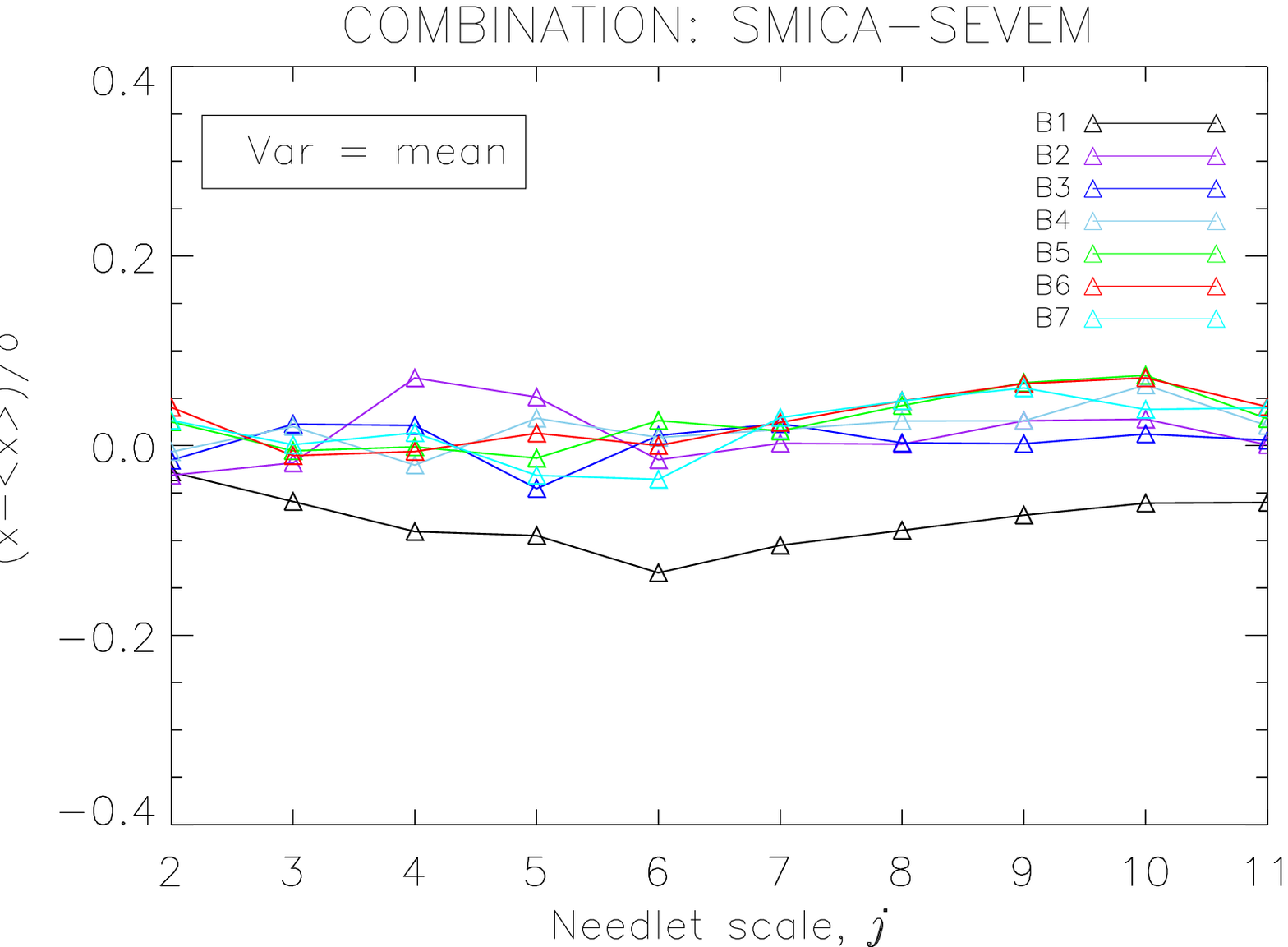}
    %\vspace{4ex}
    \end{minipage}
    \begin{minipage}[b]{0.33\linewidth}
      \centering
      \includegraphics[width=\linewidth]{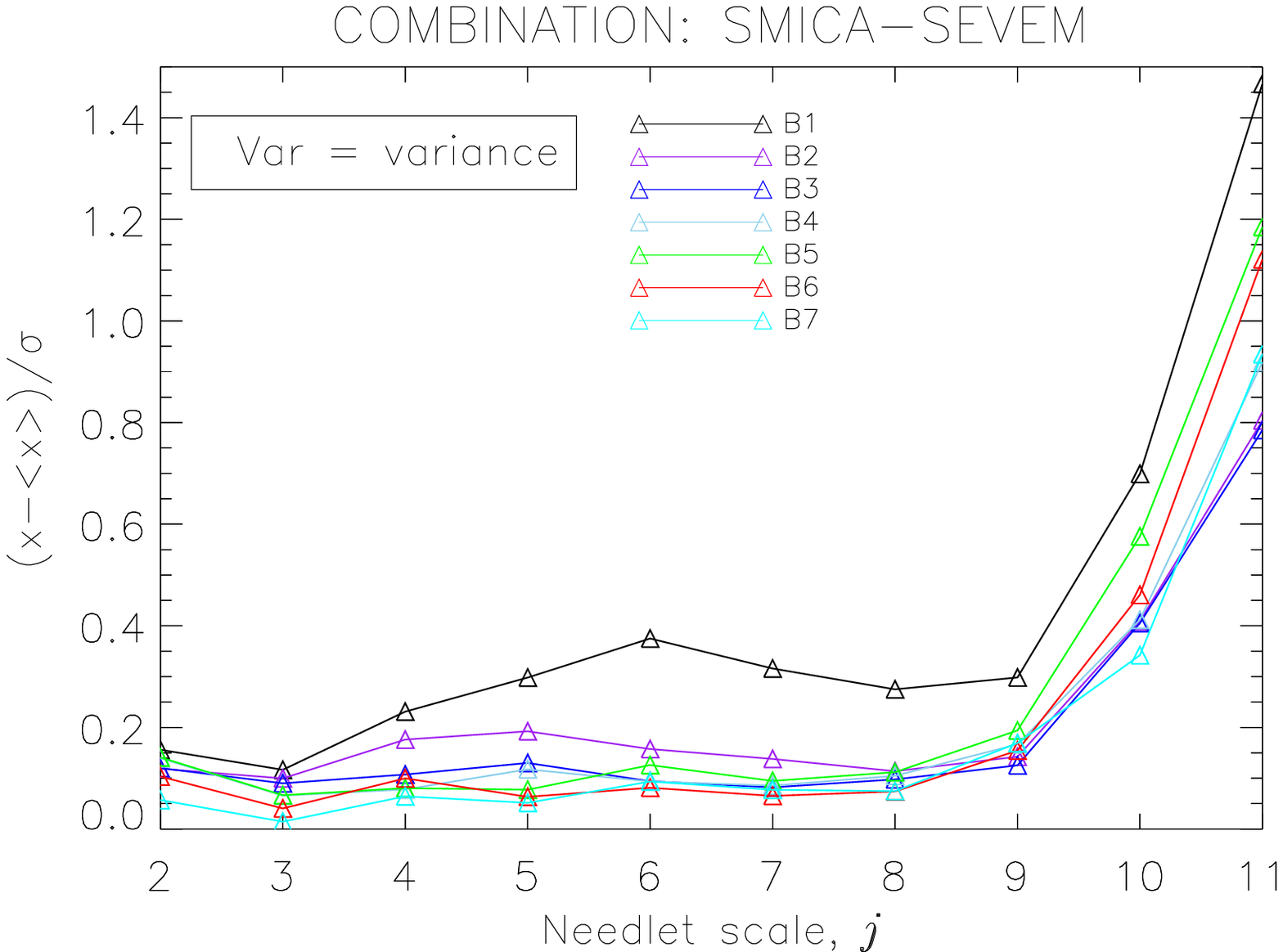}
      %\vspace{4ex}
    \end{minipage}
    \begin{minipage}[b]{0.33\linewidth}
    \centering
    \includegraphics[width=\linewidth]{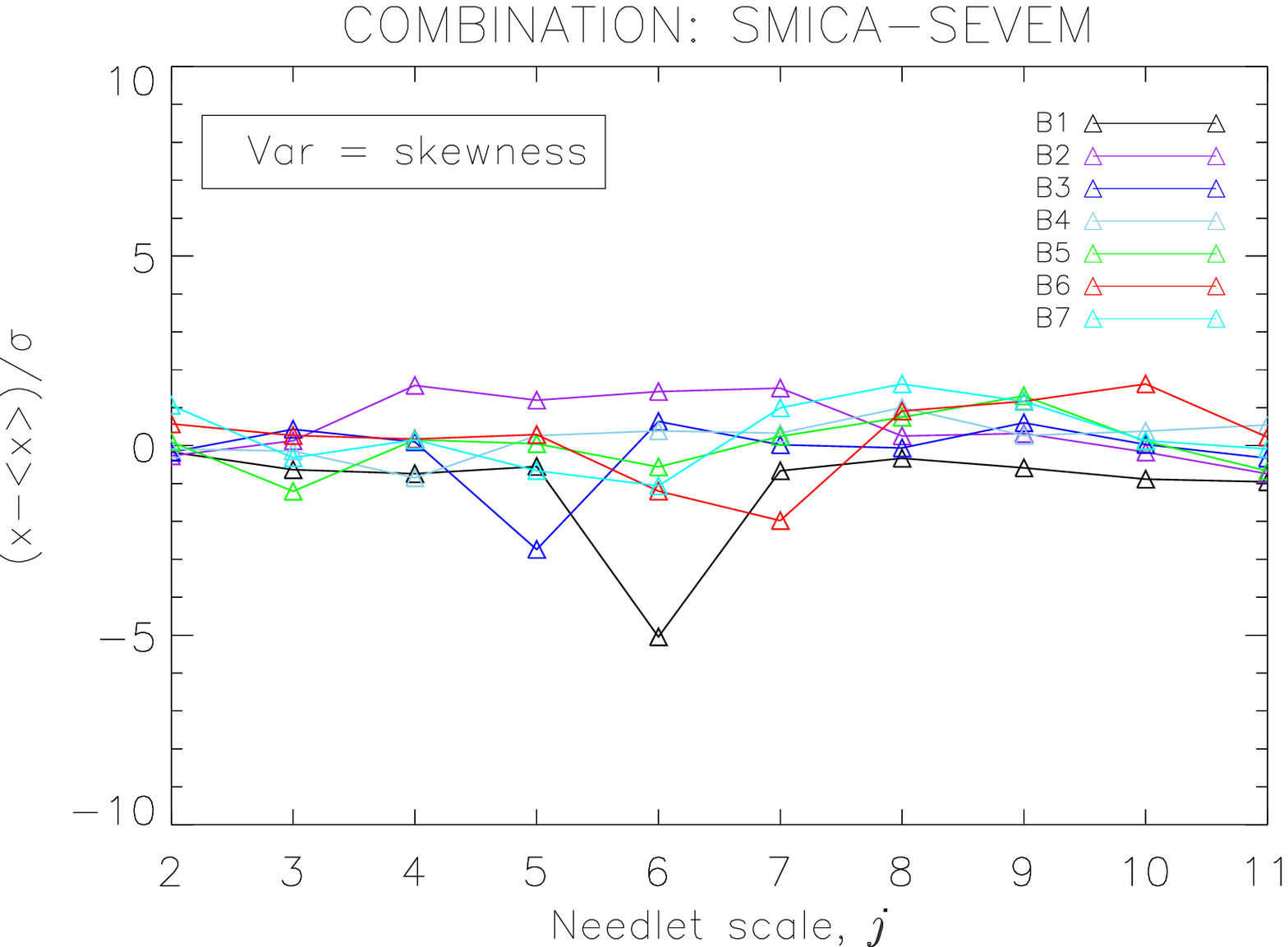}
    %\vspace{4ex}
    \end{minipage}
   \begin{minipage}[b]{0.33\linewidth}
      \centering
      \includegraphics[width=\linewidth]{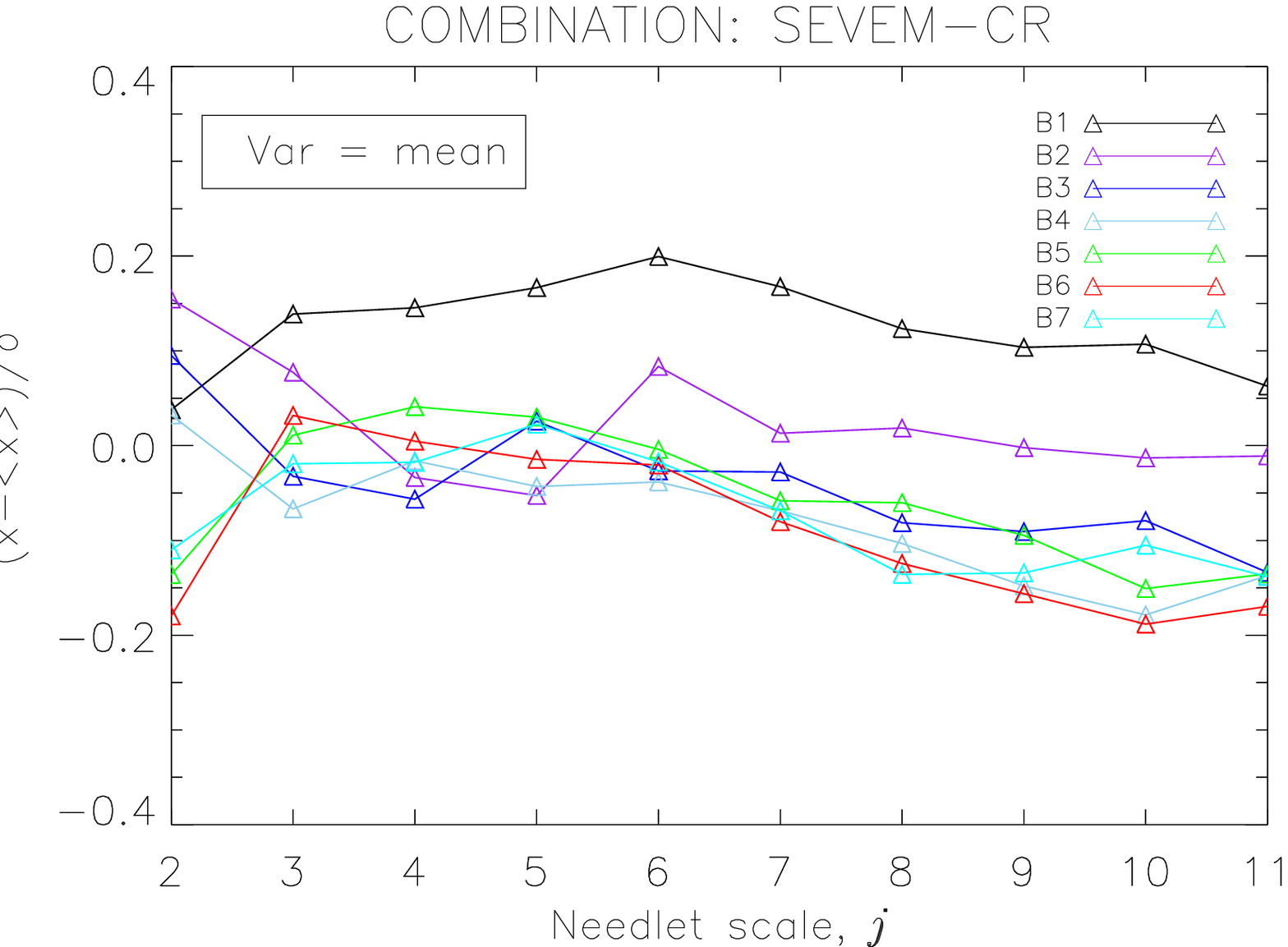}
    %\vspace{4ex}
    \end{minipage}
    \begin{minipage}[b]{0.33\linewidth}
      \centering
      \includegraphics[width=\linewidth]{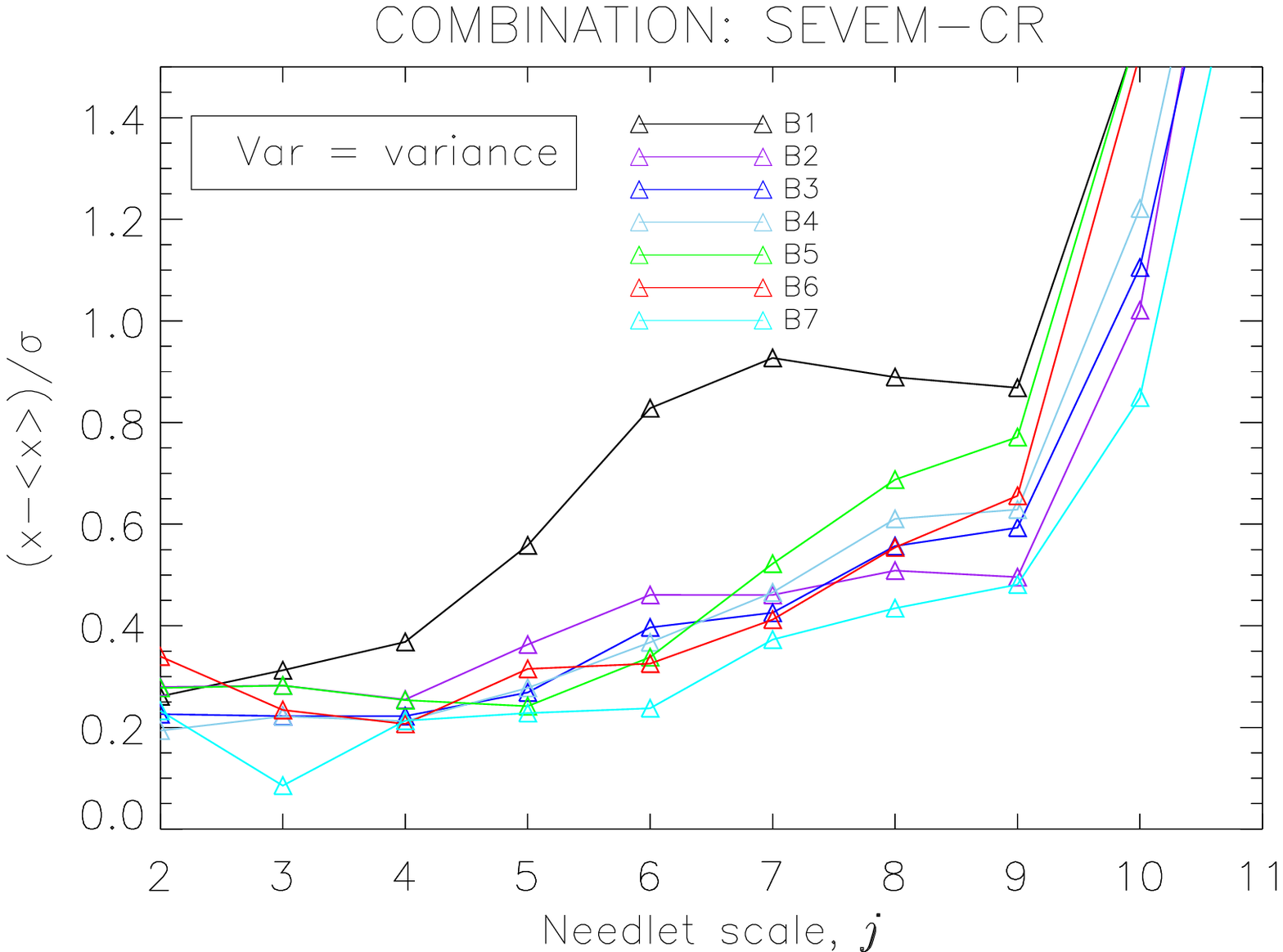}
      %\vspace{4ex}
    \end{minipage}
    \begin{minipage}[b]{0.33\linewidth}
    \centering
    \includegraphics[width=\linewidth]{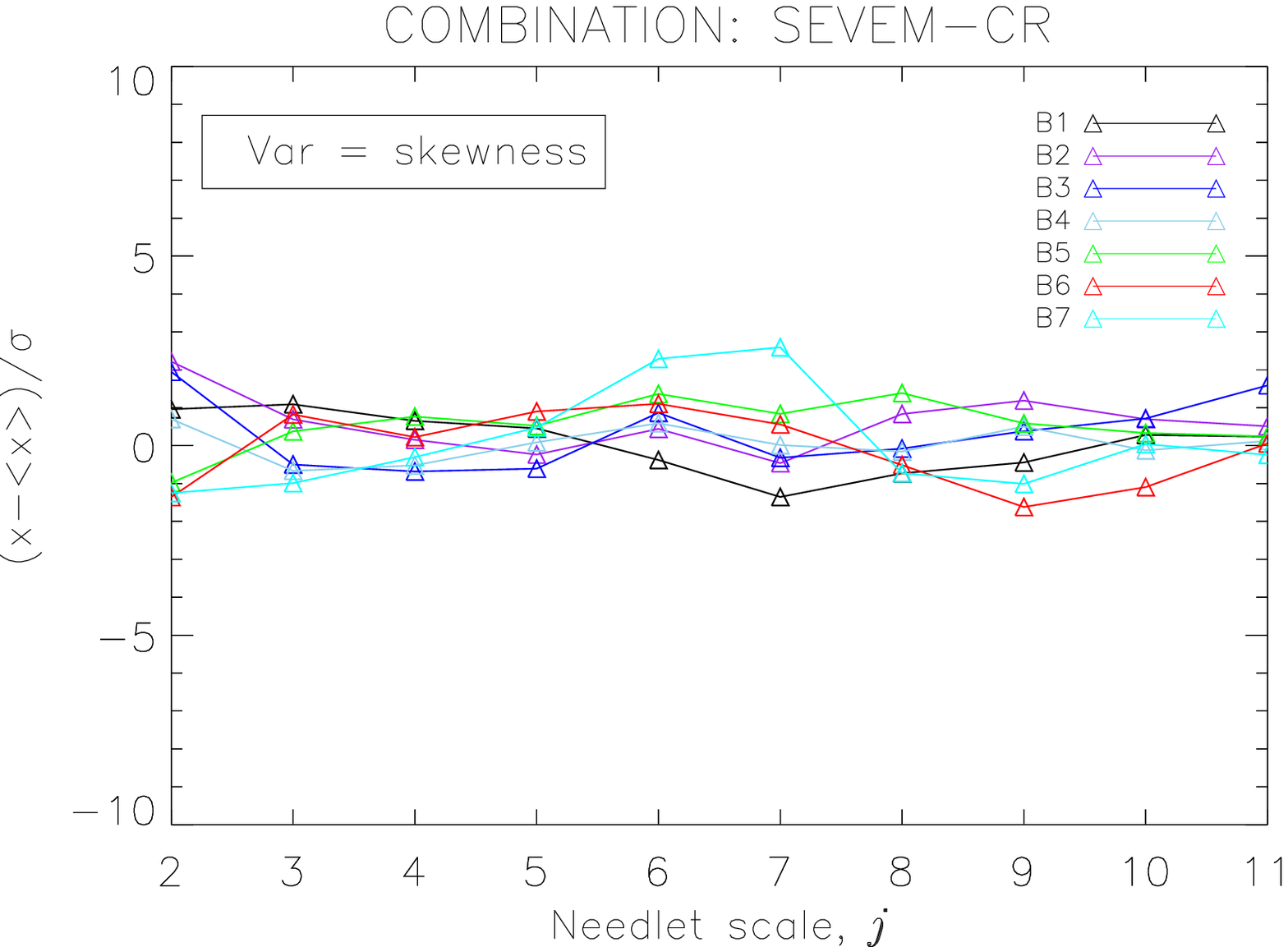}
    %\vspace{4ex}
    \end{minipage}
    \caption{ Plot of $(x-\langle x\rangle)/\sigma$ where $x$ corresponds to mean (left column), variance (middle column) and skewness (right column) of needlet coefficients computed computed on \csevem-\cnilc, \cnilc-\csmica, \csmica-\sevem and \csevem-\ruler. 
      Notice also that the standard deviation used in the skewness plots is the pure noise standard deviation while for mean and variance, the standard deviation is the one from CMB plus noise. The various bands, B1 to B7 are shown in \fig\ref{bands}. \label{diffchivals} }
\end{figure*}

\begin{figure}[htb] 
  \begin{minipage}[b]{0.8\linewidth}
    \centering
    \includegraphics[width=\linewidth]{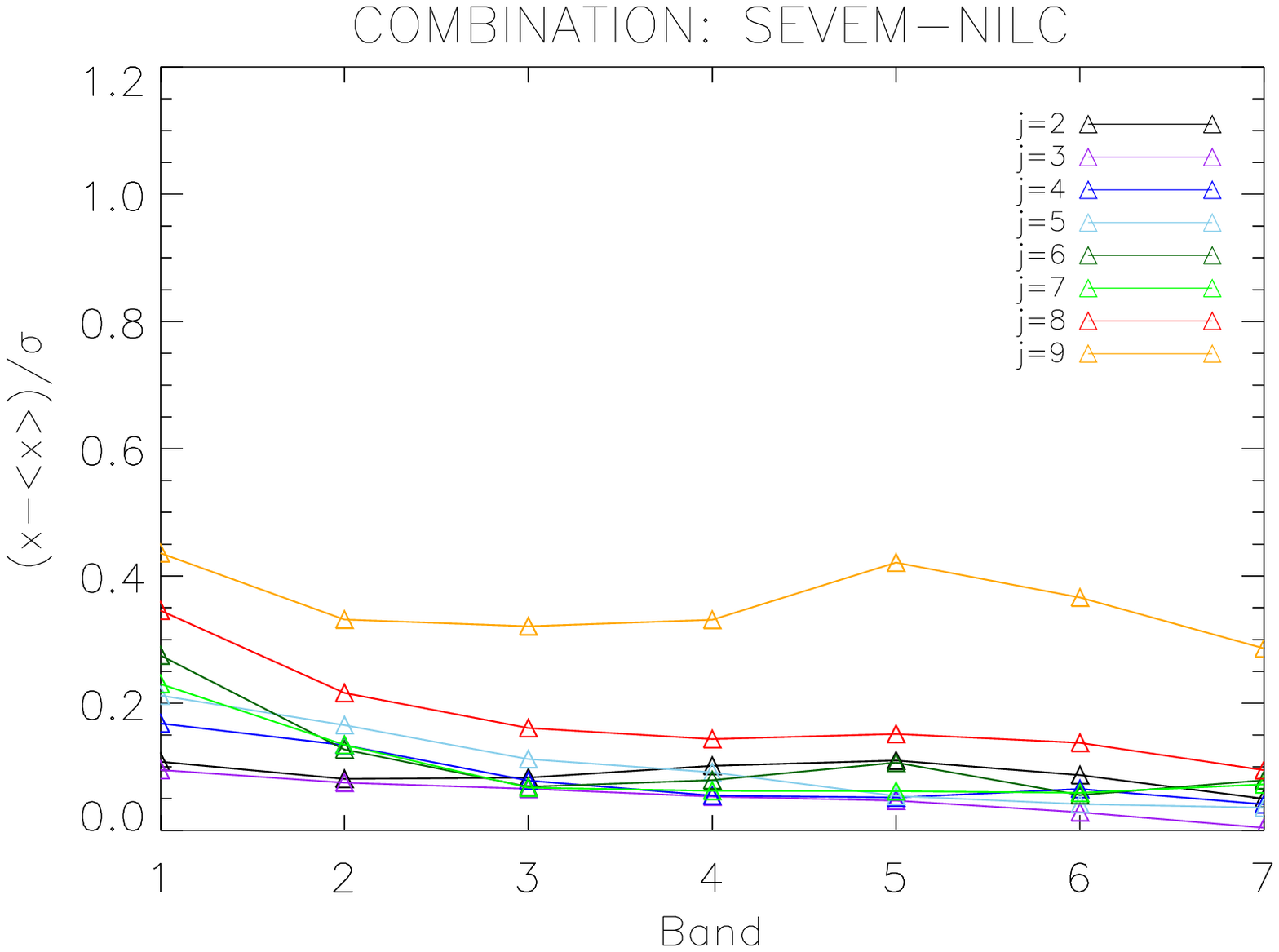}
    %\vspace{4ex}
  \end{minipage} 
   \begin{minipage}[b]{0.8\linewidth}
    \centering
    \includegraphics[width=\linewidth]{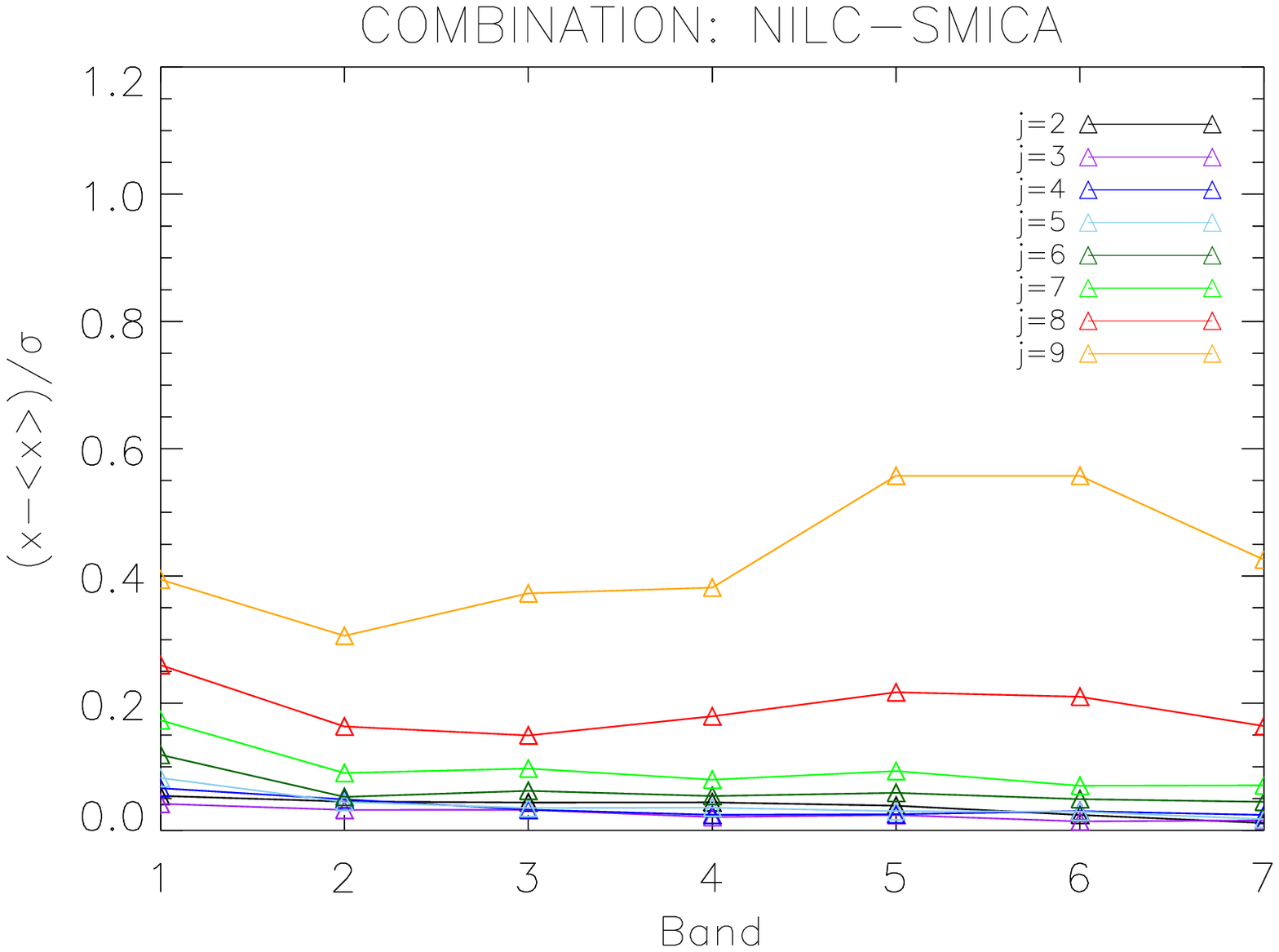}
    %\vspace{4ex}
  \end{minipage}
    \begin{minipage}[b]{0.8\linewidth}
      \centering
      \includegraphics[width=\linewidth]{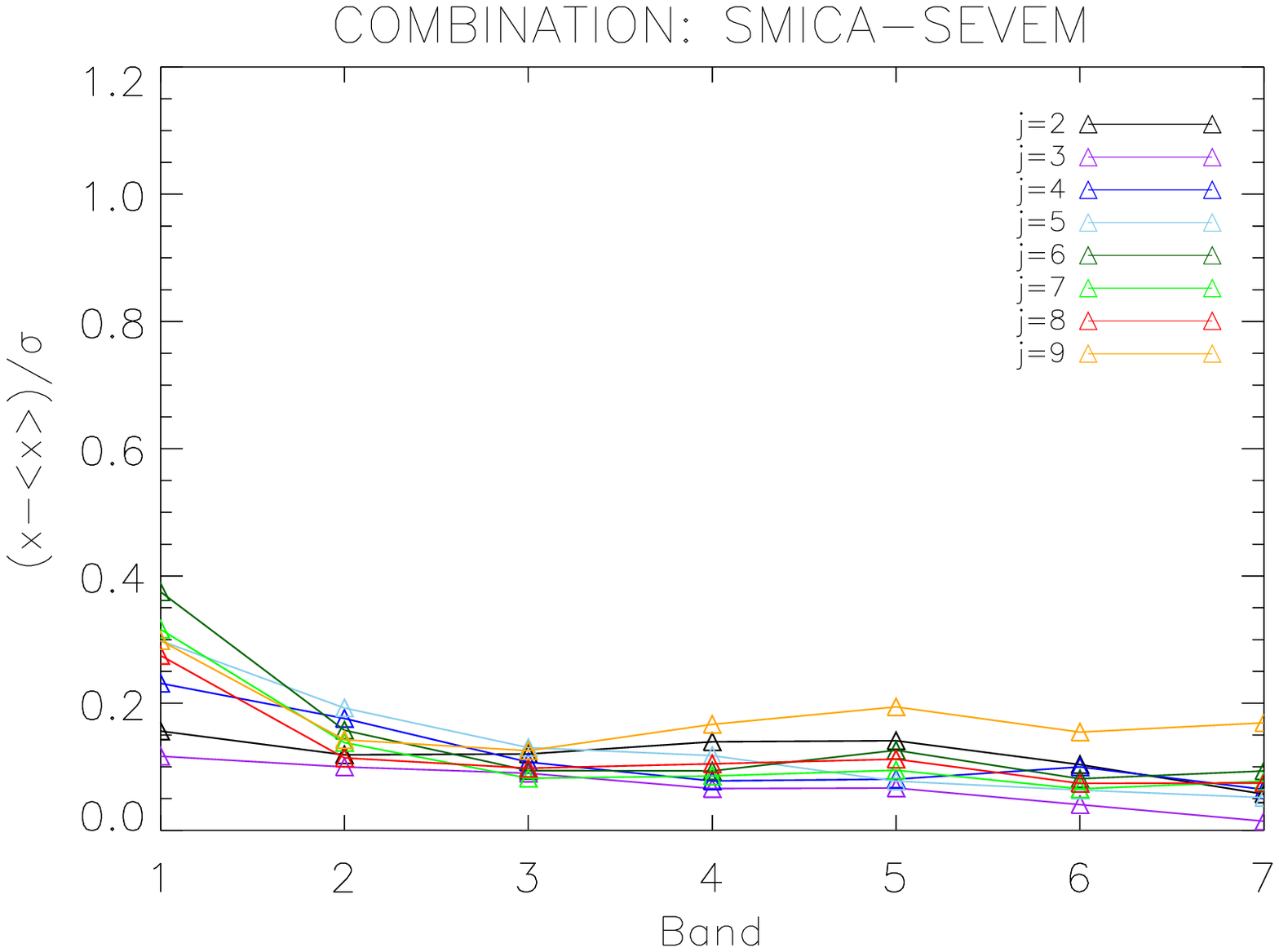}
      %\vspace{4ex}
    \end{minipage}  
    \begin{minipage}[b]{0.8\linewidth}
      \centering
      \includegraphics[width=\linewidth]{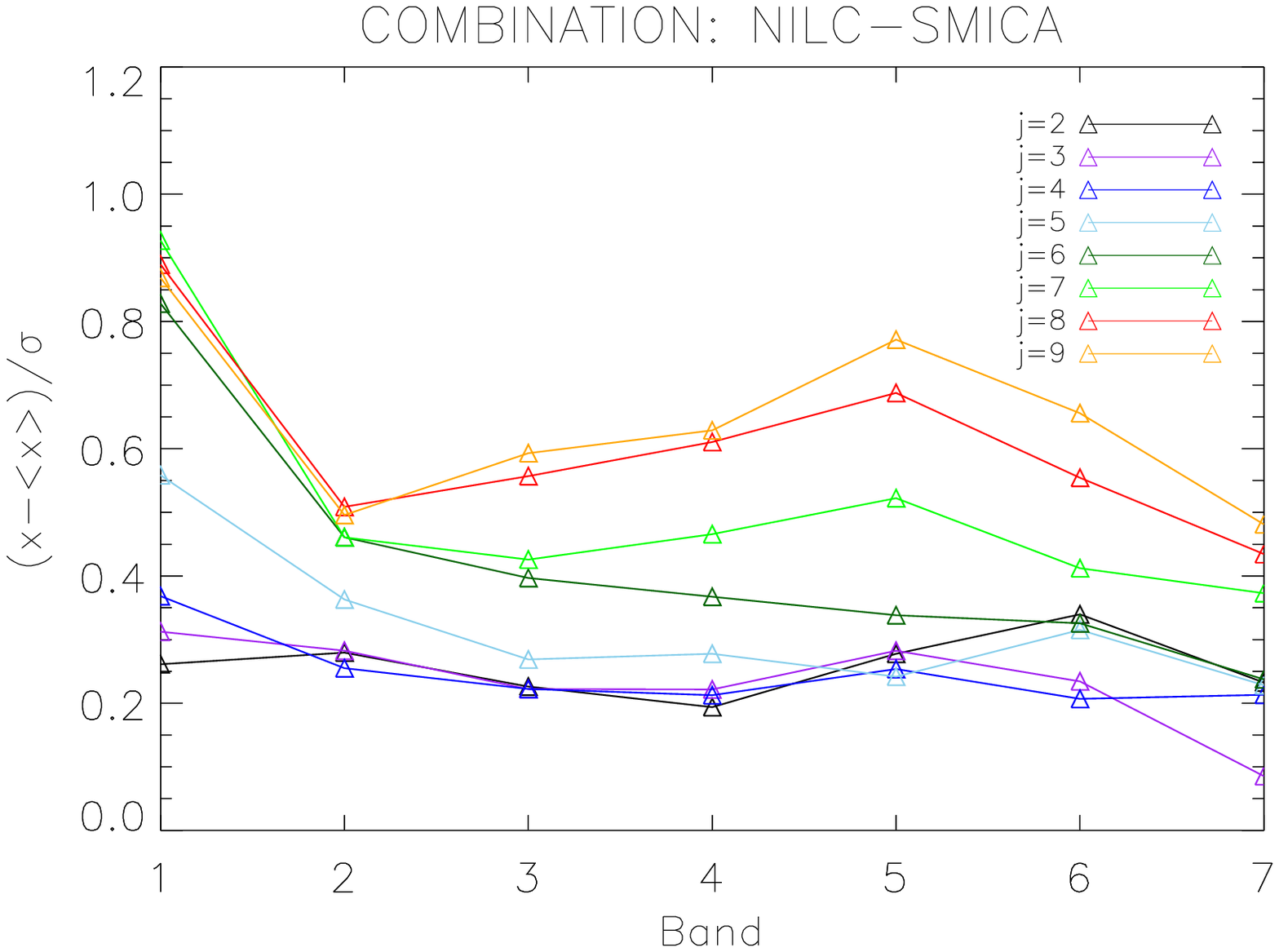}
      %\vspace{4ex}
    \end{minipage}
    \caption{Same as \fig\ref{diffchivals} for variance only but now plotted with the band number on the x-axis and with color codes indicating needlet scales. \label{diffchivals_j} }
\end{figure}

In \fig\ref{diffchivals} and \ref{diffchivals_j} we show the results for the moments of the needlet coefficients. Looking at the variance measure for larger scales, one important observation in variance is that while for \csevem, \nilc and \smica combinations, the residuals are $<0.2\sigma_\mathrm{CMB}$ for bands $>2$, \csevem-\ruler (and all other \ruler combinations, not shown) have residuals $>0.2\sigma_\mathrm{CMB}$ for all bands. In fact, even when extending the mask as described in the next section, we are unable to improve results with \ruler combinations significantly. We conclude that the \ruler map has larger differences compared to the other three maps, than the other three maps have between themselves. We are therefore, as detailed in the next section, capable of creating an extended mask with improved results using \cnilc, \smica and \sevem only. The \ruler map however is too different to allow for construction of a common mask which brings all four maps in full agreement. We therefore decided to exclude the \ruler map from the work in the next section.

However, we want to point out (1) the fact that while the \ruler map is different from the other three maps, these differences are still so tiny that they did not show up when single channel analysis including CMB was performed in the previous section. Furthermore (2) we cannot conclude from this that \ruler is the map with the highest foreground residuals. The approach used in the construction of the \ruler map takes better into account variation of foreground properties across the sky compared to the other three methods. It can therefore not be excluded that there are common residuals in the other three maps which give rise to the larger difference between \ruler and other methods.

In the following we will consider only combinations with \csmica, \nilc and \csevem. Looking at the mean of needlet coefficients we find again that \nilc and \smica are very similar with differences $<0.1\sigma_{\mathrm{CMB}}$ for all bands, while \sevem shows differences $>0.1\sigma_{\mathrm{CMB}}$ for band 1 (close to the galactic plane) compared to the other two maps. 
 
We find that the variance measure is the measure most sensitive to foreground residuals. First of all the strong increase at the last 2-3 needlet scales due to unresolved point sources is now very visible. We observe that band 1 again shows strong deviation ($>0.2\sigma_{\mathrm{CMB}}$) between methods, now also visible in the difference \cnilc-\csmica. The skewness measure also supports the fact that there are large differences between methods in band 1. Band 4 shows a very strong outlier in skewness only for the \cnilc-\smica difference map. We have not been able to identify the source of this latter difference, however, with the new mask which is derived in the following sections, we find that the skewness outliers previously present on band 1 now disappear. Looking at \fig\ref{diffchivals_j} we can clearly see the increase towards the galactic plane for the large scales, in particular for bands 1 and 2.

These results provide an incentive to further study the bands closest to the galactic center, bands 1 and 2. It is already clear
from the inferences made so far, that these bands are not consistent between foreground reduction algorithms, however, we
may not infer from the obtained data, which of the methods, if not all, have residuals causing these inconsistencies. The strategy now, is to examine 
the needlet coefficients belonging to difference maps \csevem-\cnilc, \cnilc-\csmica, and \csmica-\sevem and use these to construct a new confidence mask.

\section{Improving the mask}
\label{mask}

The U73 mask is defined to be the union of all individual foreground method masks, meaning that if one of the method masks excludes a given pixel,
then the combined mask excludes it as well, even if the other masks include it. One might contemplate if it can be made smaller, or given the
results from the previous section, be extended. The current galactic mask, including masking of point sources, allows a fraction 
$f_{\mathrm{sky}} = 73.7\%$ of the sky to be used for cosmological analysis, deeming $26.3 \%$ of the sky improper. This stands in stark contrast
to, for example, the \smica mask which has an $f_{\mathrm{sky}} \sim 88\%$. To our knowledge, no 
analysis using all three foreground method maps simultaneously has been done in order to construct a joint confidence mask. Such an analysis is the topic of this section.

Our methodology is to examine the needlet coefficients scale by scale in wavelet space. This will allow construction of "scale masks", $M_{j}(i)$, 
where $j$ is the needlet scale, and $i$ is a pixel index. The advantage we have over methods that use pixel maps is that we can examine each scale 
individually and thus be more flexible. From these masks we can then define the complete mask as the product of scale masks over all 
relevant scales:
\be{newmask}
M(i) = \prod_j M_j(i)
\ee
where $M(i)$ is the total mask for a given difference map and the product runs over all relevant scales $j$.

We obtain the scale mask for a given pixel $i$ from the 
needlet coefficients of the difference map divided by their standard deviation obtained from simulations:
\be{scalemask}
M_j(i) = \left \{
\begin{array}{cl}
1 &~\mathrm{if}~\frac{\abs{\beta_{j}(i)}}{\sigma_{j}^\mathrm{CMB}(i)} \le \mathrm{threshold} \\
\\
0 &~\mathrm{else}
\end{array} \right.
\ee
where $\beta_j(i)$ is the needlet coefficent from a difference map and $\sigma_{j}^\mathrm{CMB}(i)$ is the corresponding
standard deviation including the expected standard deviation from CMB. We use difference maps for $j \in [3,11]$. In order to minimize the influence of 
foreground residuals, we require these to have values less than $0.1\sigma_{j}^\mathrm{CMB}$ for $j\le 7$. For $j>7$ the noise level is higher than 
$0.1\sigma_{j}^\mathrm{CMB}$ in some pixels and we therefore use the maximum value in the jack-knife difference map as a threshold. For the pixels 
exceeding the threshold we zero all pixels within a disc with scale dependent radius ranging between $24^\circ$ and $0.18^\circ$
at $j=3$ and $j=11$ respectively. The disk radius is calculated according to the recommended procedure in \cite{Scodeller2012}.
In this way we obtain a new and more conservative mask.

\begin{figure}[!t]
\includegraphics[width=\linewidth,angle=0]{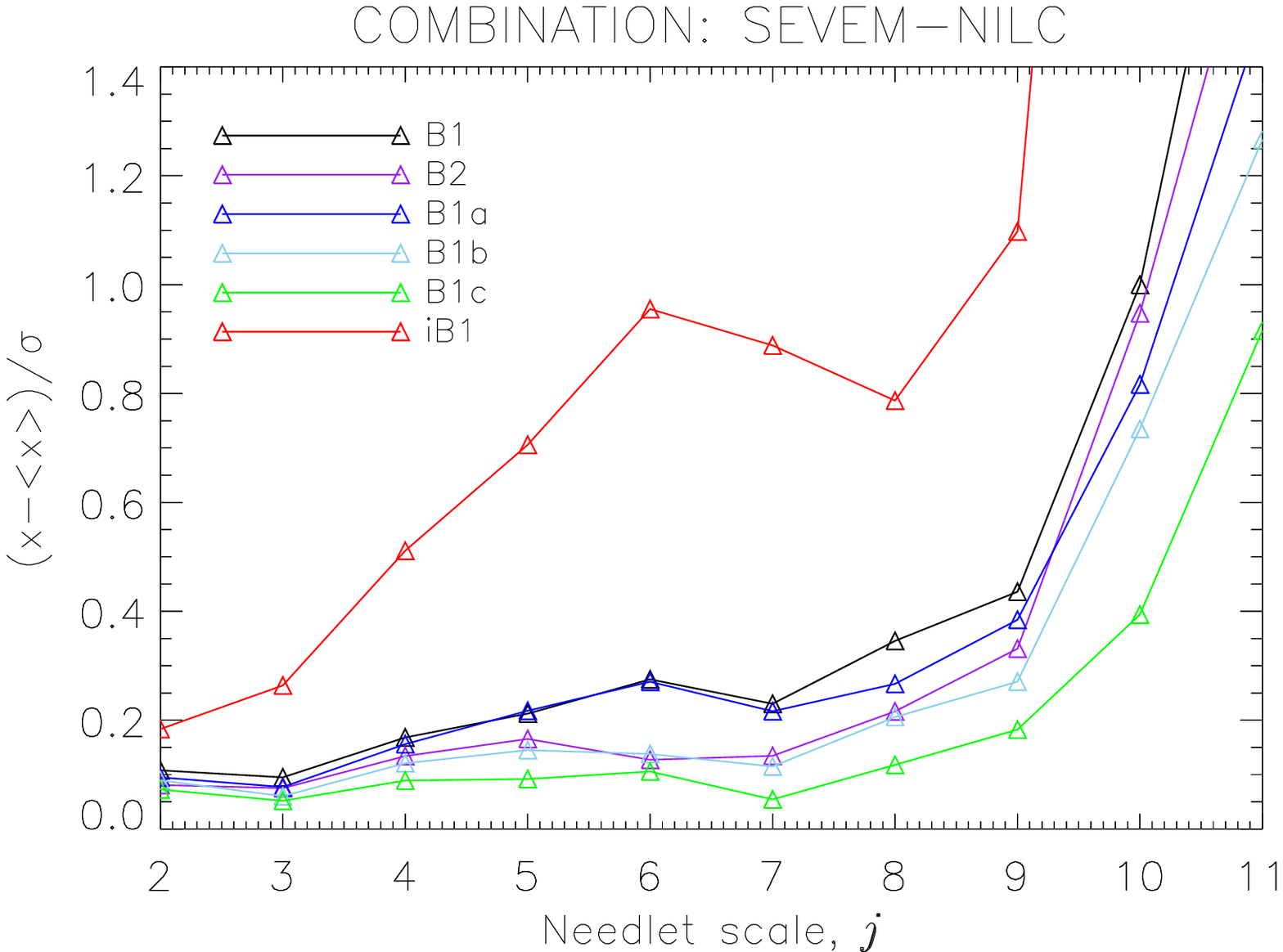}
\includegraphics[width=\linewidth,angle=0]{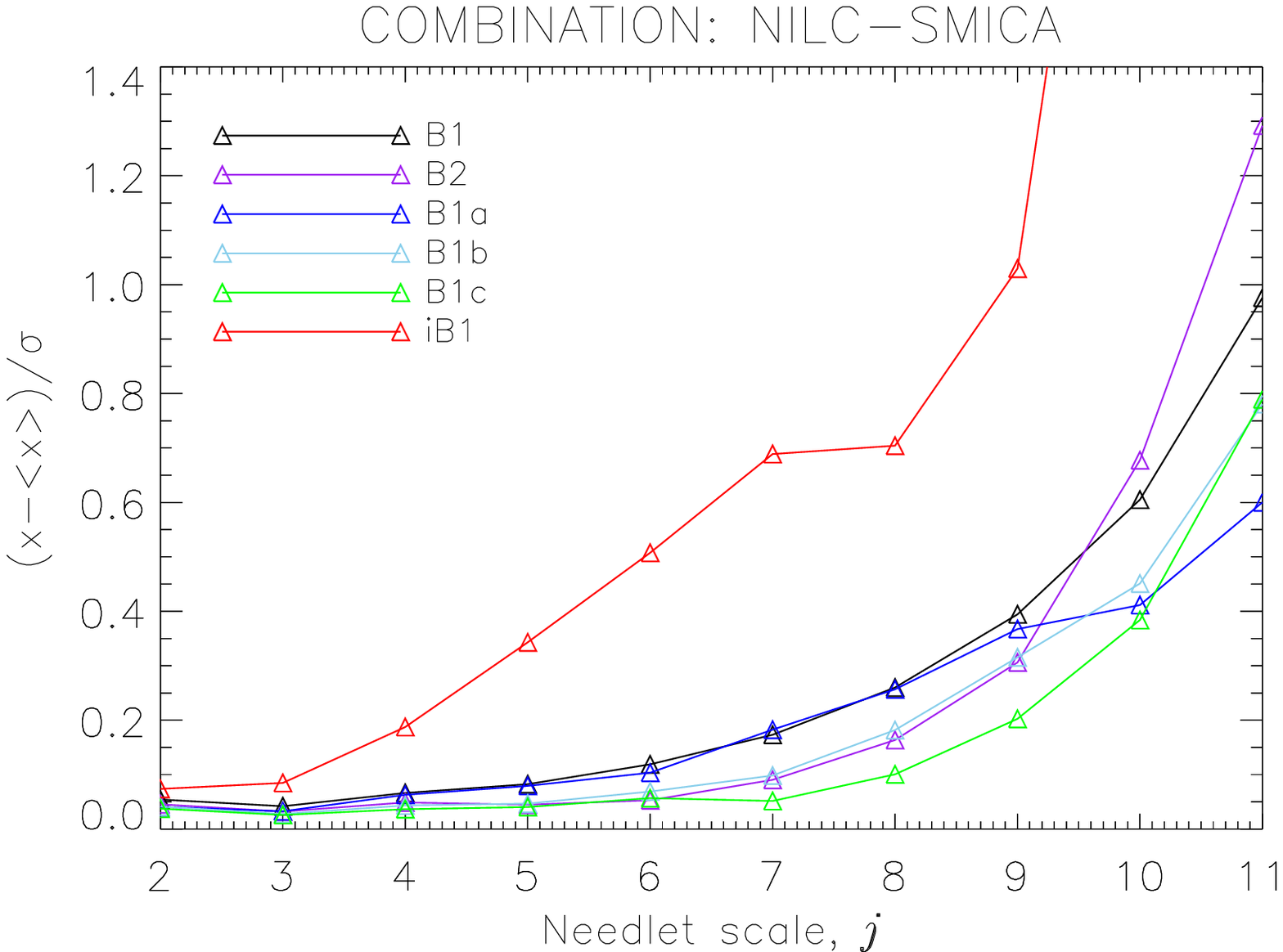}
\includegraphics[width=\linewidth,angle=0]{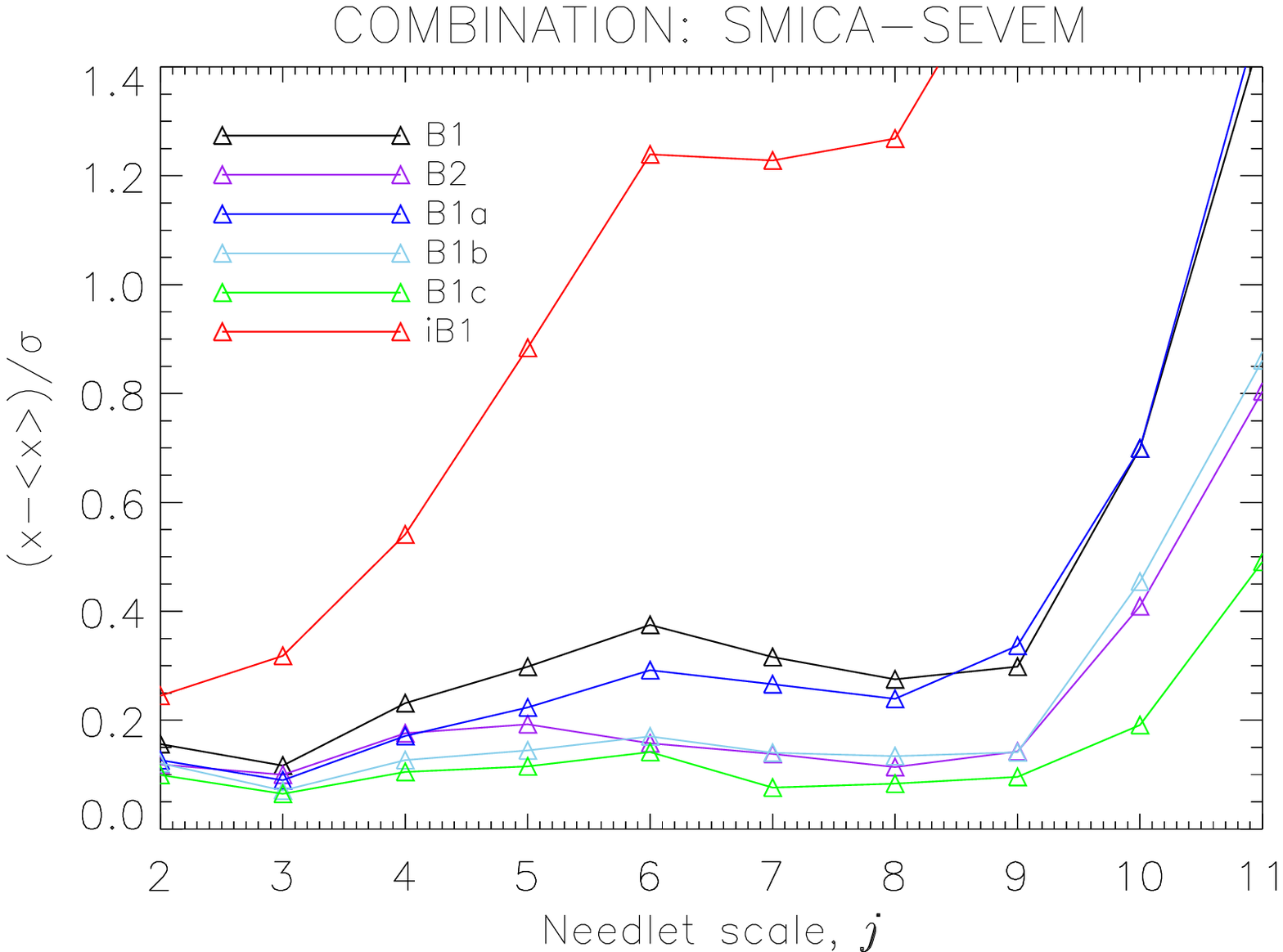}
\caption{\emph{From top}:  $(x-\langle x\rangle)/\sigma_\mathrm{CMB}$ for \csevem-\cnilc, \cnilc-\csmica,
  and \csmica-\sevem for variance after applying the extended mask. We show results on the full bands 1 and 2 using the old U73 mask. Band 1 has been divided into B1a, B1b and B1c, We show results for these smaller bands as well as for the first band (iB1) inside the U73 mask after the mask extension described in the text has been applied. 
\label{chivals2}}
\end{figure}

\begin{figure}[!t]
\includegraphics[width=\linewidth,angle=0]{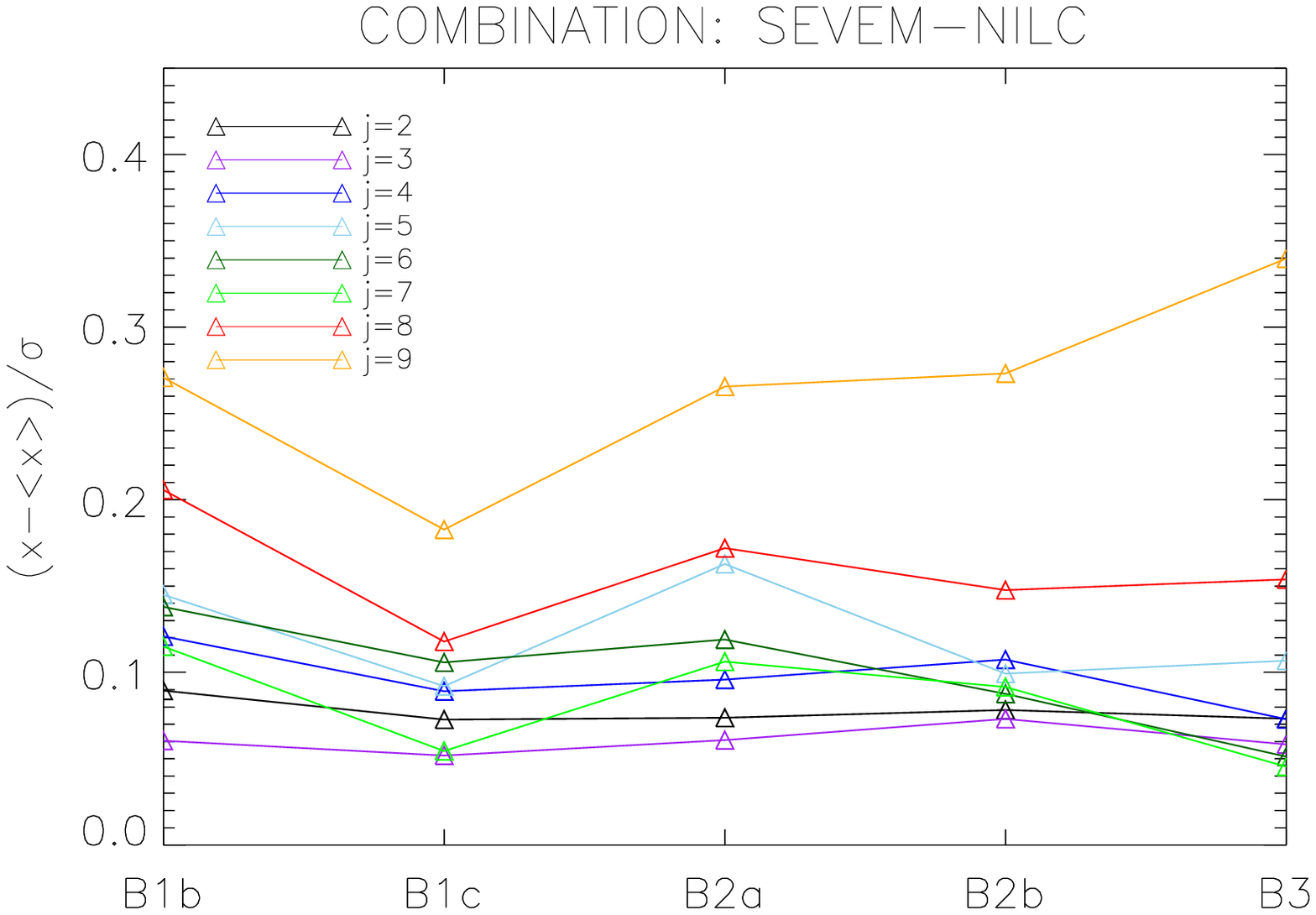}
\includegraphics[width=\linewidth,angle=0]{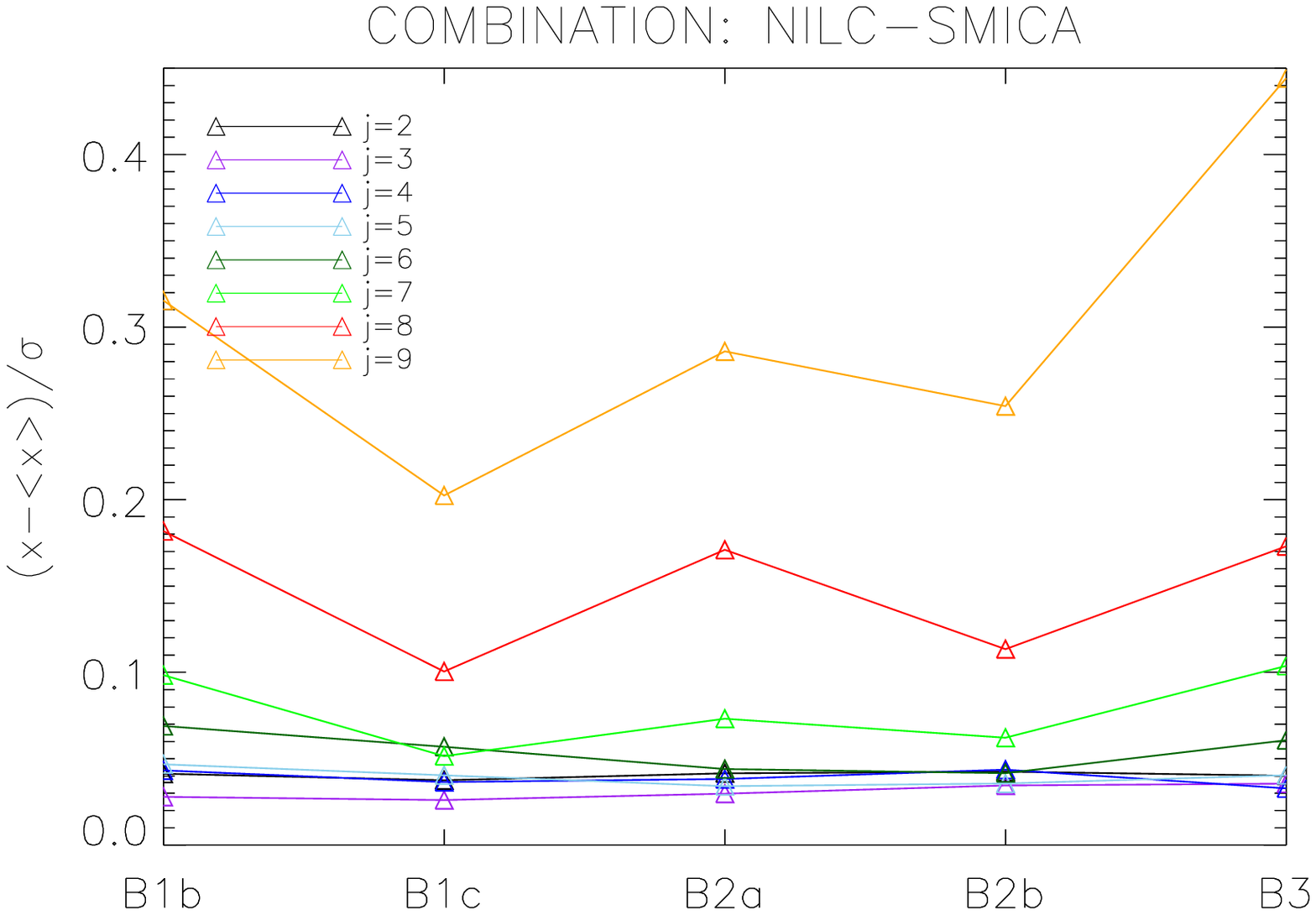}
\includegraphics[width=\linewidth,angle=0]{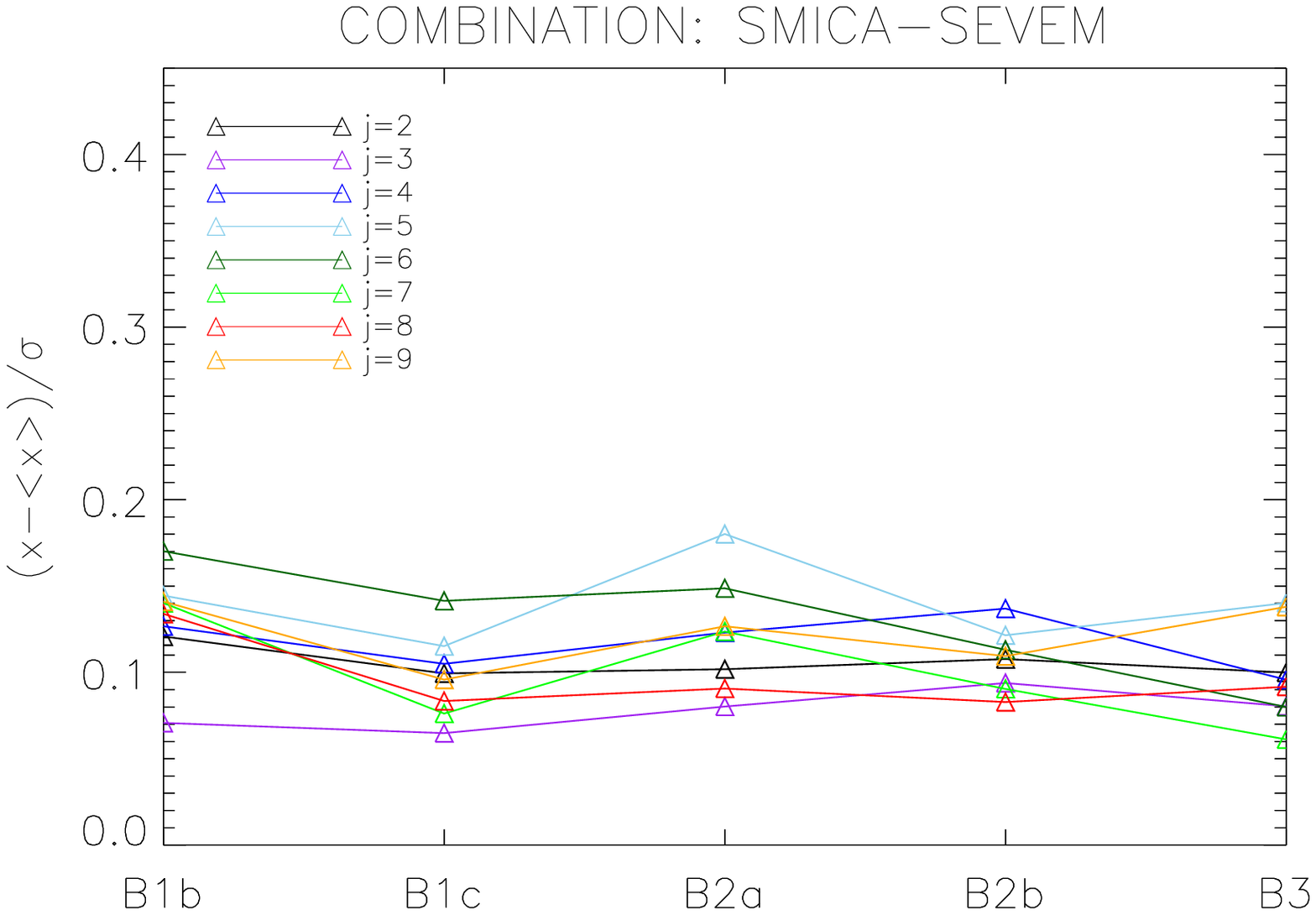}
\caption{Same as \fig\ref{chivals2} but now plotted with the band number on the x-axis and with color codes indicating needlet scales. Band 2 has been
  divided into B2a and B2b. In this figure we only show results based on the new extended mask.
 \label{chivals2_j}}
\end{figure}

As it turns out that large regions of band 1 are removed in the extended mask, we found that these bands needed a further subdivision into bands 
1a, 1b and 1c, where band 1a lies closest to the galactic plane, and band 1c lies farthest away from it.  
We use these smaller bands in order to test the results with the extended mask close to the borders of the U73 mask.  In \fig\ref{chivals2} and \ref{chivals2_j} we show results on the variance of needlet coefficients from the analysis with the extended mask. Included in the plots are results from analysis on the first band inside the U73 mask, defined in figure \ref{inside}, and labeled iB1. Pixels analyzed on the inside bands have undergone the same mask extension procedure, as the bands on the outside of U73. After this mask extension only a sky fraction of $2.2\%$ of the original $4.1\%$ remains in band iB1. Still, we clearly see from the figure that this band is unsuitable for cosmological analysis, the same conclusion is valid for all five inside bands. Note that in \fig\ref{chivals2} we show the full band 1 and 2 analyzed with the U73 mask whereas bands 1a,b,c and iB1 were analysed with the extended mask. We find that band 1a has to be fully discarded in order to achieve residuals $<0.2\sigma_\mathrm{CMB}$ for the large scales whereas bands 1b and 1c can be kept with this new extended mask. From \fig\ref{chivals2_j} we can see how the new extended mask has removed the increase in variance towards the galactic plane. In fact, only in band 1b are there signs of an increase, but it is well below $<0.2\sigma_\mathrm{CMB}$. Also notice that band 2 has been subdivided
into bands 2a and 2b, as done previously with band 1, in order to examine if band 2 may be fully used with the new mask. Band 2a lies closest to the 
galactic plane while band 2b lies farthest away. From the plots shown we conclude that entire band 2 may be kept. 

 We have thus arrived at a further extended mask which equals the mask obtained above but with the further extension that all pixels in band 1a are set to zero. This new mask gives satisfactory results for all measures used in this paper allowing $f_{\mathrm{sky}} = 65.9\%$ of the sky for cosmological analysis.

\section{Point source extensions}
\label{ps}

\begin{figure}[t]
  \includegraphics[width=0.65\linewidth,angle=90]{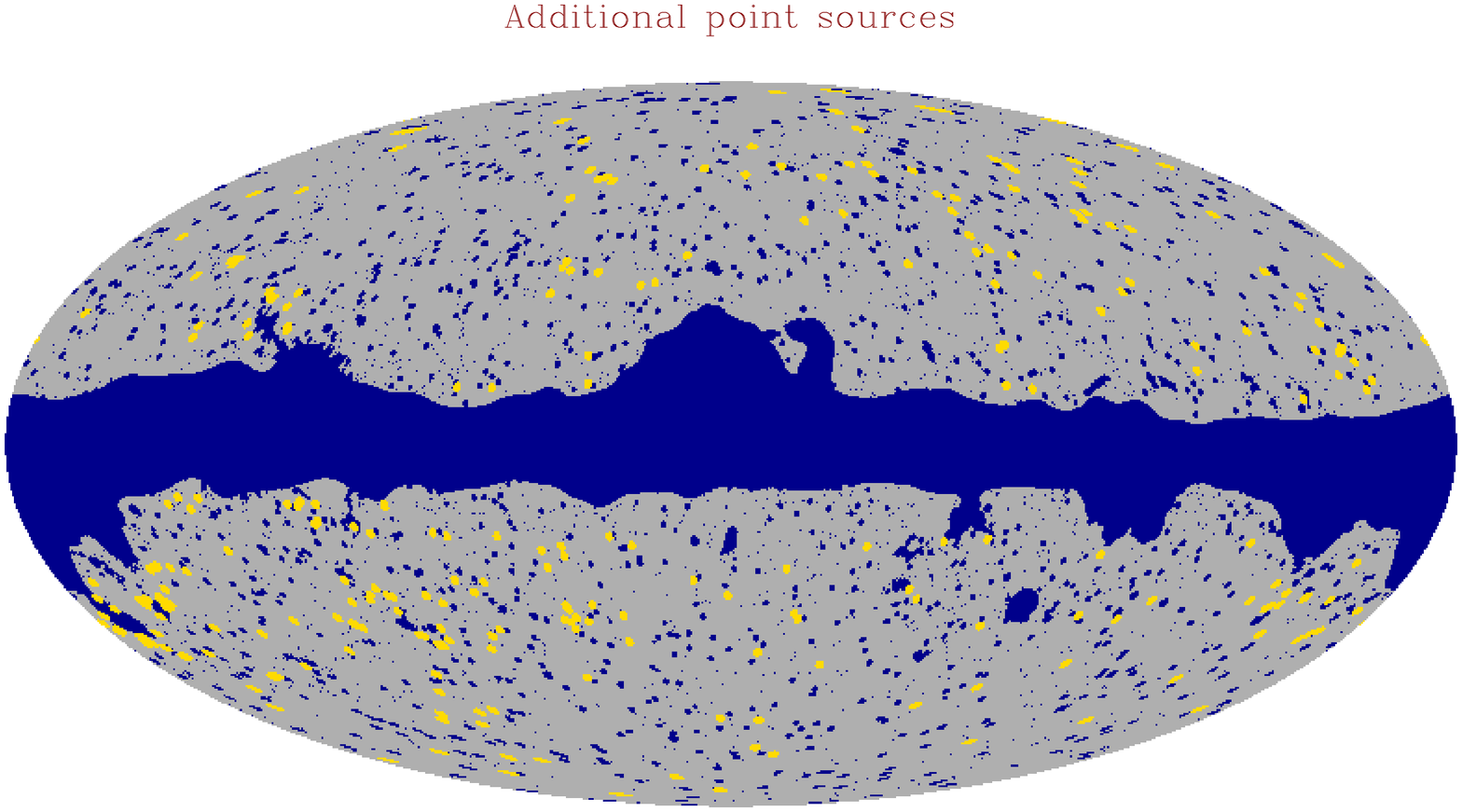}
%%\mbox{\epsfig{figure=figs/srcmask.ps,angle=90,width=\linewidth,clip=}}
  \caption{The U73 mask with the 276 new point sources indicated by large discs for illustration, the actual holes
    are much smaller. \label{src}}
\end{figure}

\begin{figure}[t]
\includegraphics[width=0.65\linewidth,angle=90]{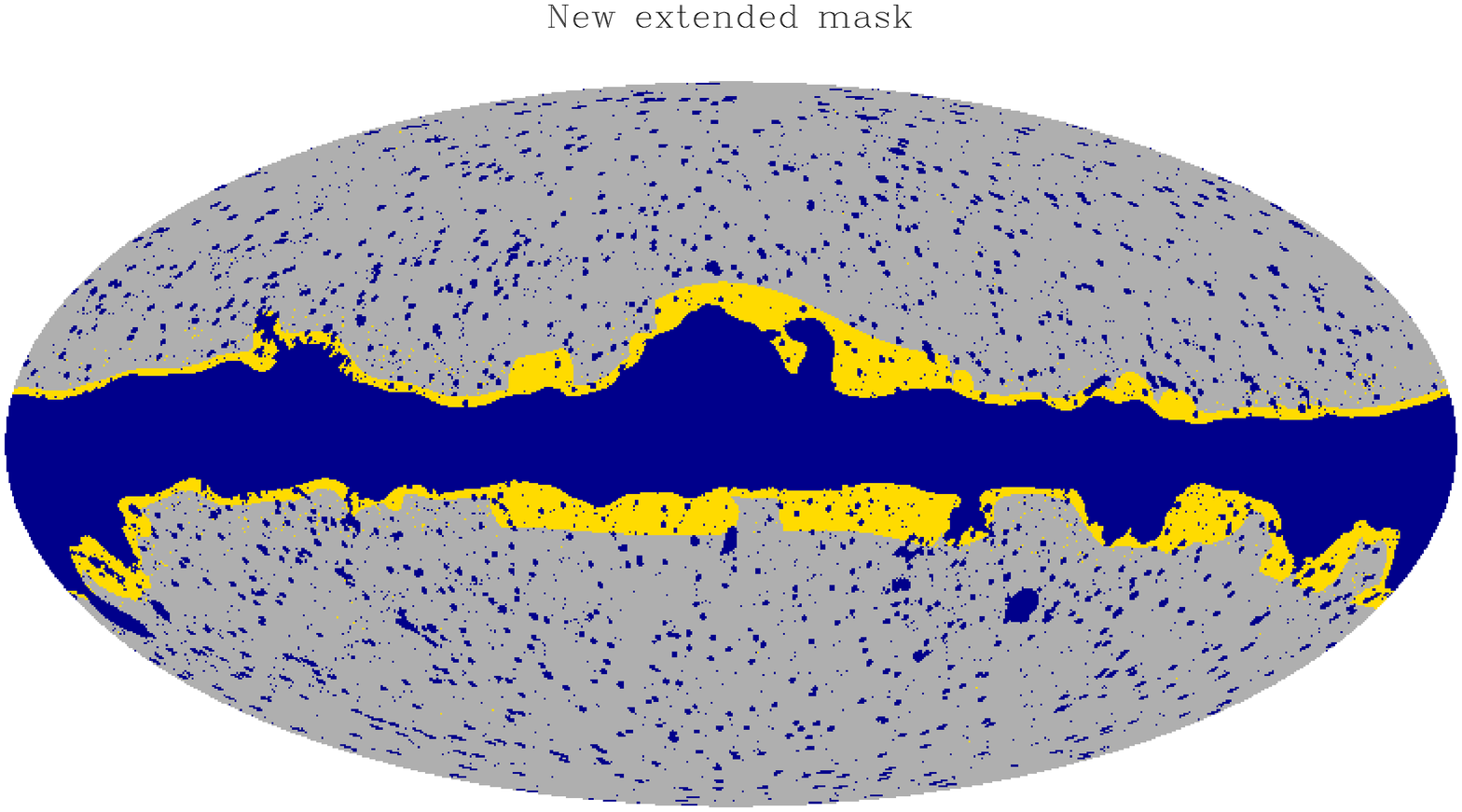}
%%\mbox{\epsfig{figure=figs/newmask.ps,angle=90,width=\linewidth,clip=}}
  \caption{New extended U66 mask (yellow) with the U73 mask (blue). \label{finalmask}}
\end{figure}

We have seen in the previous plots that unresolved point sources give rise to large discrepancies between the methods on smaller angular scales. We cannot do much to remove the unresolved sources, but we will check if there are sources left in the difference maps which can be resolved and therefore masked.

Following the approach in \cite{scodeller}, we find 276 point sources at $>5\sigma$ in the difference maps. These point sources include only those which are not already masked by the above described extended U73 mask. Many of these are common to several difference map combinations, others are detected only in one combination but is present but slightly below the detection limit in others. We include point source holes with radius 0.1 degrees for all these sources in our extended U73 mask. The final extended mask now has a usable sky fraction of $65.9\%$.  In \fig\ref{src} we show the position of these 276 sources and in \fig\ref{finalmask} we show the final extended U66 mask which we have made publicly 
available\footnote{http://folk.uio.no/frodekh/PS\_catalogue/ \\ planck\_extended\_mask.fits}.

\section{Conclusions}
\label{conclusion}
In this work, the \csmica, \csevem, \nilc and \ruler foreground cleaned \Planck data maps have been compared to 
simulated data. It is known that the current maps are recommended for joint cosmological analysis up to $\ell_{\mathrm{max}}=1500$, but
for smaller scales, the complex foregrounds and noise properties of the maps are not yet fully understood. We have therefore limited our study to $\ell<1500$.

The aim of this work was to test for foreground residuals in the cleaned maps outside and inside the U73 mask, checking whether the U73 mask needs extension or if it can be made smaller and still be suitable for cosmological analysis. We divided the sky outside U73 into 7 bands north and south of the galactic equator and tested for foreground residuals in these bands by analyzing their local power spectra as well as mean, variance and skewness of needlet coefficients at several scales. We performed this test, both on the individual foreground cleaned maps as well as on difference maps constructed from pairs of these maps. We found that in particular the variance of needlet coefficients on difference maps was highly sensitive to residuals.

Based on the needlet variance test, we found that all the difference maps where the \ruler foreground cleaned map was present, the differences to the other maps were so large that we decided to exclude the \ruler map from further analysis. Even with a highly extended mask we were unable to make the \ruler map agree with the other maps at a satisfactory level. Note that this difference was not seen in the full maps including CMB, only in the difference maps, and only at a level of 0.3 to 0.4 CMB standard deviations. This may influence some cosmological analyses, but is too small to significantly influence \eg the power spectrum. Note however that it is not clear whether this difference comes from large residuals in the \ruler map or in the other three maps.

 The other three methods however were found to agree with differences less than $20\%$ of the standard deviation of the CMB over most scales after an extended U73 mask was applied. This extended U73 map was constructed by removing pixels where the needlet coefficients were found to be higher than a certain scale dependent threshold. Analysis of bands inside the U73 mask revealed such high levels of foreground contamination that we can confirm that areas which are currently masked by U73 cannot be reliably used for cosmologial analysis. Our extended mask was finally further extended by point source holes for point sources detected in the difference maps. 276 point sources which are not masked in U73 were detected in the difference maps. Our final extended U66 mask, including point source holes for the additional sources has a usable sky fraction of $65.9\%$ and is publicly available (see previously specified url). We recommend the use of this mask rather than the U73 mask for cosmological analysis of the foreground cleaned \Planck maps, in particular for analyses which are performed on smaller patches on the sky rather than on the full sky. The effect of the foreground residuals which we detected outside the U73 mask, mostly close to the mask borders, will probably be small for any CMB analysis using the full sky. We also point out that we did not detect these residuals in the individual foreground cleaned maps, only in the differences between these.

We further note that the method presented here can easily be extended to polarization. We have seen that simply taking the product of the individual method specific masks does not necessarily yield a common mask which masks all residuals. By using the differences between the cleaned maps we can extend this simple common mask according to the desired acceptance level of foreground residuals. This may be of even higher importance for the soon-to-be-released \Planck polarization maps as the properties of polarized foregrounds are much less known than for temperature.

\begin{acknowledgements}
Maps and results have been derived using the Healpix\footnote{http://healpix.jpl.nasa.gov} software 
package developed by \cite{Gorski:2004by}. 
This work was performed on the Abel Cluster, owned by the University of Oslo and the Norwegian metacenter
for High Performance Computing (NOTUR), and operated by the Department for Research Computing at USIT,
the University of Oslo IT-department, http://www.hpc.uio.no/. 
This work is based on observations obtained with \Planck (http://esa.int/Planck), an ESA science mission with instruments and contributions
directly funded by ESA Member States, NASA, and Canada. The development of \Planck has been supported by: ESA; CNES and CNRS/INSU-IN2P3-INP (France);
ASI, CNR, and INAF (Italy); NASA and DoE (USA); STFC and UKSA (UK); CSIC, MICINN and JA (Spain); Tekes, AoF and CSC (Finland); DLR and MPG (Germany);
CSA (Canada); DTU Space (Denmark); SER/SSO (Switzerland); RCN (Norway); SFI (Ireland); FCT/MCTES (Portugal); and PRACE (EU). We acknowledge the use of 
the Planck Legacy Archive.
\end{acknowledgements}

\bibliography{prefs}

\end{document}